\documentclass[iop,numberedappendix]{./emulateapj}
\usepackage{amssymb,amsmath,amsthm}
\usepackage{longtable}
\usepackage{array,multirow,graphicx}
\usepackage{xspace}
\usepackage{tabularx}
\usepackage{graphics,graphicx} 
\usepackage{rotating}
\renewcommand{\thefootnote}{\arabic{footnote}}

\usepackage{color}
\usepackage{booktabs}
\usepackage{sidecap}
\usepackage{color}
\newcommand{\xmm}{{\sl XMM-Newton}\xspace}

\newcommand{\esas}{{\sl XMM-ESAS}\xspace}
\newcommand{\sas}{{\sl SAS}\xspace}
\newcommand{\xspec}{{\sl XSPEC}\xspace}

\newcommand{\asca}{{\sl ASCA}\xspace}
\newcommand{\astroh}{{\sl Astro-H}\xspace}

\newcommand{\chandra}{{\sl Chandra}\xspace}
\newcommand{\suzaku}{{\sl Suzaku}\xspace}
\newcommand{\apec}{{\sl apec}\xspace}
\def\apj{\texttt{ApJ}}
\def\apjs{\texttt{ApJS}}
\def\apjl{\texttt{ApJL}}
\def\prd{\texttt{Phys. Rev.}}

\def\physrep{\texttt{PhysRep}}
\def\mnras{\texttt{MNRAS}}
\def\araa{\texttt{ARAA}}
\def\aap{\texttt{A\&A}}
\def\jcap{\texttt{JCAP}}

\renewcommand{\tabcolsep}{0.2cm}
   
\submitted{Submitted to ApJ, 2014 February 10, Accepted 2014 April 28}
\begin{document}


\title{Detection of An Unidentified Emission Line
in the Stacked X-ray spectrum of Galaxy Clusters }


\author{
Esra~Bulbul\altaffilmark{1,2},
Maxim~Markevitch\altaffilmark{3},
Adam~Foster\altaffilmark{1},
Randall~K.~Smith\altaffilmark{1}
Michael~Loewenstein\altaffilmark{2,4},
and Scott~W.~Randall\altaffilmark{1}
}
\affil{$^1$ Harvard-Smithsonian Center for Astrophysics, 60 Garden Street, Cambridge, MA, USA\\
$^2$ CRESST and X-ray Astrophysics Laboratory, NASA Goddard Space Flight Center, Greenbelt, MD, USA\\
$^3$ NASA Goddard Space Flight Center, Greenbelt, MD, USA\\
$^4$ Department of Astronomy, University of Maryland, College Park, MD, USA\\
}
\email{ebulbul@cfa.harvard.edu}

\begin{abstract}
We detect a weak unidentified emission line at $E=(3.55-3.57)\pm0.03$ keV in
a stacked \xmm spectrum of 73 galaxy clusters spanning a redshift range
$0.01-0.35$. MOS and PN observations independently show the presence of the
line at consistent energies. When the full sample is divided into three
subsamples (Perseus, Centaurus+Ophiuchus+Coma, and all others), the line is
seen at $>3\sigma$ statistical significance in all three independent MOS
spectra and the PN ``all others'' spectrum. The line is also detected at the
same energy in the \chandra\ ACIS-S and ACIS-I spectra of the Perseus
cluster, with a flux consistent with \xmm\ (however, it is not seen in the
ACIS-I spectrum of Virgo). The line is present even if we allow maximum
freedom for all the known thermal emission lines. However, it is very weak
(with an equivalent width in the full sample of only $\sim 1$ eV) and
located within 50--110 eV of several known faint lines; the detection is at
the limit of the current instrument capabilities and subject to significant
modeling uncertainties. On the origin of this line, we argue that there
should be no atomic transitions in thermal plasma at this energy. An
intriguing possibility is the decay of sterile neutrino, a long-sought dark
matter particle candidate. Assuming that all dark matter is in sterile
neutrinos with $m_s=2E=7.1$ keV, our detection in the full sample
corresponds to a neutrino decay mixing angle $\sin^2(2\theta)\approx 7\times
10^{-11}$, below the previous upper limits. However, based on the cluster
masses and distances, the line in Perseus is much brighter than expected in
this model, significantly deviating from other subsamples. This appears to
be because of an anomalously bright line at $E=3.62$ keV in Perseus, which
could be an Ar \textsc{xvii} dielectronic recombination line, although
its emissivity would have to be 30 times the expected value and 
physically difficult to understand. In principle, such an anomaly might
explain our line detection in other subsamples as well, though it would
stretch the line energy uncertainties. Another alternative is the
above anomaly in the Ar line combined with the nearby 3.51 keV K line also
exceeding expectation by a factor 10--20. Confirmation with \chandra and
\suzaku, and eventually \astroh, are required to determine the nature of
this new line. (APJ HAS THE ABRIDGED ABSTRACT)

\end{abstract}

\section{Introduction} 

Galaxy clusters are the largest aggregations of hot intergalactic gas and
dark matter. The gas is enriched with heavy elements
(\citet{mitchell76,serlemitsos77} and later works) that escape from galaxies
and accumulate in the intracluster/intergalactic medium (ICM) over billions
of years of galactic and stellar evolution. The presence of various heavy
ions is seen from their emission lines in the cluster X-ray spectra. Data
from large effective area telescopes with spectroscopic capabilities, such
as \asca, \chandra, \xmm\ and \suzaku, uncovered the presence of many
elements in the ICM, including O, Ne, Mg, Si, S, Ar, Ca, Fe, and Ni
\citep[for a review see, e.g.,][]{bohringer2010}. Recently, weak
emission lines of low-abundance Cr 
and Mn were discovered \citep{werner2006,tamura2009}. Relative abundances of
various elements contain valuable information on the rate of supernovae of
different types in galaxies \citep[e.g.,][]{loewenstein2013} and illuminate the enrichment
history of the ICM \citep[e.g.,][]{bulbul2012b}. Line ratios of various ions can also provide
diagnostics of the physical properties of the ICM and uncover the presence of
multi-temperature gas, nonequilibrium ionization states, and nonthermal
emission processes such as charge exchange \citep[CX,][]{paerels03}.

As for dark matter, 80 yr from its discovery by \citep{zwicky1933,zwicky1937}, its
nature is still unknown (though now we do know for sure it exists, from
X-ray and gravitational-lensing observations of the Bullet Cluster;
\citet{clowe06}, and we know accurately its cosmological abundance, e.g.,
\citet{hinshaw13}). Among the various plausible dark matter candidates, one
that has motivated our present work is the hypothetical sterile neutrino
that is included in some extensions to the standard model of particle
physics (\citet{dodelson1994} and later works; for recent reviews see, e.g.,
\citet{abazajian2007,boyarsky2009}). Sterile neutrinos should decay
spontaneously with the rate
\begin{equation}
  \Gamma_\gamma(m_s,\theta) = 1.38\times 10^{-29}{\rm\ s}^{-1}\ 
  \left(\frac{\sin^2 2\theta}{10^{-7}}\right)
  \left(\frac{m_s}{1\rm\ keV}\right)^5,
\label{eq:rate}
\end{equation}
where the particle mass $m_s$\/ and the ``mixing angle'' $\theta$\/ are
unknown but tied to each other in any particular neutrino production model
\citep{pal1982}. The decay of sterile neutrino should produce a photon of
$E=m_s/2$ and an active neutrino. The mass of the sterile
  neutrino may lie in the keV range, which would place its decay line in the
  range accessible to X-ray observations of dark matter aggregations, such
  as clusters of galaxies, nearby galaxies, and the Milky Way 
    \citep{abazajian2001a,abazajian2001b}. So far, searches in
  various types of massive systems have resulted only in upper limits
  \citep[e.g.,][]{boyarsky2012, abazajian2012}.

Current X-ray archives of \xmm, \chandra\ and \suzaku\ contain vast
collections of galaxy cluster observations. Mining these databases can
result in significant improvement in sensitivity to faint spectral features
compared to individual cluster observations (as proposed, e.g.,
  by \citet{abazajian2001b}), with respect to both the statistical and (in
a less obvious way) systematic or instrumental uncertainties. In this paper,
we undertake a fishing expedition that combines the spectra of many bright
clusters from the \xmm\ archive in order to search for any kind of faint,
unidentified X-ray emission lines --- be they thermal lines from previously
undetected rare elements in the ICM or the elusive sterile neutrino decay
line.

To improve the sensitivity to weak spectral lines and reduce systematic
effects, we stack the X-ray spectra from clusters at different redshifts in
their rest frame, rescaling the photon energies to $z=0$. After
blueshifting each cluster spectrum to $z=0$, any background lines or
instrumental response artifacts are smeared out (since they occur in the
detector frame), but any weak line intrinsic to clusters would be amplified
and may become detectable in the combined spectrum. In this paper, we use
this method to detect a previously unknown, extremely faint emission line at
$E\approx 3.55-3.57$ keV. The line was detected in the stacked \xmm X-ray
spectra of 73 bright galaxy clusters in the redshift range $0.01< z< 0.35$,
and independently in several subsamples. The redshift range in
  our sample corresponds to an energy difference between 0 and 1.2 keV at
  the energy of the line of $E\approx 3.5$ keV. In the blueshifted
frame, the response features, background, and instrumental lines are 
smeared by this line shift, while the detected line appears at the same energy, making
it possible to detect the 3.57 keV line with high confidence.

This paper is organized as follows. Sections 2 and 3 describe the \xmm data
processing, background modeling, and spectra stacking methods. We also
determine the best-fit flux and energy of the detected spectral feature
using the \xmm MOS and PN instruments. In Section 4, we confirm the existence of
this spectral line in the \chandra ACIS-I and ACIS-S spectra of the Perseus
cluster, and obtain an upper limit from the ACIS-I observations of
the Virgo Cluster. In Section 5, we discuss the possible interpretations of this new emission line. All
errors quoted throughout the paper correspond to 68\% (while in parentheses,
90\%) single-parameter confidence intervals; upper limits are at 90\%
confidence, unless otherwise stated.

\section{Cluster Sample Selection and Data Analysis}
\label{sec:analysis}
 
\subsection{Sample Selection }
\label{sec:sample}
 
\begin{figure}[ht!]
\centering
\vspace{3mm}
\hspace{-4mm}\includegraphics[width=8.9cm, angle=0]{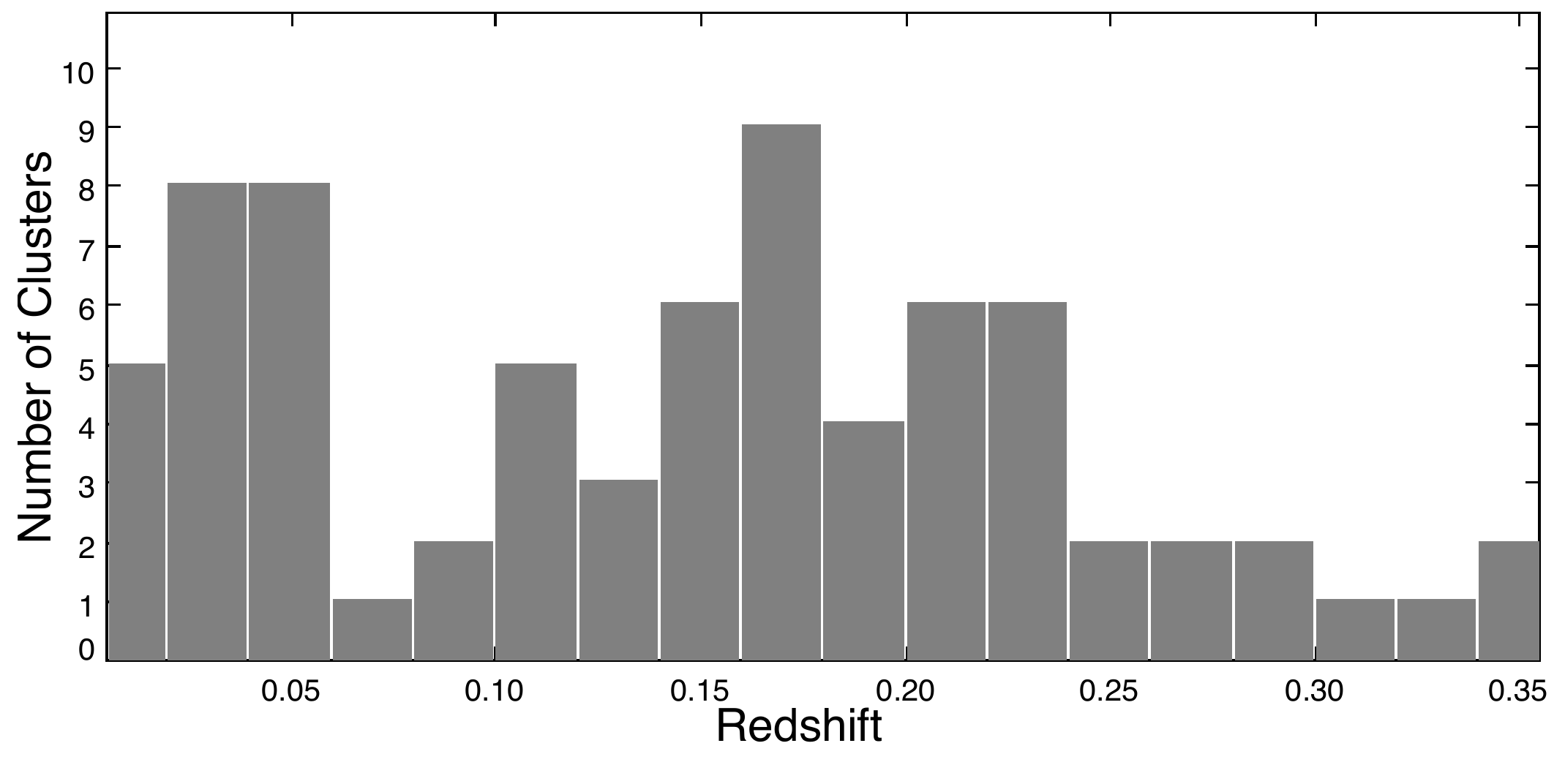}
\caption{Redshift histogram of the total of 73 galaxy clusters in the sample, selected from the \xmm archive.}
\label{fig:hist}
\end{figure}

We searched the \xmm archive for galaxy cluster observations that yielded
large numbers of X-ray counts. We first selected clusters below a redshift
of 0.4; higher-redshift clusters are too faint to contribute
significantly to the stacked spectrum. 
We then calculated the total X-ray counts expected from these \xmm
observations using the \textit{ROSAT} count rates reported in eBCS
\citep{ebeling2000}, NORAS \citep{boehringer2000}, REFLEX
\citep{boehringer2004}, XBACs \citep{ebeling1996}, and MACS catalogs
\citep{ebeling2001} and \xmm exposures. To prevent nearby clusters from
dominating the stacked spectrum, we used different cluster count limits for
different redshift ranges. We chose clusters with a minimum of $10^{5}$
counts per cluster for clusters with $z < 0.1$, and $10^{4}$ counts per
cluster for clusters with redshifts $0.1< z < 0.4$, to have a wide enough
range for the redshift-smearing effect.  Offset pointings were
excluded from the sample.  In the end, a sample of 73 clusters was
selected. Included in Table \ref{table:sample} 
are the \xmm observation identification (ObsID) numbers, total MOS and PN
clean exposure times, count rates, and our best-fit redshifts (see
Section \ref{sec:analysis1}). The redshift histogram of the sample is given in
Figure \ref{fig:hist}. The count rates reported in Table \ref{table:sample} have been used
only for sample selection.
%

\subsection{XMM-Newton Data Analysis}
\label{sec:analysis1}
 
The EPIC data processing and background modeling were carried out with the
\xmm Extended Source Analysis Software 
\citep[\esas;][]{kuntz2008,snowden2008}. We reduced MOS and PN data with the \xmm
Science Analysis System (\sas) version 12. Our \xmm analysis is described
fully in \citet{bulbul2012a,bulbul2012b}; here we give relevant details.

The light-curve filtering was applied to
eliminate periods of elevated background. Cleaned events files were created
using the good time interval file (GTI) produced by this process. The net
exposure time after filtering the event files for good time intervals is
given in Table \ref{table:sample}.

Images were created in the $0.4-7.0$ keV band for MOS and PN
observations and used for detecting point sources with the CIAO tool
\textit{wavdetect}.\ The images 
were examined carefully for any missed point sources, as well as for individual
MOS CCDs operating in an anomalous state.\ The CCDs in an
anomalous state and all point sources were excluded from further analysis.

{
{\renewcommand{\arraystretch}{1.2}
\begin{longtable*}{llllcccc}
\caption{\xmm Galaxy Cluster Sample \label{table:sample}}\\
\hline \hline
 \multicolumn{1}{l}{\textbf{Cluster}} & \multicolumn{1}{c}{\textbf{ RA }} & \multicolumn{1}{c}{\textbf{Decl.}}& \multicolumn{1}{c}{\textbf{ObsID }} & \multicolumn{1}{c}{\textbf{ \xmm}}  & \multicolumn{1}{c}{\textbf{ \xmm}} & \multicolumn{1}{c}{\textbf{ Count Rate  }} & \multicolumn{1}{c}{\textbf{ Best-fit }} \\ 

\multicolumn{1}{c}{\textbf{}} & \multicolumn{1}{c}{\textbf{ }} & \multicolumn{1}{c}{\textbf{}}& \multicolumn{1}{c}{\textbf{ }} & \multicolumn{1}{c}{\textbf{MOS Exp}} &\multicolumn{1}{c}{\textbf{PN Exp }} & \multicolumn{1}{c}{\textbf{(ROSAT)}} & \multicolumn{1}{c}{\textbf{ Redshift }}\\

\multicolumn{1}{c}{\textbf{}} & \multicolumn{1}{c}{\textbf{ }} & \multicolumn{1}{c}{\textbf{}}& \multicolumn{1}{c}{\textbf{ }} & \multicolumn{1}{c}{\textbf{(ks)}} &\multicolumn{1}{c}{\textbf{(ks)}} & \multicolumn{1}{c}{\textbf{ (cnts s$^{-1}$ )}} & \multicolumn{1}{c}{\textbf{  }}\\ \hline 

\endfirsthead

\multicolumn{3}{l}
{{\bfseries \tablename\ \thetable{} -- continued from previous page}} \\\\
\hline
 \multicolumn{1}{c}{\textbf{Cluster}} & \multicolumn{1}{c}{\textbf{ RA }} & \multicolumn{1}{c}{\textbf{Decl.}}& \multicolumn{1}{c}{\textbf{ObsID }} & \multicolumn{1}{c}{\textbf{\xmm}}& \multicolumn{1}{c}{\textbf{  \xmm  }}  & \multicolumn{1}{c}{\textbf{ Count Rate  }} & \multicolumn{1}{c}{\textbf{ Best-fit }}\\
 
 \multicolumn{1}{c}{\textbf{}} & \multicolumn{1}{c}{\textbf{ }} & \multicolumn{1}{c}{\textbf{}}& \multicolumn{1}{c}{\textbf{ }} & \multicolumn{1}{c}{\textbf{MOS Exp}} & \multicolumn{1}{c}{\textbf{PN Exp}}& \multicolumn{1}{c}{\textbf{}} & \multicolumn{1}{c}{\textbf{Redshift  }}\\ 
 
 \multicolumn{1}{c}{\textbf{}} & \multicolumn{1}{c}{\textbf{ }} & \multicolumn{1}{c}{\textbf{}}& \multicolumn{1}{c}{\textbf{ }} & \multicolumn{1}{c}{\textbf{(ks)}} &\multicolumn{1}{c}{\textbf{(ks)}} & \multicolumn{1}{c}{\textbf{ (cnts s$^{-1}$ )}} & \multicolumn{1}{c}{\textbf{  }}\\ \hline 

 \hline 
\\
\endhead
\\
\hline \multicolumn{1}{r}{{Continued on next page}} \\ \hline
\endfoot
\hline \hline
\endlastfoot

\hline\hline
\\

Centaurus       & 12: 48: 43.06 & -41.0: 18: 42.4& 0046340101& 292.0 & 101.9 & 8.8   & 0.009\\
			&			&		& 0406200101 & & && 0.009\\

A1060           & 10: 36: 46.22 & -27.0: 31: 26.9& 0206230101& 67.0 & 24.9  & 3.7 & 0.012\\

A262          & 1: 52: 45.95 & 36.0: 9: 14.2  & 0109980101  & 101.1& 25.5 & 2.5   & 0.016 \\
		&			&		        & 0504780101 &  &&&  0.016\\
		&			&		        & 0504780201 &  &&&  0.016 \\

Perseus           & 3: 19: 48.36 & 41.0: 30: 40.6 & 0085110101 & 317.3& 38.0 & 35.9 & 0.016\\
			&		 &			    & 	0305780101 && && 0.016\\

AWM7	& 2:54:32.2 	& +41:35:10 	&  0135950301 & 296.2 &109.7	& 12.7	& 0.016 \\
		&			&			& 0605540101 	& 	& &	& 0.016 \\
		
Coma      & 12: 59: 51.94 & 27.0: 57: 1.4& 0124711401	& 171.7 & 66.1& 25.8 &   0.022\\
		&			&		    & 0153750101 	& & && 0.022  \\
		&			&		   & 0300530101 	& &&& 0.022\\
		&			&		   & 0300530301 	& & && 0.022\\

A3581       & 14: 7: 29.68 & -27.0: 1: 8.3 & 0205990101	& 187.8  & 56.4 & 1.6  & 0.024  \\
			&		&			& 0504780301 & &&& 0.024 \\
			&		&			& 0504780401  & &&&  0.023 \\

Ophiuchus     & 17: 12: 23.08 & -23.0: 22: 44.7 & 0505150101	& 61.5 &15.9 & 6.2  & 0.028 \\

A4038           & 23: 47: 41.96 & -28.0: 9: 20.1 & 0204460101& 56.5 & 23.6  & 3.9  & 0.030 \\

A496            & 4: 33: 39.38 & -13.0: 15: 46.5 & 0135120201 & 250.2 & 81.1 & 4.1  & 0.033  \\
			&		&			& 0506260301	& & && 0.031 \\ 
			&		&			& 0506260401 & & && 0.032 \\ 
 
A2063           & 15: 23: 0.96  & 8.0: 36: 25.4 & 0200120401& 53.8 & 13.6  & 2.3   & 0.034   \\
   			&		&			& 0550360101 && &&  0.034 \\ 

A2147           & 16: 2: 21.62 & 15.0: 58: 35.8 &	0300350301 & 17.6& 3.3  & 3.2  & 0.035 \\
     			&		&			& 0505210601 & & && 0.035 \\ 			
			
A3571           & 13: 47: 22.74 & -33.0: 9: 54.3 & 0086950201 & 48.4 & 14.8 & 5.9  & 0.039 \\
		
A3558           & 13: 27: 57.2  & -31.0: 30: 18.7 & 0107260101& 83.0 & 33.1 & 3.5  & 0.048 \\

A4059           & 23: 57: 2.99  & -34.0: 45: 39.6 & 0109950101& 29.3 & 20.6 & 2.5  & 0.047 \\
			&		&			& 0109950201	& & && 0.047 \\ 	

Triangulum Australis& 16:38:20.3 & -64:30:59& 0093620101 &	18.4&	& 21.0	& 0.049\\

Hydra-A    & 9: 18: 0.97  & -12.0: 5: 11.0 & 0109980301 & 169.4& 50.9 & 2.6   & 0.052 \\
		&			&	            & 0504260101 && && 0.052 \\

A754	& 9:8:50.1		& -09:38:12	&0136740101	& 26.8	& 11.0 & 7.5	& 0.054\\

A2319       & 19: 21: 10.08 & 43.0: 55: 52.6 & 0302150101& 159.4 & 50.3 & 4.7  & 0.055  \\
		&			&			& 0302150201 &  &&&0.054 \\
		&			&			& 0600040101 & & && 0.054\\
		
Cygnus A	&19:59:30.0	& +40:44:00	& 0302800201	& 35.5 & 14.1 & 9.3 	&0.056\\		
		
AS1101          & 23: 13: 59.16 & -42.0: 43: 43.8 & 0123900101 & 167.1 & 62.5  &1.2 & 0.055  \\
   			&		&			& 0147800101 	 &  &&& 0.056\\ 
   			
			 
A3112       & 3: 17: 57.44 	& -44.0: 14: 12.8 &0105660101 & 358.3 & 109.7 & 2.3  & 0.074 \\
		&			&			& 0603050101 &  & && 0.075\\
		&			&			& 0603050201 & & && 0.075 \\

A2597           & 23: 25: 17.98 & -12.0: 6: 23.8 &	0108460201 & 29.9 & 6.3 & 1.6  & 0.081\\
			&		&			& 0147330101 	  & & && 0.081\\ 

A478            & 4: 13: 30.29 & 10.0: 28: 5.7 & 0109880101  & 123.0 & 43.1 & 1.6 & 0.083   \\

PKS0745-19  & 7: 47: 31.6  & -19.0: 17: 32.7 &	 0105870101 & 315.6 &  5.3 & 0.9  & 0.101  \\

A2811           & 0: 42: 7.06  & -28.0: 33: 32.7 & 0404520101 & 43.8  & 16.8 & 0.8  & 0.108  \\

A2034           & 15: 10: 16.86 & 33.0: 30: 18.8 &	 0149880101 & 38.2  & 11.0 & 0.9  & 0.113 \\
   			&		&			& 0303930101 & & && 0.117\\ 
  			&		&			& 0303930201 & & &&  0.117\\
 
RXC J0616.8-4748 & 6: 16: 47.14 & -47.0: 48: 12.3& 0201901101 & 43.3 & 16.0 & 0.2 & 0.116 \\
   			&				&			& 0302610401 	& &&&  0.113\\ 
			
RXC J0145.0-5300 & 1: 45: 3.38  & -53.0: 1: 57.6 &	0201900501 & 72.1& 21.3 & 0.3   & 0.120\\
  			&				&			& 0404910201 & & && 0.120 \\ 
 			&				&			& 0404911301 & & &&  0.120\\ 

RXC J1044.5-0704 & 10: 44: 37.64 & -7.0: 4: 20.8 & 0201901501 & 47.9 &  21.3 & 0.5 & 0.130 \\

A1068           & 10: 40: 38.89 & 39.0: 57: 6.4  &	0147630101 & 38.3& 13.2  & 0.7  & 0.138 \\
 
RXC J2218.6-3853 & 22: 18: 40.22 & -38.0: 53: 50.6 & 0201903001	& 35.5 & 6.8 & 0.4 & 0.140\\
 
RXC J0605.8-3518 & 6: 5: 58.31 & -35.0: 18: 5.8& 0201901001 & 38.3 & 11.6 & 0.4 & 0.132\\
 
A1413           & 11: 55: 24.17 & 23.0: 24: 17.5 & 0112230501& 346.8 & 107.1 & 1.0  & 0.138  \\
  			&		&			& 0502690101 & & && 0.142\\ 
  			&		&			& 0502690201 & & && 0.142 \\ 
  			&		&			& 0551280101 &&&& 0.141\\ 
 			&		&			& 0551280201 & & &&  0.142 \\ 
 
 A2204           & 16: 32: 47.11 & 5.0: 34: 36.1& 0112230301 & 124.4 & 35.3 & 0.9  & 0.150  \\
 			&		&			& 0306490101 & & &&  0.152\\ 
 			&		&			& 0306490201 & & &&  0.152 \\ 
 			&		&			& 0306490301 & & && 0.150 \\ 
		 	&		&			& 0306490401 & & && 0.151 \\  

A3888           & 22: 34: 27.77 & -37.0: 44: 59.6 & 0201903101& 86.1 & 29.2 & 0.6 & 0.153 \\
  			&			&			& 0404910801 & &&& 0.154 \\ 

RXC J0958.3-1103 & 9: 58: 17.69 & -11.0: 3: 29.9 &	 0201903501& 17.6 & 2.0 & 0.4 & 0.153 \\
   			&				&			& 0404910601 & &&& 0.153\\ 

A545            & 5: 32: 20.65 & -11.0: 33: 8.0 & 0304750101 & 11.3 & 3.7 & 0.4  & 0.154    \\

RXC J2014.8-2430 & 20: 14: 46.14 & -24.0: 31: 12.5& 0201902201	& 43.1&  13.5 & 0.6  & 0.155   \\

RX J1720.1+2638  & 17: 20: 4.81 & 26.0: 37: 30.3& 0500670201 & 108.8& 29.9  & 0.8   & 0.160 \\
  			&			&			& 0500670301 & & && 0.158\\ 
  			&			&			& 0500670401 & & && 0.159\\ 

RXC J0645.4-5413 & 6: 45: 25.63 & -54.0: 12: 11.2& 0201901201 & 35.5 & 6.7 & 0.5  & 0.167 \\
  			&			&				& 0201903401 & & && 0.167\\ 
  			&			&				& 0404910401 & & && 0.167\\ 			

A1201           & 11: 12: 58.9  & 13.0: 26: 19.5 &	0500760101 & 51. 5 & 41.8 & 0.4  & 0.169 \\

A1914           & 14: 26: 7.37  & 37.0: 49: 29.9 &	0112230201 & 25.5  & 9.2 & 1.2  & 0.166 \\
  
A2345           & 21: 27: 11.16 & -12.0: 9: 21.5 &	0604740101 & 93.0  &24.6 & 0.4   & 0.180  \\

A2218           & 16: 35: 41.82 & 66.0: 11: 37.6 & 0112980101 & 98.7& 24.9  & 0.4  & 0.176   \\
  			&			&			& 0112980401 & & && 0.176\\ 
  			&			&			& 0112980501 & & && 0.176\\

A2254           & 17: 17: 41.32 & 19.0: 40: 8.0& 0601080101 & 111.7 & 39.2 & 0.3   & 0.178  \\


A665            & 8: 30: 58.0  & 65.0: 50: 20.0 & 0109890401	& 43.3 & & 0.6  & 0.182  \\
  			&		&			& 0109890501 & & && 0.180\\ 
  
A1689           & 13: 11: 33.68 & -1.0: 20: 29.7 &	0093030101 & 69.3 & 26.5 & 1.0   & 0.183  \\

A383       & 2: 48: 6.7   & -3.0: 30: 57.9 & 0084230501 & 52.9&  16.1 & 0.3  & 0.186   \\
  
A520            & 4: 54: 13.14 & 2.0: 55: 55.0 & 0201510101	& 59.1 & 12.1 & 0.4  & 0.199\\

A2163  & 16: 15: 41.47 & -6.0: 9: 16.7 & 0112230601 & 20.1& 6.3 & 0.9  & 0.203 \\
  			&			&		& 0112231501 & & & & 0.203 \\

A209   & 1: 31:  50.59 & -13.0: 37: 36.4 & 0084230301 & 33.9& 11.3 &  0.5 & 0.210\\

A963    & 10: 17: 9.38 & 39.0: 2: 49.9 & 0084230701 & 44.0 & 13.1 &  0.4 & 0.203 \\

RXC J1504.1-0248  &15: 4: 7.79 &  -2.0: 48: 11.8 & 0401040101& 72.1& 27.1 & 1.0 & 0.215  \\  

MS 0735.7+7421  & 7: 41: 28.18 & 74.0: 14: 25.1 & 0303950101& 92.7 & 37.2 & 0.2 & 0.2165 \\ 

A773   & 9: 17: 45.67 & 51.0: 43: 33.2 & 0084230601 &  26.7 & 14.2  & 0.5 & 0.219\\  

AS0592  & 6: 38: 39.55 &  -53.0: 57: 59.1 & 0650860101& 56.2 & 7.5 & 0.3 & 0.220 \\

A2390   & 21: 53: 39.66 & 17.0: 42: 45.5 & 0111270101 & 19.9& 8.2 & 0.5 & 0.227\\

A2667    & 23: 51: 42.3 & -26.0: 4: 13.6 & 0148990101& 38.6& 14.0  & 0.5 & 0.230 \\

A267     & 1: 52: 38.71 & 0.0: 59: 49.4 & 0084230401 &  11.7& & 0.4 & 0.230\\

RX J2129.6+0005  & 21: 29: 34.73 & 0.0: 4: 53.7 & 0093030201 & 63.0 & 18.1 & 0.3 & 0.235\\ 

RXC J1314.4-2515 & 13: 14: 35.3 & -25.0: 15: 38.7& 0501730101 & 64.7&19.8  & 0.3 &  0.247 \\   
  			&			&		& 0551040101  &&&&  0.247  \\  
 
A1835     & 14: 1: 1.88 &  2.0: 52: 40.4 & 0147330201 & 241.8 & 62.4	& 1.0 & 0.253    \\ 
			&		&			& 0551830101 & & && 0.250\\ 
   			&		&			& 0551830201 & & && 0.252 \\ 
 
A1758     & 13: 32: 48.01 & 50.0: 32: 0.9 &0142860201&  58.9 & 17.7 &  0.3 & 0.279  \\ 

A1763     & 13: 35: 24.76 & 40.0: 59: 41.2 & 0084230901	& 24.1 & 8.2 & 0.6 & 0.222\\  
 
A689     & 8: 37: 20.93 & 14.0: 58: 6.8 & 0501010101 & 69.3 & 26.5 & 0.5 & 0.279      \\
 
ZW 3146   & 10: 23: 39.84 & 4.0: 11: 5.6 & 0108670101 & 387.6 & 146.3  & 0.5 & 0.290 \\ 
  			&		&			& 0605540201 & & && 0.291 \\ 
   			&		&			& 0605540301 & & && 0.289\\ 
			

A781    & 9: 20: 9.6  & 30.0: 29: 47.3 & 0150620201 & 118.9 & 47.5 & 0.3 & 0.290  \\  
 			&			&		& 0401170101 &  &&&  0.295 \\ 

1E 0657-558 &06: 58: 31.1 & -55: 56:49 &0112980201 &43.7 & 17.4 & 4.1& 0.296 \\ 

MS 2137.3-2353   & 21: 40: 15.28 & -23.0: 39: 43.5 & 0008830101 & 21.4 & 6.3 & 0.2    & 0.313 \\
			& 			&			& 0673830201	&  & && 0.313 \\

MACS J2229.7-2755 & 22: 29: 45.13 & -27.0: 55: 33.7 & 0651240201  & 58.3 & 14.0 & 0.2  & 0.320\\

MACS J1532.8+3021 & 15: 32: 48.98 & 30.0: 21: 14.8 & 0039340101  &21.5 & 8.0 & 0.2  & 0.350\\
   			&			&			& 0651240101 & & && 0.345 \\ 
			
AS1063  & 22: 48: 46.69 & -44.0: 30: 48.9 & 0504630101 &21.6 & 18.0 & 0.4 &  0.354\\

\\
\end{longtable*}
}
\normalsize
Subtle errors in the detector energy gain may cause a fraction of a percent
shifts of the location of the emission lines in different X-ray observations
of the same cluster. In addition, a redshift measured from the optical
observations of a cluster may differ from an X-ray redshift of the gas. To be able to
stack spectra in the same frame, we determined the best-fit X-ray
redshift for each \xmm observation using the bright Fe lines. These
redshifts (Table \ref{table:sample}), which correct  
for both of the above-mentioned effects, were then used to scale the
individual spectra in the source frame of each observation, as will be
described in Section  \ref{sec:stacking}. Our selected observations provide adequate
statistics to fit an X-ray redshift for each spectrum.

For most clusters, the spectra were extracted within the overdensity radius
$R_{500}$. The overdensity radii were calculated using the \citet{vikhlinin2009}
mass-temperature scaling relation for each cluster. Due to the large
solid angle of nearby clusters, e.g., Coma, 
Perseus and Centaurus, their spectra were extracted within the full field of
view (FOV). Redistribution matrix files (RMFs) and ancillary response files
(ARFs) were created with the \sas tools \textit{rmfgen} and \textit{arfgen},
respectively.

Although we stack the cluster spectra in this work (and end up using
only the 2--10 keV band for the line search), it is still important to accurately
subtract the background from each individual observation. 
For each extracted spectrum, we model a superposition of four main background components: quiescent particle background (QPB), cosmic X-ray background emission (including Galactic halo, local hot bubble (LHB), and unresolved extragalactic sources), solar wind CX as well as residual contamination from soft protons. 
We use the ROSAT All$-$Sky Survey (RASS)
background spectrum to model the soft X-ray background using the background
tool at the High Energy Astrophysics Science Archive Research Center
Web site. 
The RASS spectrum was
extracted from an annulus from 1$^{\circ}$ to 2$^{\circ}$ surrounding
the cluster center, with the assumption that this
spectrum reasonably represents the soft X-ray background
in the direction of the cluster.

We simultaneously modeled the soft X-ray emission from theLHB
 or heliosphere with a cool unabsorbed single-temperature thermal
component ($E\,\sim$ 0.1 keV), while the Galactic hotter halo and
intergalactic medium were modeled with an absorbed thermal component (
$E\,\sim$ 0.2 keV). The energies of the \textit{apec} model were
restricted but allowed to vary with free normalizations. The
abundances were set to 1$A_{\odot}$. We model the contamination due to
unresolved point sources 
using an absorbed power-law component with a spectral index of $\alpha
\simeq 1.46$ and normalization of $8.88 \times 10^{-7}$ photons keV$^{-1}$
cm$^{-2}$ s$^{-1}$ at $\sim$1 keV \citep{kuntz2008}.\ Soft-proton flares are
largely removed by the light-curve filtering. However, after the filtering
some soft-proton residuals may remain in the data and were modeled by
including an extra power-law model component and diagonal response matrices
provided in the \sas distribution in the final spectral analysis
\citep{snowden2008}.

The EPIC-MOS QPB spectra have two bright
instrumental fluorescent lines: the Al-K (1.49 keV) and the Si-K (1.74 keV)
lines. The PN QPB spectra have fluorescent lines of Al-K (1.49 keV), Ni-K
(7.48 keV), Cu-K (8.05, 8.91 keV), and Zn-K (8.64, 9.57 keV).  Since small
variations in the gain and the line strengths between the source and
background spectra can lead to residuals in the spectral fitting
\citep{kuntz2008} and \esas software does not include these instrumental
lines in the QPB spectra, we modeled these instrumental lines spectrally by
adding Gaussian models to our spectral fits to determine the best-fit
energies, widths, and normalizations.  The total background was constructed
by adding the models for the Al-K, Si-K, Ni-K, Cu-K, and Zn-K lines with the
best-fit energies, widths, and normalizations to the QPB produced in the
\esas analysis for all pointings.  These total QBP spectra were directly
subtracted from the summed observation to obtain source spectra.

The fitting of the source spectra was done with the spectral fitting package
XSPEC 12.8.0 \citep{arnaud1996}. The 0.3$-$10 keV energy interval was used
for MOS spectra, whereas the 0.4 $-$ 10.0 keV band was used for the PN fits.
To determine the best-fit cluster redshifts for each observation
(given in Table \ref{table:sample}), the
cluster spectra were fit with a standard absorbed
multi-temperature collisional equilibrium plasma model
\citep[\apec;][]{smith2001} and AtomDB v2.0.2 \citep{foster2012}. We did not
observe any differences beyond a fraction of a percent in terms of the
detector gain variations. 

\subsection{Spectra Stacking Methods}
\label{sec:stacking}

The best way of distinguishing a real spectral feature in a class of distant
objects from instrumental artifacts and the X-ray background features is to
detect that feature in multiple objects at different redshifts in their rest
frame, in which case the line coming from an object will stay at the same
energy, unlike the detector artifacts. To accomplish this, we stacked the
spectra of our selected 73 clusters, blueshifting them to the source frame
using the best-fit X-ray redshift of each observation determined above.

Technically, the energies of the source and background X-ray events
were rescaled to the source 
frame using the best-fit redshifts. The scaled event files were then used
to extract the source within $r=R_{500}$ or
the full FOV of MOS, and the same extraction region was used for 
PN observations for nearby clusters that fill the 
FOV. The particle background spectra were extracted using the scaled filter wheel closed data.
Counts from each individual spectrum were co-added into a single
stacked spectrum using the FTOOL \textit{mathpha} to produce the stacked
source and the particle background spectra. At the
end of the stacking process, we obtained spectra with $\sim$ 6 Ms of good cluster exposure
with MOS 1 and MOS 2 (that were co-added) and $\sim$ 2 Ms with PN for
the full \xmm sample. 

\begin{figure}[ht!]
\centering
\hspace{-5mm}\includegraphics[width=9cm, angle=0]{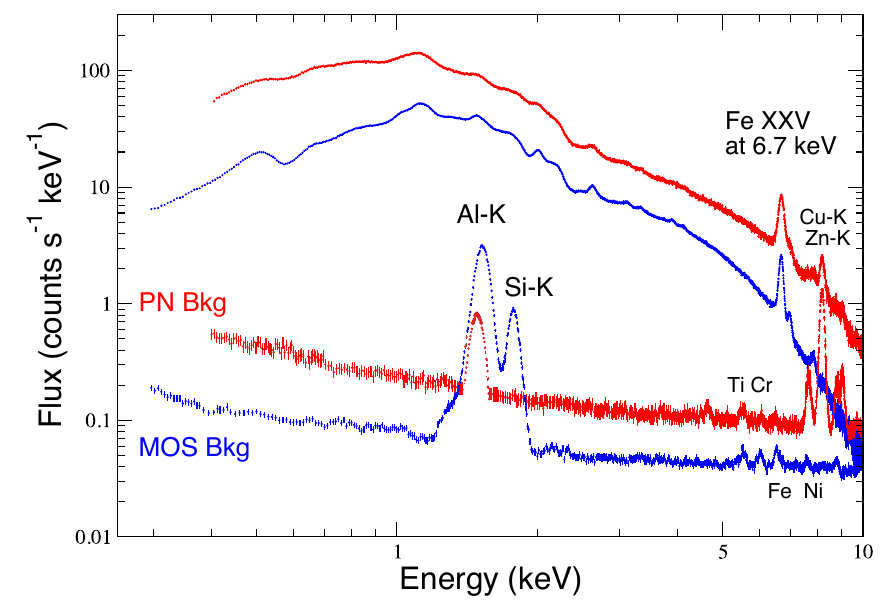}
\caption{ \xmm MOS and PN background-subtracted source spectra and
  particle background spectra for the Perseus Cluster. The spectra
  were obtained by co-adding the observations of the cluster
  in the cluster's rest frame. In the co-added scaled spectra, the Fe
  \textsc{xxv}  line is located at its rest energy, $\sim6.7$
  keV. Energy of background and instrumental lines is blue-shifted
  according to the cluster's redshift. 
}
\label{fig:perseus}
\end{figure}

The RMF and ARF to be used with the stacked spectrum were constructed by averaging
the responses for individual observations with proper weighting.
The individual RMFs and ARFs were first remapped to the source frame using the best-fit
redshifts. The weighing factors for stacking RMFs and ARFs were calculated using the
total counts in the energy band we will use for our line search (2--10
keV). These factors ($\omega_{\rm{cnt}}$) are given in Table
\ref{table:weight}. The weighted and normalized ARFs and RMFs were stacked
using the FTOOLS \textit{addarf} and \textit{addrmf}. These X-ray count-weighted
response files were used to model the continuum and the known plasma
emission lines; we will also try a different weighting of responses for the possibly
nonthermal new line, as will be described below.

For a check, each background-subtracted, blue-shifted, single-cluster
spectrum was fit with an \textit{apec} model using the corresponding scaled ARF and
RMF to verify that the best-fit redshifts were consistent with zero. For
illustration, the co-added MOS and PN source and background spectra of the
Perseus Cluster in its source frame are shown in Figure \ref{fig:perseus}.
We note that the Fe \textsc{xxv} line is located at its rest energy $\sim$ 6.7 keV,
while the background and instrumental lines are blue-shifted.

The stacked MOS and PN source and background spectra of the clusters in the
sample are shown in Figure \ref{fig:stackedSpec}. The background
spectra show the smearing effect on the background
lines, e.g., Al-K (1.48 keV), Si-K (1.75 keV), Cr (5.4 keV), Mn (5.8 keV),
Fe-K (6.4 keV), Cu-K (8.05 keV, 8.91 keV), Zn-K (8.64 keV, 9.61 keV) and Au
(9.1 keV).  They are much less prominent in the stacked spectrum compared
with the single-source spectrum shown in Figure
\ref{fig:perseus}. Similarly, any residuals from inaccurate
background subtraction are smeared. We will see other advantages of
this smearing below.

\begin{figure*}[ht!]
\centering
\hspace{-4mm}\includegraphics[width=8.8cm, angle=0]{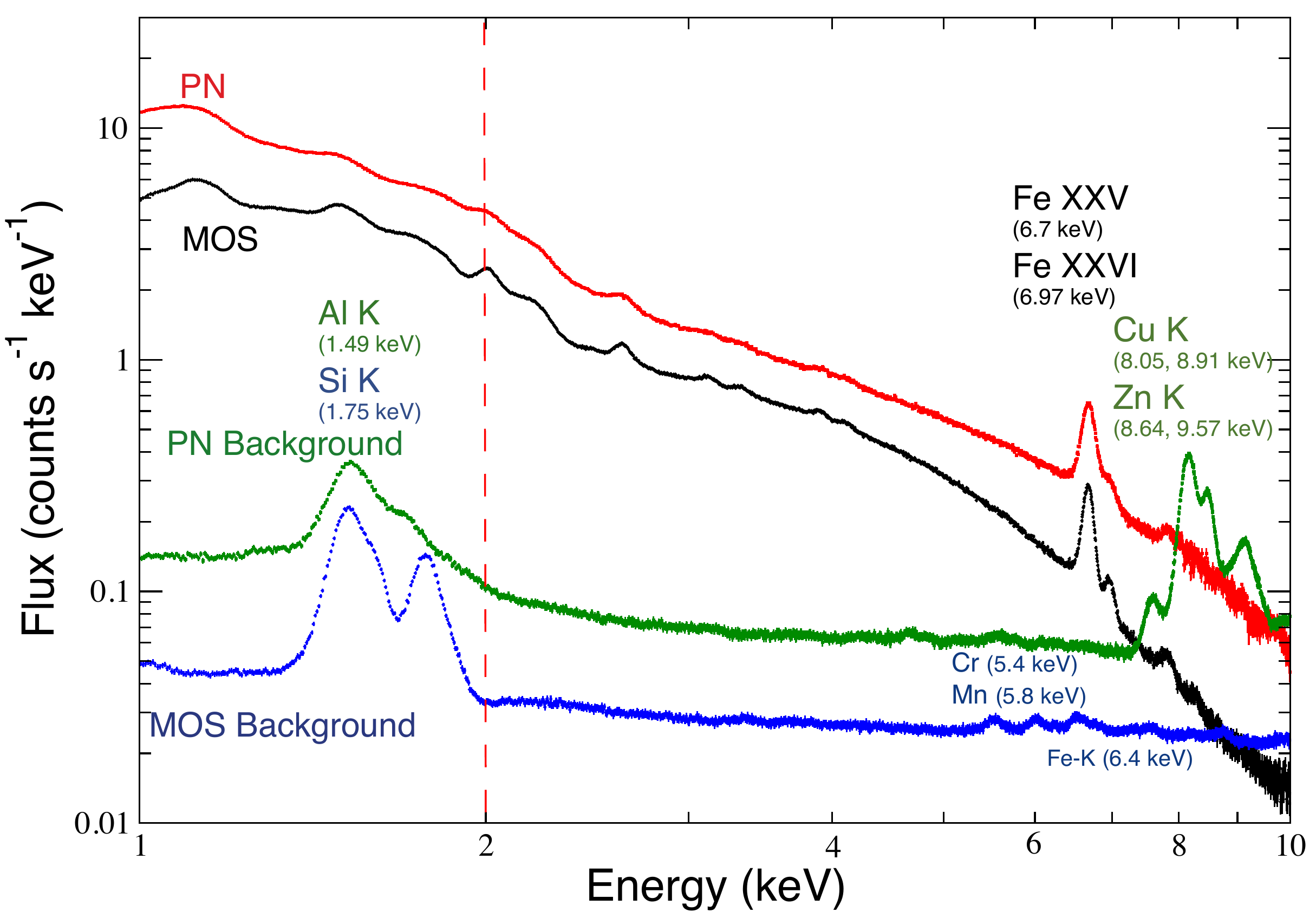}
\hspace{2mm}\includegraphics[width=8.7cm, angle=0]{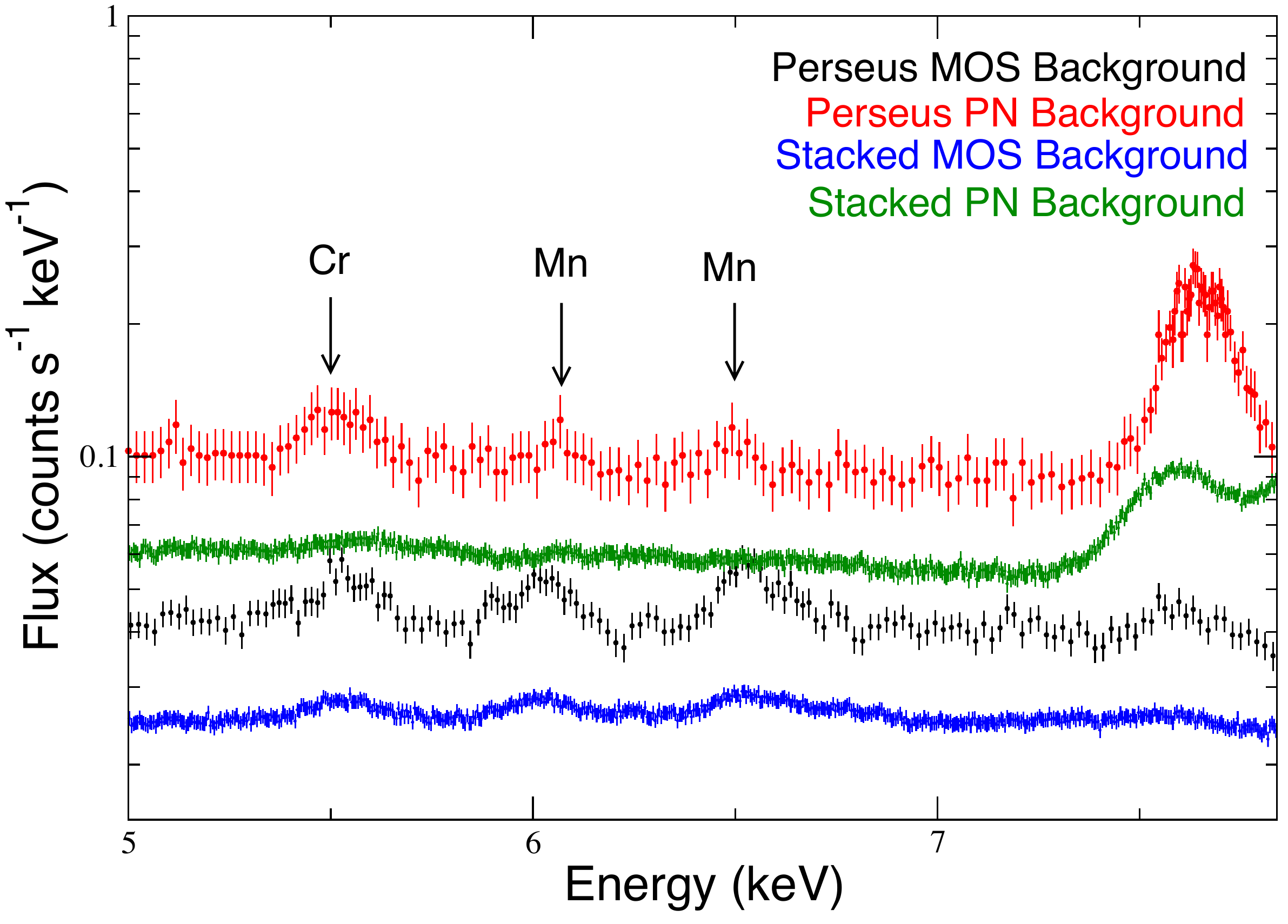}
\caption{Left Panel: stacked \xmm  MOS and PN background-subtracted 
source spectra and particle 
background spectra of the full sample. The spectrum of each observation was scaled to the rest 
frame prior to stacking. The total filtered exposure time was 6 Ms for MOS and 2 Ms 
for PN. The background MOS (in blue) and PN (in green) spectra show the effect 
of smearing of instrumental lines, such as Cr, Mn, Fe, and Ni, as well as Al-K 
and Si-K fluorescent lines. The effect is due to the stacking of background 
spectra that are scaled by different cluster redshifts. Right Panel: close-up view 
of 5.0 $-$ 8.0 keV band of the background 
\xmm MOS and PN spectra of the Perseus Cluster compared to the stacked 
\xmm MOS and PN background spectra. The background lines 
are less prominent in the stacked background spectra than in the single-source 
background spectra.
}
\label{fig:stackedSpec}
\end{figure*}
%

\section{Analyzing the Stacked XMM-Newton Spectra}

We will limit our line search to the $2-10$ keV energy band. After
looking at the stacked spectra, we concluded that the band below
2 keV is hopelessly crowded with  lines, such as the strong Ne \textsc{x} (1.21
keV), Fe \textsc{xxiv} (1.55 keV), Mg \textsc{xii} (1.74 keV), and Si
\textsc{xii} (1.86 keV) features, making the detection of any weak emission
features between them difficult, given the $\sim 100$ eV energy
resolution of \xmm and other CCD detectors.

To search for any unidentified spectral lines in the stacked spectra, we
need to model the known lines and the continuum emission to a very good
precision. We do not necessarily need to obtain a physically meaningful
model (which would be a mixture of all the thermal components in all the clusters), but one
that allows enough freedom to account for all known lines and the possible
errors in their theoretical emissivities. To this end, we fit the
background-subtracted stacked source spectra with a \textit{line-free}
multi-temperature \textit{apec} model to represent the continuum emission
with high accuracy, and then we add individual lines. We start with four
continuum components to represent the multi-temperature nature of the
stacked spectra. The \textit{line-free apec} model accounts for the continuum due to
thermal bremsstrahlung, 
radiative recombination, and two-photon emissions. The best-fit temperature
and normalization parameters of \textit{line-free} \apec models are shown in
Table \ref{table:temp}. The best-fit temperatures in the table do not have
physical meaning, since they are obtained by fitting the stacked
blue-shifted spectra. (We note that the continuum of a redshifted
thermal model can be well represented by a continuum with a different
redshift and a different temperature.) The abundance was set to 0.3
in order to include the recombination edges in the fitting
process. The abundance parameter does not 
affect the line modeling, since the \textit{line-free apec} model does not
include lines.

In order to account for the known plasma emission lines in a
model-independent way, for each known line in the $2.0-10.0$ keV band, we
added a Gaussian line to the model. Initially we have added Gaussian
models for the known strong emission lines from the AtomDB database%
\footnote[1]{http://www.atomdb.org/Webguide/webguide.php}
with emissivities $>\ 5\times10^{-19}$ photons cm$^{3}$ s$^{-1}$ for the lowest temperature
given in Table \ref{table:temp}. The strong emission lines (which can be resolved with a CCD
detector) included in our model at their rest energies are Al \textsc{xiii} (2.05
keV), Si \textsc{xiv} (2.01 keV and 2.51 keV), Si \textsc{xii} (2.18 keV, 2.29
keV, and 2.34 keV), S \textsc{xiv} (2.62 keV), S \textsc{xv}
(complex at 2.45 keV, 2.88 keV), Ar \textsc{xvii} (triplet at 3.12 keV, 3.62
keV, 3.68 keV), K \textsc{xviii} (3.47 keV and 3.51 keV), K \textsc{xix}
(3.71 keV), Ca \textsc{xix} (complex at 3.86 keV, 3.90 keV, 4.58 keV), Ar
\textsc{xviii} (3.31 keV, 3.93 keV), Ca \textsc{xx} (4.10 keV), Cr
\textsc{xxiii} (5.69 keV), Fe \textsc{xxiv} (complex at 6.62 keV), Fe
\textsc{xxv} (complex at 6.70 keV, 8.29 keV, 7.81 keV, 7.88 keV), Fe
\textsc{xxvi} (6.95 keV, 8.3 keV, and 8.70 keV), and Ni \textsc{xxvii} (7.79 keV).\
Initially,
a total of 28 Gaussian model components were included in the 2--10 keV energy band.
Individual Gaussian components were then removed if they were not
required to accurately model the spectra (to improve convergence of
the fit). The widths of Gaussians
were left free, but restricted to the range 0 $<\Delta E/E< 10^{-2}$. The
energies of the Gaussian components were allowed to vary by up to 5 eV to 
account for residual uncertainties in the gain and in the energies in the atomic database.
This way, we were able to model the continuum emission and strong known
emission lines accurately, leaving a clean residual spectrum to search for
any unidentified lines.

We also fit a power-law model in the full band to represent the residual
soft-proton background contamination (see Section \ref{sec:analysis1}), and we used 
these power law indices and normalizations for further narrower-band fits
(see Section  \ref{sec:all}). The spectral counts in each energy bin were
sufficiently high to allow the use of the Gaussian statistics in this
analysis \citep{protassov2002}.

\subsection{Stacked Spectra of the Full Cluster Sample}
\label{sec:all}

After the stacking process, we obtained a total of 8.5$\times \,10^{6}$ source
counts in the 6 Ms MOS spectra, while the 2 Ms PN stacked spectra have a
total  of 5.1$\times \,10^{6}$ source counts. The \textit{line-free apec}
model with Gaussian lines produces an acceptable fit to the stacked MOS and
PN spectra with $\chi^{2}$ values of 564.8 for 566 dof (MOS) and 510.5 for 564 degrees of freedom (dof)
(PN).  After modeling all the known thermal plasma lines in the stacked
spectrum, we examined the residuals in each 1 keV band carefully. We found
one significant unidentified residual emission feature at $E\approx
3.55-3.57$ keV, which is 
not associated with any plasma emission lines in the band. Near this line,
there are four tabulated weak thermal emission lines of K \textsc{xviii}
(1s$^1$ 2s$^1$ $\rightarrow$ 1s$^2$) at a rest energy of 3.47 keV, K
\textsc{xviii} (1s$^1$ 2p$^1$ $\rightarrow$ 1s$^2$) at 3.51 keV, a
dielectronic recombination (DR) line of Ar \textsc{xvii} at 3.62 keV, Ar
\textsc{xvii} (1s$^1$ 3p$^1$ $\rightarrow$ 1s$^2$) at 3.68 keV, and K
\textsc{xix} (2p$^1$ $\rightarrow$ 1s$^1$) at 3.72 keV. 

In order to separate the excess emission feature from these weak
contaminating K and Ar lines, we make conservative estimates of their flux
using AtomDB. Ideally, line flux measurements would be based on other
lines of the same ions; however, there are no other strong K \textsc{xviii} and 
K \textsc{xix} lines in the spectrum. Therefore, we use the lines from
the relatively clean part of the band, namely, the S \textsc{xvi} (2p$^1$
$\rightarrow$ 1s$^1$), Ca \textsc{xix} (1s$^1$ 2p$^1$ $\rightarrow$ 1s$^2$),
and Ca \textsc{xx} (2p$^1$ $\rightarrow$ 1s$^1$) lines at 2.63 keV, 3.90 keV,
and 4.11 keV, respectively, to estimate the flux of the 3.47 keV, 3.51 keV,
3.68 keV and 3.72 keV lines. The best-fit flux measurements of these S
\textsc{xvi}, Ca \textsc{xix}, and Ca \textsc{xx} lines are given in Table
\ref{table:temp}.

{
\begin{table*}[ht!]
\begin{center}
\caption{Best-fit Temperature and Normalizations of \textit{line-free apec} Model in 2 $-$ 10 keV Fit to the Stacked MOS and PN Spectra for Various Samples. The temperature ($kT_{i}$) normalization ($N_{i}$) are in the units of keV and ($10^{-2}$ cm$^{-5}$), respectively and the line fluxes of S \textsc{xvi}, Ca \textsc{xix},  and Ca \textsc{xx} are in the units of $10^{-5}$ photons cm$^{-2}$ s$^{-1}$ at rest energies 2.63 keV, 3.90 keV, and 4.11 keV, respectively.  }
\renewcommand{\arraystretch}{1.5}
\begin{tabular}{lcc|cc|cc|cc}
\hline\hline\\
 & Full   &&  Coma&  & Excluding  &   \\
&Sample&&+ Centaurus&  &Nearby &&Perseus  \\
&&&+ Ophiuchus &&Clusters&\\
Parameters      & MOS  & PN & MOS & PN &MOS & PN &MOS &PN\\
\\\hline
\\
kT$_{1}$ & 5.9 $\pm$ 0.1 &  7.3 $\pm$ 0.2 &  3.9 $\pm$ 0.1 &  2.5 $\pm$ 0.2 & 3.5 $\pm$ 0.2 & 2.0 $\pm$ 0.3 & 3.6 $\pm$ 0.6 & 2.17 $\pm$ 0.9 \\

$N_{1}$   & 2.2 $\pm$ 0.1 &  1.1 $\pm$ 0.1 &  6.5 $\pm$ 0.1  & 5.4 $\pm$ 0.1 & 0.6 $\pm$ 0.1 &  0.3 $\pm$ 0.1 & 15.7 $\pm$ 7.8 & 10.2 $\pm$ 6.9\\

kT$_{2}$  & 6.1 $\pm$ 0.1  &  2.3 $\pm$  0.3 & 6.8 $\pm$ 0.1 & 6.5 $\pm$  0.2 & 6.8 $\pm$ 0.1 & 9.4 $\pm$ 0.2 & 7.6 $\pm$ 0.7 & 6.25 $\pm$ 0.8\\

$N_{2}$ &  1.8 $\pm$ 0.1   &  0.6 $\pm$ 0.1 &  8.9 $\pm$ 0.1  & 6.1 $\pm$ 0.1  & 0.8 $\pm$ 0.1 &  0.1 $\pm$ 0.1 & 44.0 $\pm$ 6.8 & 50.2 $\pm$ 14.1\\

kT$_{3}$  &  7.3 $\pm$ 0.2  & 18.7  $\pm$ 0.2   & 10.7 $\pm$ 0.2  & 15.4 $\pm$ 0.6  &10.3 $\pm$ 0.3 & 4.4 $\pm$ 0.7&  $-$&  $-$  \\

$N_{3}$  &  1.6 $\pm$  0.1 &  0.4 $\pm$ 0.1  & 8.9 $\pm$ 0.1  & 7.2 $\pm$ 0.2 & 0.7 $\pm$ 0.1 &  0.1 $\pm$ 0.02 &  $-$&  $-$  \\

kT$_{4}$ & 10.9 $\pm$ 0.5   &  6.9 $\pm$ 0.1  & 7.4 $\pm$ 0.2 &  4.0 $\pm$ 0.2  & 6.9 $\pm$ 0.2 &  $-$ &  $-$&  $-$ \\

$N_{4}$ & 0.9 $\pm$ 0.1  & 1.0 $\pm$  0.1 &  6.9 $\pm$ 0.1 &  4.6 $\pm$ 0.2 &  0.6 $\pm$ 0.1 &$-$&  $-$ &  $-$   \\

Flux of S \textsc{xvi} & 7.9 $\pm$ 0.1 & 3.9 $\pm$ 0.1 &  39.1 $\pm$ 6.6 &  13.1 $\pm$  0.9 & 2.9 $\pm$ 0.1 & 2.8 $\pm$ 0.1 &
49.1 $\pm$ 7.3 & 55.5 $\pm$ 4.9\\

Flux of Ca \textsc{xix}& 2.4  $\pm$  0.1 & 0.9 $\pm$ 0.2& 13.5 $\pm$ 4.8  & 4.6 $\pm$ 0.6 & 0.7 $\pm$ 0.1 & 0.6 $\pm$ 0.1 & 25.6 $\pm$ 1.5 &11.9 $\pm$ 2.9 \\

Flux of Ca \textsc{xx}& 1.7 $\pm$  0.1 & 0.4 $\pm$ 0.2& 8.5 $\pm$ 0.5 &  1.8 $\pm$ 0.6  &  0.5 $\pm$ 0.1 & 0.4 $\pm$ 0.1 &14.7 $\pm$ 1.2 & 11.1 $\pm$ 7.3\\

\\
\hline\hline
\end{tabular}
\label{table:temp}
\end{center}
\end{table*}
}

We assume that the relative abundances of S, Ca, Ar, and K are proportional to
their abundances in the solar photosphere \citep{andersgrevesse1989}. While
this may not be exactly true, it gives a reasonable starting point (we
will relax this assumption below). Then, using AtomDB,
we calculated the relative emissivity of the K \textsc{xviii}, K
\textsc{xix}, and Ar \textsc{xvii} lines compared to the S \textsc{xvi},
Ca \textsc{xix}, and Ca \textsc{xx} lines based on the equilibrium
collisional plasma conditions at the various temperatures of our
\textit{line-free apec} components. In practice, the emissivities of K
\textsc{xviii}, K \textsc{xix}, and Ar \textsc{xvii} lines are 
stronger at the lowest temperatures of each model, so the other
components can be  ignored. The curves in Figure
\ref{fig:atomdb} represent the emissivities of K and Ar lines 
as a function of plasma temperature for the normalizations of the lowest temperature 
components measured in our spectra.

\begin{figure*}[]
\centering \vspace{3mm} 
\hspace{-4mm}\includegraphics[width=8.9cm,angle=0]{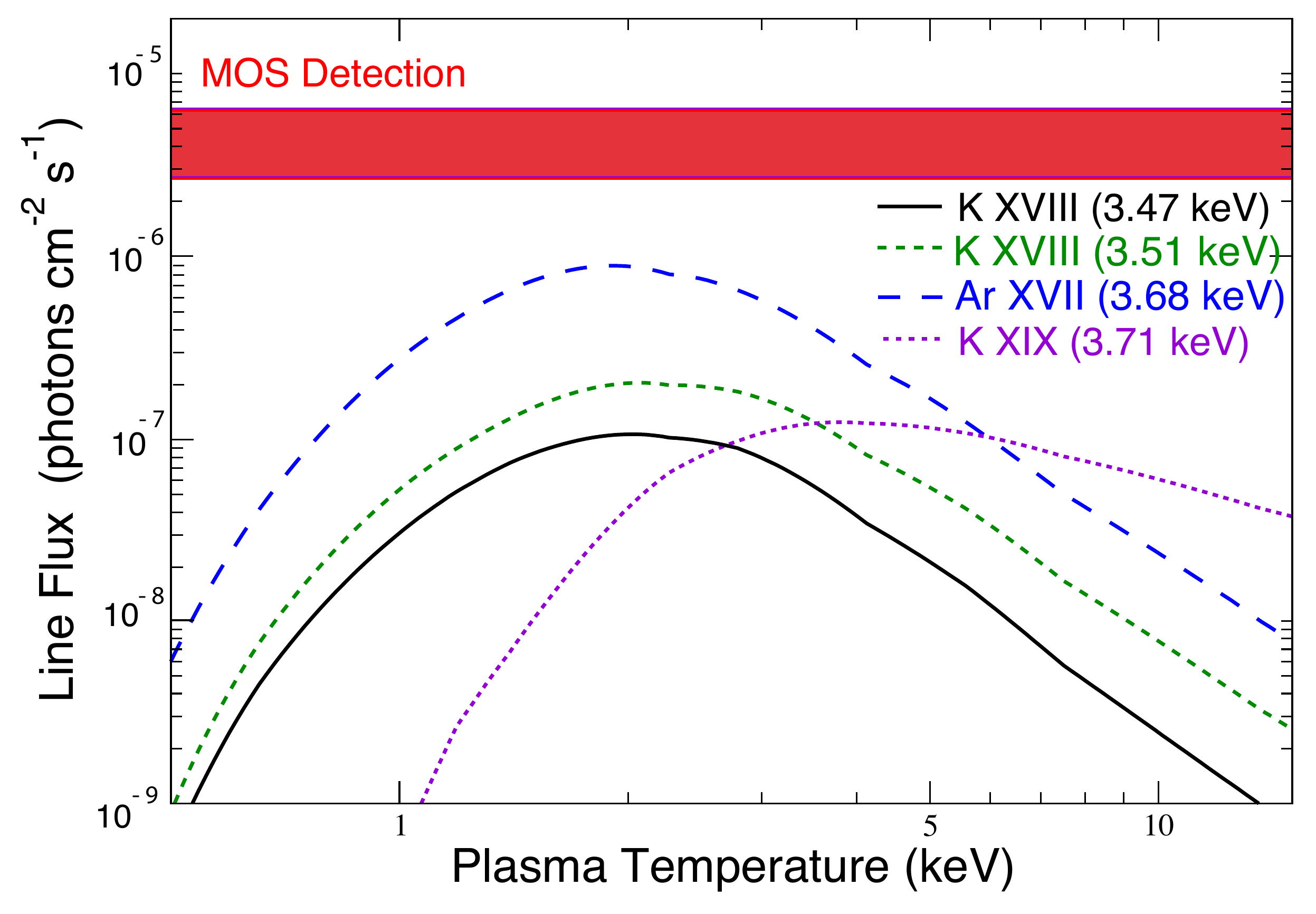}
 \includegraphics[width=8.9cm, angle=0]{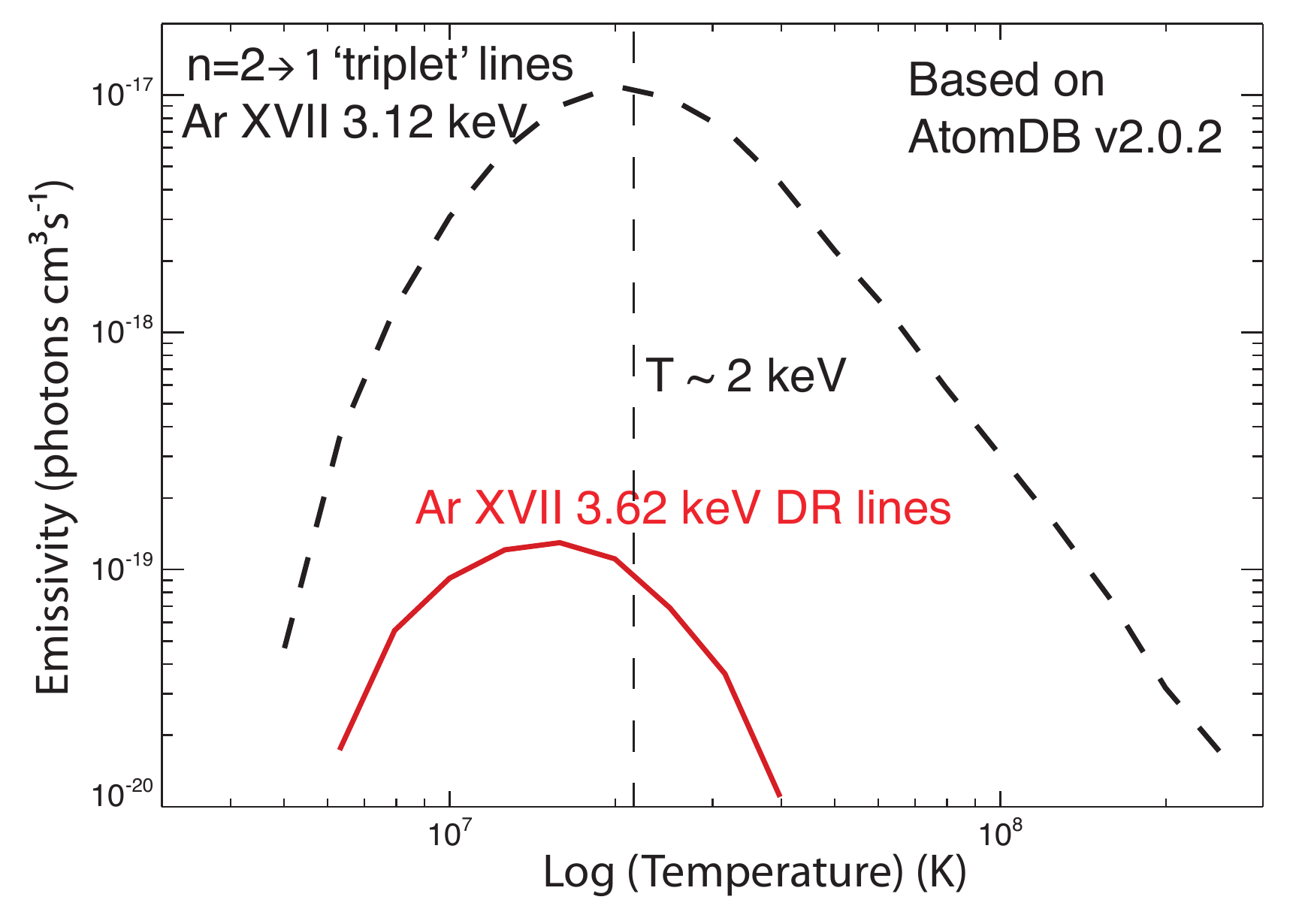}
\caption{Left Panel: Eetimated line fluxes of the K \textsc{xviii} at the
  rest energies 3.47 keV, 3.51 keV, the Ar \textsc{xvii} at the rest
  energy 3.68 keV, and the K \textsc{xix} at the rest energy 3.71 keV as a
  function of plasma temperature.The line fluxes are calculated based on the observed fluxes of S
\textsc{xvi}, Ca \textsc{xix}, and Ca \textsc{xx} from the stacked
\xmm MOS observations of the full sample. The flux
  detection and 90\% errors on the flux of the unknown spectral feature
  measured from the stacked MOS observations of the full sample are shown
  with the red shaded area. Right Panel: Comparison of
  emissivities of the Ar \textsc{xvii} triplet lines at 3.12 keV and Ar
  \textsc{xvii} DR line at 3.62 keV. The figure shows that the ratio of the Ar \textsc{xvii} DR line at 3.62 keV to the Ar \textsc{xvii} line at 3.12 keV could at most be 1\% at the lowest temperature we observe in our fits ($T\,\sim$ 2 keV indicated
  with the dashed line). 
  This fraction was used as an upper limit to the
  flux of the Ar \textsc{xvii} DR line in our spectral fits and given in Table \ref{table:fluxArK}
  for each sample.}
\label{fig:atomdb}
\end{figure*}

{
{
\begin{table*}[ht!]
\begin{center}
\caption{Estimated Maximum Fluxes of K \textsc{xviii} at the Rest
  Energies 3.47 keV, 3.51 keV,  Ar \textsc{xvii} at the Rest Energies 3.68
  keV, and K \textsc{xix} at the Rest energy 3.71 keV Lines Obtained from
  AtomDB in the Units of photons cm$^{-2}$ s$^{-1}$. Estimates were performed based on best-fit fluxes obtained from the fluxes of S \textsc{xvi}, Ca \textsc{xix}, and Ca \textsc{xx} lines in the line-free \apec model. The maximum flux for the Ar \textsc{xvii} DR line at 3.62 keV line was initially set to 1\% of the Ar xvii line at 3.12. The fits were allowed to go a factor of 3 three above these estimates."
  }
\begin{tabular}{lccccccc}
\hline\hline\\
&Sample & Inst. & Flux & Flux& Flux &Flux & Flux \\
&&& K \textsc{xviii}  & K \textsc{xviii} & Ar \textsc{xvii}  & Ar \textsc{xvii}&K  \textsc{xix} \\
&&& (3.47 keV)  & (3.51 keV) & (3.62 keV) & (3.68 keV) & (3.71 keV)\\
&&& ( $10^{-7}$ )  & ( $10^{-7}$ ) & ( $10^{-7}$ ) & ( $10^{-6}$ )  & ( $10^{-6}$ )  \\
\\
\hline
\\

&& MOS    &  1.3 $\pm$ 0.7 & 3.5 $\pm$ 1.8 & 0.12 & 1.0 $\pm$ 0.5 & 1.2 $\pm$ 0.6 \\ 
&Full & \\
&Sample & PN  & 0.9 $\pm$ 0.4  & 1.8 $\pm$ 0.9 &  0.14& 0.7 $\pm$ 0.3 & 0.3 $\pm$ 0.1\\
\\
\cmidrule{2-8}
\\
&Coma + & MOS &  2.7 $\pm$ 2.1 & 8.2 $\pm$ 6.3 & 7.0 &  2.5 $\pm$ 1.9 & 5.2 $\pm$ 4.1\\
&Centaurus + \\
&Ophiuchus& PN & 3.3 $\pm$ 2.3  & 6.8 $\pm$ 4.7 &  1.4& 2.5 $\pm$ 1.8 &  0.8 $\pm$ 0.6\\
\\
\cmidrule{2-8}
\\
& Perseus& MOS & 18.5 $\pm$ 9.9 & 45.7 $\pm$ 24.4 & 6.4& 15.1 $\pm$ 8.1 & 11.6 $\pm$ 6.2 \\
\\
&& PN & 13.8 $\pm$ 6.8 & 36.0 $\pm$ 17.8 & 1.99 & 10.8 $\pm$ 5.4 & 9.15 $\pm$ 4.5 \\
&\\
\\
\cmidrule{2-8}
\\
&All & MOS   & 0.5 $\pm$ 0.2 & 1.3 $\pm$ 0.5  & 0.10  & 0.4 $\pm$ 0.1 & 0.29 $\pm$ 0.1 \\
&Other &\\
&Clusters & PN  & 1.3 $\pm$ 0.5 & 2.6 $\pm$ 0.9 & 0.90& 1.1 $\pm$ 0.4 & 1.2 $\pm$ 0.4\\

\\
\hline\hline
\end{tabular}
\label{table:fluxArK}
\end{center}
\end{table*}
}

Having obtained the relative theoretical emissivity of the lines from
AtomDB, we estimated the flux as

\begin{equation}
\Gamma_{\it{l}} = \Gamma_{\it{r}} \sum\limits^{\it{i}} \rm{Norm}_{\it{i}}\varepsilon_{\it{l}}(T_{\it{e}})/ \varepsilon_{\it{r}}(T_{\it{e}}),
\end{equation}

where subscripts $l$ and $r$ represent the lines of interest (K
\textsc{xviii} and Ar \textsc{xvii}) and reference lines (S \textsc{xvi}, Ca
\textsc{xix}, and Ca \textsc{xx}), respectively, $\Gamma$ is the flux in the
line, $\varepsilon(T_e)$ is the calculated emissivity from AtomDB at the
electron temperature $T_e$, and the sum over $i$ represents the different
temperature components listed in Table \ref{table:temp} with their
normalizations $\rm{Norm}_i$.
We use 0.1 and 3 times the
maximum values of these fluxes as lower and upper bounds, respectively, for the
normalizations of the Gaussian lines in the \xspec fitting. The lower
limits of 0.1 is set to
avoid the lines vanishing and posing problems for the minimization
routine. The factor three represents a conservative 
allowance for variation of the relative elemental abundances between 
S and Ca (the measured lines on which the predictions are based)
on one hand and K and Ar on the other. (This factor of three is not included in Table
\ref{table:fluxArK}.)

Since our detected emission line is only 50 eV away from the Ar \textsc{xvii}
DR at rest energy 3.62 keV, we calculated
the emissivity of the Ar \textsc{xvii} DR line in a conservative way, using
AtomDB v2.0.2. The He-like argon triplet including four lines (known
either as \textit {w, x, y, z or R, I1, I2, \rm{and} F}) was summed, since the components
cannot be distinguished at the CCD resolution. The two Ar \textsc{xvii} DR
lines at 3.62 keV, known in AtomDB as 10077 $\rightarrow$ 2 and 10078
$\rightarrow$ 3, and which are the result of an He-like Ar ion recombining to
Li-like Ar and emitting a photon at 3.62 keV, were similarly extracted and
summed. The right panel of Figure \ref{fig:atomdb} shows the comparison of the
emissivity of Ar \textsc{xvii} DR and the He-like Argon triplet at
$E\approx 3.12$ keV. To model the flux of the Ar \textsc{xvii} DR line in our spectral fits
in a conservative way, we set the lower and upper limits of the flux to be
0.001 and 0.01 times the flux of the He-like Ar, respectively. The upper limit corresponds
to the highest flux that Ar \textsc{xvii} DR can have for the ICM plasma
temperatures that we see in our spectra (this will be further discussed in 
Section \ref{sec:perseusxmm}). For a 
plasma temperature of $\sim$2 keV, the lowest temperature observed in 
our samples, the ratio of the flux of the Ar XVII DR line to He-like Ar line
corresponds to 0.01 and thus was chosen as an upper limit (see Figure 4 right panel). 
The lower limit has been set to avoid
problems with the fitting procedure.

Once the lower and upper limits on flux estimates of the K \textsc{xviii},
K\textsc{xix}, and Ar \textsc{xvii} lines were set, we performed the fit in a
narrower 3 $-$ 6 keV energy band (to avoid strong S and Si lines below
3 keV and Fe lines above 6 keV). This band is sufficiently wide to measure
the continuum accurately (to better than 1\%). The weak residual emission
line at $E\approx 3.57$ keV was detected in the fits. The excess emission after
the Gaussian K and Ar lines were included in the model at their
maximum fluxes (as described above) in MOS and PN spectra is shown in Figure
\ref{fig:epicfits}. Figure \ref{fig:all-rebin} shows the excess in the rebinned MOS spectrum
of the full sample.
We then added a Gaussian model to fit the remaining
residuals, leaving its flux and energy to vary. The fit was improved by
$\Delta\chi^{2}$ of 22.8 for MOS and $\Delta\chi^{2}$ of 13.9 for PN for an
additional 2 dof (energy and normalization). The best-fit
energy of the added Gaussian line is $3.57\, \pm\, 0.02\, (0.03)$ keV in the
stacked MOS and $3.51\, \pm\, 0.03\, (0.04)$ keV in the stacked PN
observations. The line energies from MOS and PN are in significant tension, $2.8\sigma$
apart (Figure\ \ref{fig:steppar}). However, given the systematic
uncertainties of the fitting procedure, we consider it acceptable;
this tension disappears once another level of complexity is introduced
in modeling (see Section \ref{sec:perseusafter} below). The width of the new line is
unresolved and broadened only by the instrumental response. This is
the only significant unidentified feature we have detected  
in the 2--10 keV band of MOS and PN spectra.

To measure the flux of this line, we have to use a statistically proper
response file, which will depend on the physical interpretation of the
line. If the line were coming from the thermal plasma, then the same
spectral responses that were used for the thermal components are
appropriate. However, there are no known thermal 
plasma lines at this energy, so we explore a possible interpretation of the
detected line as a decay signature of the sterile neutrino (see Section 1). In
this interpretation, the spectral fitting procedure has to be slightly
modified. In particular, when co-adding the instrumental responses used for
the DM line component, the individual cluster responses should be weighted
by the factor $\omega_{\rm{dm}}$ proportional to the estimated dark matter photon
flux from each cluster (as opposed to the X-ray flux used for the
response averaging so far).

{
\LTcapwidth=\textwidth
{\renewcommand{\arraystretch}{1.2}
\begin{longtable*}{lcccccc}
\caption{\it Columns (1)  and (2) show the estimated projected dark matter masses in the spectral extraction radii $M_{DM}^{proj}$ ($R_{ext}$) and the extraction radii $R_{ext}$ in Mpc, Column (3) is the projected dark matter masses per distance squared, and column (4) shows the ratio of the exposure time to the total exposure stacked for each cluster, column (5) is the weighting factors ($\omega_{dm}$) calculated based on the predicted dark matter flux used in the stacking of ARFs and RMFs of each cluster in the sample. These stacked ARFs and RMFs were then used to determine the flux of the detected line, and column (6) shows the weighting factors ($\omega_{cnt}$) calculated based on the total counts in the fitting band. The response files which were stacked using these factors were utilized to model plasma emission lines. \label{table:weight}}\\

\hline \hline
 \multicolumn{1}{l}{\textbf{}} & \multicolumn{1}{c}{\textbf{(1)}} & \multicolumn{1}{c}{\textbf{(2)}}& \multicolumn{1}{c}{\textbf{(3)}} & \multicolumn{1}{c}{\textbf{(4)}}  & \multicolumn{1}{c}{\textbf{(5) }}&\multicolumn{1}{c}{\textbf{(6) }}   \\ 

 \multicolumn{1}{l}{\textbf{Cluster}} & \multicolumn{1}{c}{\textbf{$M^{proj}_{DM}$ ($R_{ext}$) }} & \multicolumn{1}{c}{\textbf{$R_{ext}$}}& \multicolumn{1}{c}{\textbf{$M^{proj}_{DM}$ /D$^{2}$ }} & \multicolumn{1}{c}{\textbf{ Exp/Exp$_{tot}$}}  & \multicolumn{1}{c}{\textbf{$\omega_{dm}$ }}&\multicolumn{1}{c}{\textbf{$\omega_{cnt}$}}   \\ 

\multicolumn{1}{c}{\textbf{}} & \multicolumn{1}{c}{\textbf{ ($10^{14}$ $M_{\odot}$) }} & \multicolumn{1}{c}{\textbf{ (Mpc) }}& \multicolumn{1}{c}{\textbf{ ($10^{10}$ $M_{\odot}$/ Mpc$^{2}$) }} & \multicolumn{1}{c}{\textbf{}} &\multicolumn{1}{c}{\textbf{}}&\multicolumn{1}{c}{\textbf{}} \\ \hline 

\endfirsthead

\multicolumn{3}{l}
{{\bfseries \tablename\ \thetable{} -- continued from previous page}} \\\\
\hline
 \multicolumn{1}{l}{\textbf{Cluster}} & \multicolumn{1}{c}{\textbf{$M^{proj}_{DM}$ ($R_{ext}$) }} & \multicolumn{1}{c}{\textbf{$R_{ext}$}}& \multicolumn{1}{c}{\textbf{$M^{proj}_{DM}$ /D$^{2}$ }} & \multicolumn{1}{c}{\textbf{ Exp/Exp$_{tot}$}}  & \multicolumn{1}{c}{\textbf{$\omega_{dm}$ }}  & \multicolumn{1}{c}{\textbf{$\omega_{cnt}$ }}  \\ 

\multicolumn{1}{c}{\textbf{}} & \multicolumn{1}{c}{\textbf{ ($10^{14}$ $M_{\odot}$) }} & \multicolumn{1}{c}{\textbf{ (Mpc) }}& \multicolumn{1}{c}{\textbf{ ($10^{10}$ $M_{\odot}$/ Mpc$^{2}$) }} & \multicolumn{1}{c}{\textbf{}} &\multicolumn{1}{c}{\textbf{}}&\multicolumn{1}{c}{\textbf{}} \\ \hline 
\\
\endhead
\\
\hline \multicolumn{1}{r}{{Continued on next page}} \\ \hline
\endfoot

\hline \hline
\endlastfoot

\hline\hline
\\

Centaurus		 &   0.63	 &  0.17  &  2.41  &   0.044 & 	0.139  & 0.074 \\     

A1060		 &  0.59 	 &  0.21  &  1.82  &  0.010 & 	0.024 & 0.009  \\

A262 		 &  0.52 	 &  0.24  &  1.24  &  0.015 & 	0.025  & 0.011  \\

Perseus		 &  1.49  &  0.24  &  2.89  & 0.048 & 	0.181  &  0.39 \\      

AWM7		 &  0.86 	 &  0.24  & 1.82  &  0.045	 & 0.106  & 0.061  \\

Coma   		 &  2.72  & 	0.33  &  2.78  & 0.026	 & 0.094 &  0.06 2 \\

A3581     		 &  1.32  &  0.27  & 1.35  &  0.028	 & 0.050 &  0.013 \\    

Ophiuchus     	 &  4.14   &  0.38  &  3.05  &  0.009 & 	0.037 &  0.032  \\     

A4038          	 &  1.31 	 &  0.39  &  0.91  & 0.008  & 	0.010 &  0.007   \\    
  	
A496    		 &  2.29   &  0.44  &  1.24 &  0.038 & 	0.061 &  0.044 \\ 
 
A2063		 & 1.92	 &  0.48  &  0.88  & 0.008 & 	0.009 &  0.0057  \\

A2147    		 & 2.06	 &  0.47  &  0.96  & 0.002 & 	0.003 &  0.0016	 \\   
			
A3571          	 &  3.94 	 &  0.53  &  1.42  & 0.007 & 	0.013 &  0.012	 \\
		
A3558           	 &  3.40  &  0.64  & 0.82   &  0.012	 & 0.013 &  0.01 \\

A4059        	 & 2.75  &  0.61  &  0.27   &  	0.010 & 	0.003 &  0.007	 \\

Triangulum Australis  &  7.58	 &  0.68  &  1.66  &  0.002 & 	0.006  &  0.003 \\

Hydra A   		 &  2.68	 &  0.72 &  0.51  & 0.025 & 0.016  & 0.023 \\

A754		 &  7.91  &  0.72  &  1.48  &  0.004	 & 0.008 &  0.0032 \\ 

A2319       	 &  6.93  &  0.72  &  1.31  &  0.003	 & 0.0004 &  0.033 \\

Cygnus A		 &  3.81 	 &  0.72  &  0.72  & 0.005 & 	0.005 &  0.004 \\
 
AS1101   		 &  1.95	 &  0.79  &  0.34  & 0.025 & 	0.011 &  0.0136 \\       
   			
A3112     		 &  4.44	 &  0.96  &  0.45  &  0.054 & 	0.034 &  0.0337	  \\  
			 
A2597        	 &  3.61	 &  0.96 &  0.29   &  0.004 & 	0.002 &  0.002	 \\     

A478      		 &  8.30	 &  1.10  &  0.61  &  0.019  & 	0.014 &  0.017 \\

PKS0745$-$19  	 &  10.03  &  1.27 &  0.52   &  0.005 & 	0.003 &  0.003  \\   

A2811     		 &  5.29	 &  1.08  &  0.15  &  0.007	 & 0.001 &  0.0018 \\

A2034   		 &  8.07    &  1.25  &  0.35  &  0.005	 & 0.002 &  0.002 \\

RXC J0616.8$-$4748	 &  3.97 	 &  0.99  &  0.16  & 0.0069 & 	0.001 &  0.0007  \\	

RXC J0145.0$-$5300	 &  6.11	 &  1.14  &   0.25  & 0.011  & 	0.003 &  0.0025  \\		

RXC J1044.5$-$0704 	 &  3.05	 &  0.89  & 0.09  &   0.007	 & 0.0009 &  0.0014 \\

A1068        		 &  4.44	 &  1.01  &  0.12  &  0.005	 & 0.0009 	 &  0.0012  \\

RXC J2218.6$-$3853 	 &  6.68	 &  1.16  &  0.20  &  0.005 & 	0.001 &  0.0013 \\
 
RXC J0605.8$-$3518   &  4.91	 &  1.05  &  0.14  & 0.005 & 	0.001 & 0.0013  \\
 
A1413		 &  9.09	 &  1.29  &  0.24  &  0.053  & 	0.016 &   0.018	 \\

A2204          	 & 8.86	 &  1.27  &  0.21  &  0.019 & 	0.005  &  0.010 \\ 

A3888       	 &  8.57	 &  1.26  &  0.20  & 0.013 & 	0.003 &  0.004	 \\

RXC J0958.3$-$1103  &  6.58 &  1.15  &  0.15  & 0.002 & 	0.0005 &  0.0006		 \\

A545    &  10.79	 &  1.36  &  0.25  & 0.002 & 	0.0005  &  0.0004 \\

RXC J2014.8-2430	 &  6.18  &  1.12 &  0.13  &  0.006  & 0.001  &  0.002	 \\

RX J1720.1+2638  	 &  6.64	 &  1.25 &  0.13   &  0.016	 & 0.003  &  0.004 \\

RXC J0645.4-5413 	 &  8.55	 &  1.47 &  0.16  &  0.005 & 	0.001 & 0.001 \\

A1201   &  5.78	 &  1.10  &  0.11 &  	0.015 & 	0.002 &  0.0017 \\

A1914  &  13.93	 &  1.67  & 0.26 &  	0.006	 & 0.002 &  0.002  \\

A2345 	 &  7.65  &  1.20  &  0.14  &  	0.015 & 0.003	 & 0.002 \\

A2218	 &  7.48	 &  1.19  &  0.13  & 	0.012 & 	0.002	 &  0.002 \\

A2254	 &  7.47	 &  1.19  &  0.13  & 0.017 & 	0.003 & 0.002 \\

A665   &  9.50	 &  1.29  &  0.16 &  	0.006 & 	0.001 & 0.0015 \\	
 
A1689    &  12.55  &  1.42  &  0.20 & 0.010 & 	0.002 &  0.004	 \\
  	
A383 	 &  3.48  &  0.92  & 	 0.05  &  0.008  & 	0.0005  & 0.009	 \\	
  
 A520  &   7.75  & 1.20   &  0.11  &  0.009 & 	0.001 &  0.001 \\
   
A2163  &  26.34	 &   1.80  &  0.34  &  0.003 & 	0.001 &  0.001 \\

A209  &  8.82	 &  1.25  &  0.11  &  0.005	 & 0.0007  &  0.0007 \\
     
A963	 &  6.81	 &  1.15  &  0.07  &  0.006	 & 0.0006  &  0.001 \\
 
RXC J1504.1-0248  &  8.87 &  1.25  &  0.09 & 0.011 & 	0.003 &  0.004 \\

MS 0735.6+7421  & 3.89  &  0.95  &  0.04 &  0.014  & 	0.0008  &  0.001	 \\

A773  &  9.34 	 &  1.27  &  0.11  & 0.004 & 	0.0005  &  0.0004  \\

AS0592 	 &  13.27	 &  1.42  &  0.14  &  0.008	 & 0.002  &  0.0017 \\

A2390   & 12.07  &  1.38  &  0.13  &  0.003	 & 0.0005  & 0.0008 \\

A2667	 &  9.66 	 &  1.28  & 0.10  &  0.006	 & 0.0007 &  0.0011 \\

A267	 &   4.83	 &  1.01  &  0.05  &  0.002 & 	0.0001 &  0.0005 \\

RXC J2129.6+0005  &  3.06	 &  0.87  &  0.03  & 0.0097 & 	0.0004  &  0.001 \\
      
RXC J1314.4-2515  &  8.61 	 &  1.22   &  0.07  & 0.010 & 	0.0009 &  0.004 \\
      
A1835  &  12.15	 &  1.37  & 0.10 &  	 0.037 & 	0.005 &  0.009 \\

A1758		 &  4.54	 &  1.04  &  0.03 &  0.009 & 	0.0004 &  0.0008 \\

A1763	 &  10.47 &  1.32  &  0.11 &  0.004	 & 0.0005  &  0.0005 \\

A689 	 &  22.51	 &  1.66 &   0.15  &  0.002  &  0.0001	 &  0.0001	  \\

ZW 3146   	 &  6.72 &  1.11  & 0.04  & 	0.059 & 	0.003  &  0.010 \\

A781   	 &  5.57	 &  1.04  &  0.03  &  0.018 & 	0.0007  &  0.001 \\

Bullet	 & 	15.24	 & 	1.45	 & 	0.09	 & 	0.006 &	0.0007	 &  0.001 \\

MS 2137.3-2353   &  4.31  &  0.95  &  0.02 & 	0.003 & 	0.0001 & 0.0002 \\

MACS J2229.7-2755  & 3.51  &  0.88  &  0.02  & 0.009	 & 0.0001  & 0.0006 \\

MACS J1532.8+3021 	 &  4.85	 &  0.97  &  0.02  & 0.003 & 	0.0007  & 0.0003 \\

AS1063   &  16.80  &  1.48  &  0.07 &  0.004 & 0.0004   & 0.0008 \\\\
\end{longtable*}
}

These response files will be 
solely used to measure the flux of the detected 3.57 keV line; for the rest
of the components, clearly originating in the ICM, the X-ray flux weighting
is correct. The dark matter response weighting was done using the following
approach.

The surface brightness of the DM decay signal is proportional to the DM
column density $S_{DM} = \int _{l.o.s.} \rho_{DM} (r) dr$. The observed
photon flux from the DM decay into a solid angle $\Omega_{FOV}$ is given by

\begin{equation}
F_{DM} = \frac{M_{DM}^{FOV} }{4\pi D_{L}^{2} } \frac{\Gamma_{\gamma}}{m_{s}} (1+z)\ \  \ \rm{photons\ cm^{-2}\ s^{-1}}.
\label{eqn:dmflux}
\end{equation}

\noindent where $\Gamma_{\gamma}$ and $m_{s}$ are the decay rate and mass of the
sterile neutrino (see Equation\ \ref{eq:rate} and \citet{pal1982}), $M_{DM}^{FOV}$ is the
projected DM mass within the spectral extraction region ($R_{ext}$,
which is either $R_{500}$ or $R_{FOV}$), and $D_{L}$ is the luminosity distance.

The DM mass projected along the line of sight is

\begin{equation}
M_{DM}^{FOV} = \int_{los} \rho_{DM} (r) dr,
\end{equation}
\noindent where $\rho_{DM}(r)$ is the distribution of dark matter 
determined by the Navarro--Frenk--White (NFW) profile \citep{navarro1997} and given by

\begin{equation}
\rho_{DM}(r) = \frac{\rho_{c}}{(r/r_{s})(1+r/r_{s})^{2}},
\end{equation}

\noindent where $\rho_{c}$ is a characteristic density and $r_s$ is a scale radius.
The integration of the dark matter distribution within the extraction radius 
(given in Table \ref{table:weight}) is along the line of sight. An extraction radius
of 700$^{\prime\prime}$ was used for the clusters larger than the field of view (FOV)
of \xmm, while an extraction radius of 
$R_{500}$ was used for the clusters smaller than the FOV.

The expected contribution of each cluster $i$ to the total DM line flux in the
stacked spectrum is
\begin{equation}
\omega_{i,dm} = \frac{M_{i, DM}^{proj}(<R_{ext})(1+z_{i})}{4\pi D_{i,L}^{2}}\ \frac{e_{i}}{e_{tot}} 
\label{eqn:weight}.
\end{equation}

\noindent where $z_{i}$ is the redshift of the \textit{i}th cluster, and $e_{i}$ and $e_{tot}$
are the exposure time of the \textit{i}th cluster and the total exposure time of
the sample.

The dark matter mass within the extraction radius is estimated as
\begin{equation}
M_{DM}(R_{ext})  = M_{tot} (R_{ext}) - M_{gas}(R_{ext})-M_{*}(R_{ext}),
\end{equation}

\noindent where $M_{tot} (R_{ext})$, $M_{gas} (R_{ext})$, and $M_{*} (R_{ext})$ are
the total mass, gas mass, and stellar mass in the extraction radius
$R_{ext}$, respectively. The observed \citet{vikhlinin2009}
temperature$-$mass scaling relation was used to infer total masses for the
intracluster gas temperatures measured from the \xmm observations.  The gas
mass is determined following the method described in \citet{bulbul2010}.
The contribution of stars to the total baryon budget is modest at large
radii but more important in the cluster centers because of the presence of
cD galaxies.\ At large radii ($\geq R_{500}$), $M_{*}$ is 10\%$-$15\% of the
gas mass \citep{mohr2004,vikhlinin2006}. Stellar masses of each cluster were
determined using the stellar mass $-$ total mass scaling relation
\citep{gonzalez2013}. The calculated dark matter masses were corrected
using this factor. The projected dark matter masses within $R_{ext}$ were
then determined by projecting NFW profiles
\citep{bartelmann1996,golse2002,loewenstein2009}. We used a concentration
parameter $c_{500}=3$ from the \citet{vikhlinin2006} $c-M_{500}$ scaling
relation and the median total mass within $R_{500}$ of the full sample,
which is $\sim \, 6 \times\, 10^{14}\ M_{\odot}$. The projected dark matter
mass within each spectral extraction radius is given in Table
\ref{table:weight}.

Weights for the responses to be included in the stacked-spectrum response
were calculated as follows. The number of dark matter decay photons in each
cluster spectrum is
\begin{equation}
S_{i} = \alpha\ \omega_{i,dm}\ e_{tot} \ A_{i},
\end{equation}
where $A_{i}$ is the ancillary response (the instrument effective area) at
photon energy $E/(1+z_{i})$, and $\alpha$ is the ratio of the decay rate of
sterile neutrinos to the sterile neutrino mass $m_{s}$ (here we denote
$\alpha\equiv \Gamma_{\gamma}/m_{s}$). The total number of dark matter
photons in the stacked line is
\begin{equation}
\begin{split}
S_{line}&=\sum_{i=0}^{i=73}S_{i}\\
&= \alpha\  \omega_{tot} \ e_{tot}\  A_{\omega},
\end{split}
\end{equation}
where the weighted ARF $A_{\omega}$ is a function of the total weight
$\omega_{tot}$,

\begin{equation} 
A_{\omega}= \sum_{i} \frac{\omega_{i}}{\omega_{tot}}A_{i},
\end{equation}
\noindent and
\begin{equation} 
\omega_{tot}= \sum_{i}\omega_{i}.
\end{equation}

The weighted responses $A_{\omega}$ were used to model our new line,
while X-ray count-weighted response files were used to model the 
other known emission lines and the continuum components.

For MOS, the flux in the 3.57 keV line was $4.0^{+0.8}_{-0.8}$
($^{+1.8}_{-1.2}$) $\times\, 10^{-6} \ \rm{photons}\ \rm{cm}^{-2}\ \rm
s^{-1}$, where the errors are 68\% (90\%). For PN, at the best-fit
energy of 3.51 keV, the line flux is $3.9^{+0.6}_{-1.0}$ 
($^{+1.0}_{-1.6})\,\times\, 10^{-6} \ \rm{photons}\ \rm{cm}^{-2}\ \rm
s^{-1}$. If we fix the line energy from the MOS fit, for PN we obtain
the flux $2.5^{+0.6}_{-0.7}$ ($^{+1.0}_{-1.1})\,\times\, 10^{-6} \ \rm{photons}\ 
\rm{cm}^{-2}\ \rm s^{-1}$.

We note that the line energy detected in the stacked PN observations of the
full sample is consistent with the K \textsc{xviii} line at 3.515 keV.
However, the measured flux from this line is a factor of 20 above the expected
flux of the K\textsc{xviii} line,estimated based on the 
measured fluxes of the S \textsc{xvi}, Ca
  \textsc{xix}, and Ca \textsc{xx} lines and assuming a consistent relative 
  abundance for K\textsc{xviii} along with the plasma temperature from AtomDB. 
  In addition, the
detected energy in the stacked MOS observations of the full sample is
$3.5\sigma$ away from the K \textsc{xviii} line. This will be further
discussed later.

Since this is a blind search, in which the line energy is not known
a priori, to estimate the significance of our detection, we must take
into account the``look elsewhere'' effect. We have examined $\sim70$
independent energy resolution elements in our search
band, and would accept a line detection in any of these bins. Taking
this into account, our $4\sigma-5\sigma$ detections correspond to the 
probability of falsely detecting a line at an unknown energy of 0.004\% for
MOS and 0.4\% for PN. However, the line is found at a consistent energy (or
at least in the same independent resolution element) in these two completely
independent samples coming from different instruments. The statistical
chance of such a false detection at the same energy is negligibly low. We
caution that these are just the rough estimates of the statistical
probabilities; systematic uncertainties are also important
(Section \ref{sec:caveats}).

Because estimating statistical significance of faint line features is a notoriously
ill-behaved problem, we have verified the above estimate
with a Monte Carlo simulations. We used the PN detection for this
test, because its significance is lower and a Monte Carlo estimate can
be done using a reasonable number of trials. We generated 1000 random
realizations of a spectrum using a model in
\xspec with no extra emission line and fit each of them with a model
that included an additional line at an arbitrary location and flux. We then counted
the realizations in which the model with the additional line improved
the fit by $\Delta \chi > 11.2$, which corresponds to our PN
detection. This false detection occurred in 4 cases out of 1000, in
agreement with the above 0.4\% probability of false detection in the
stacked PN spectrum.

\begin{figure*}[ht!]
\centering
\hspace{-5mm}\includegraphics[width=8.9cm, angle=0]{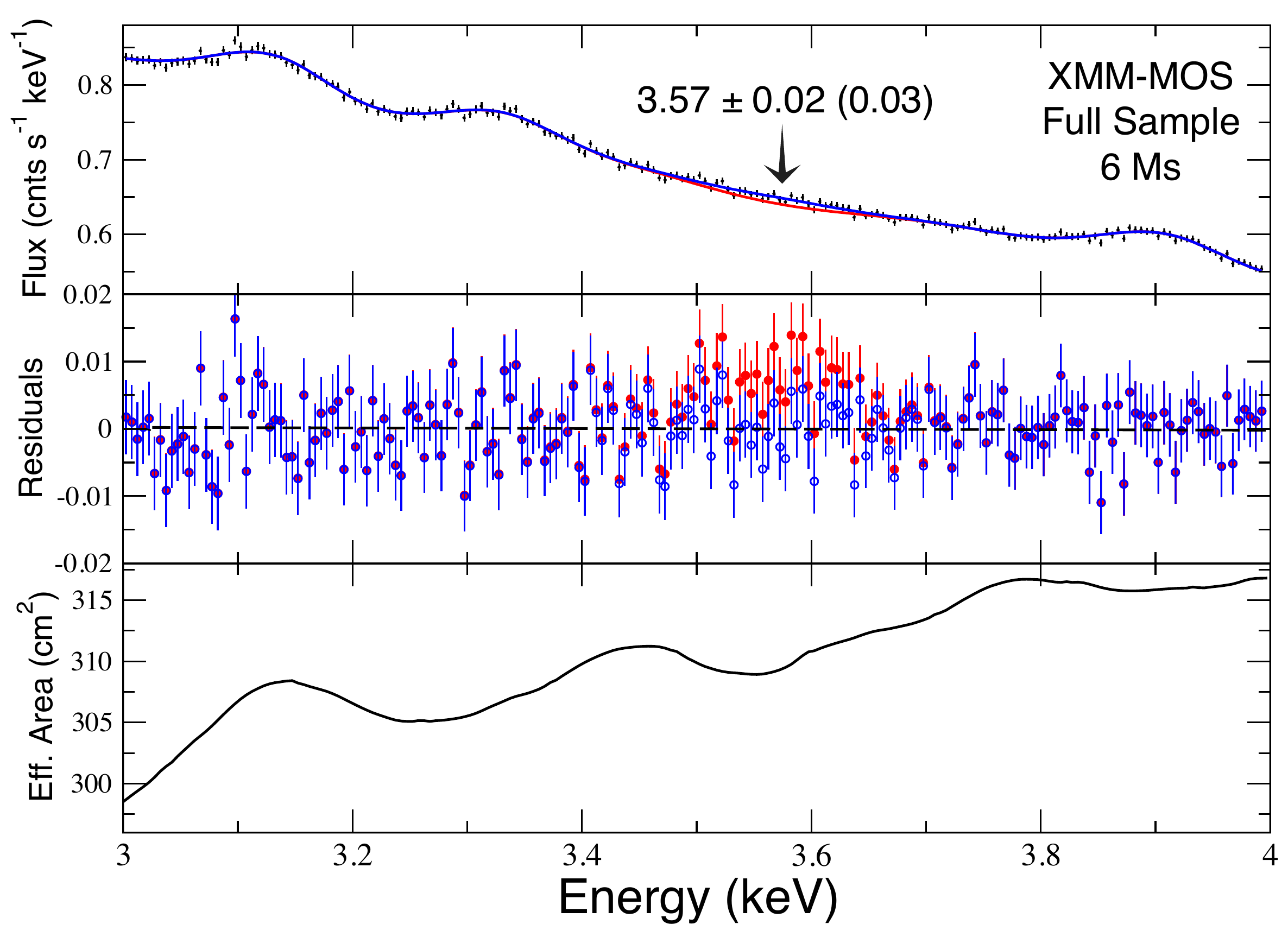}
\includegraphics[width=8.9cm, angle=0]{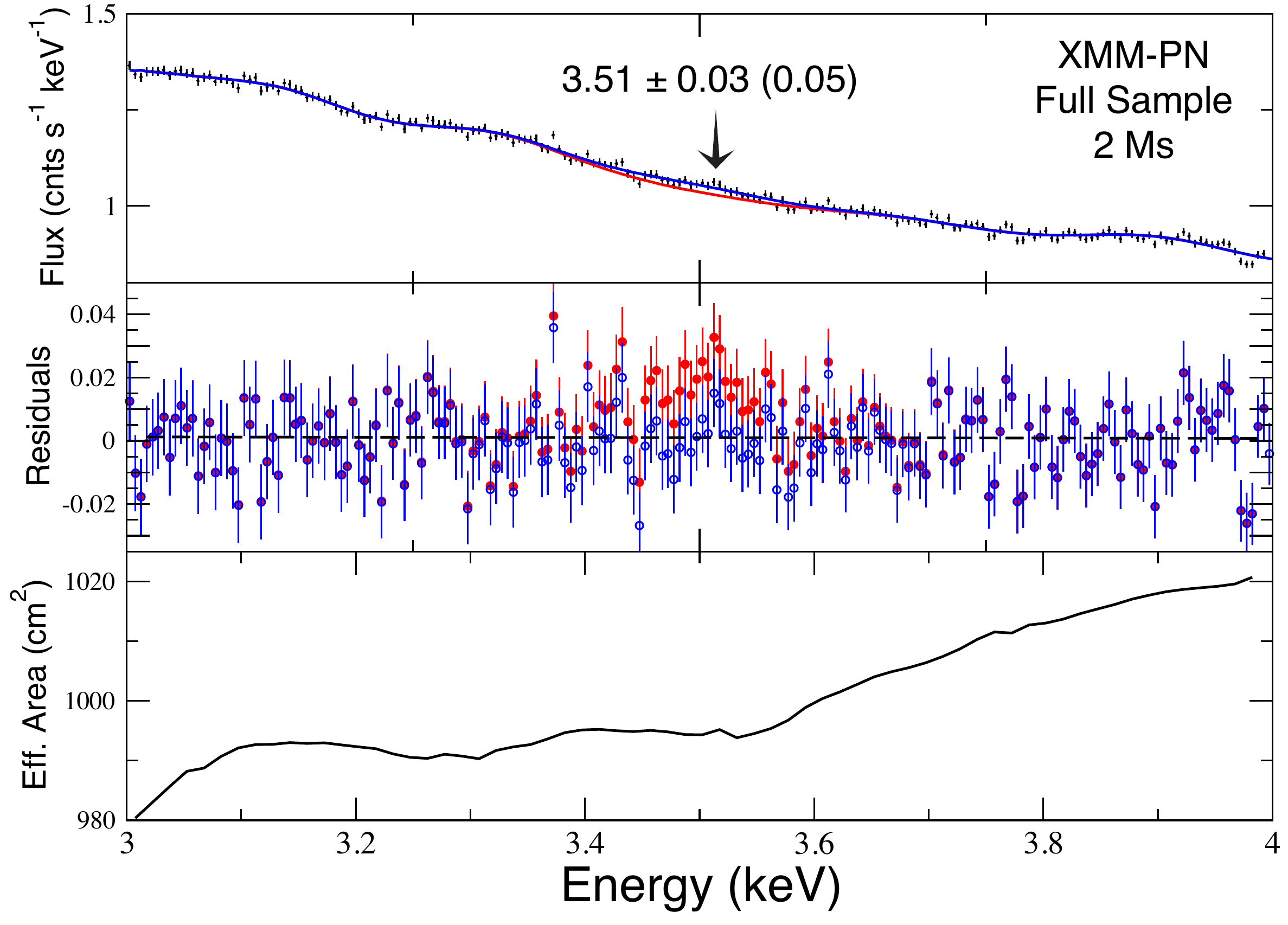}
\hspace{-5mm}\includegraphics[width=8.9cm, angle=0]{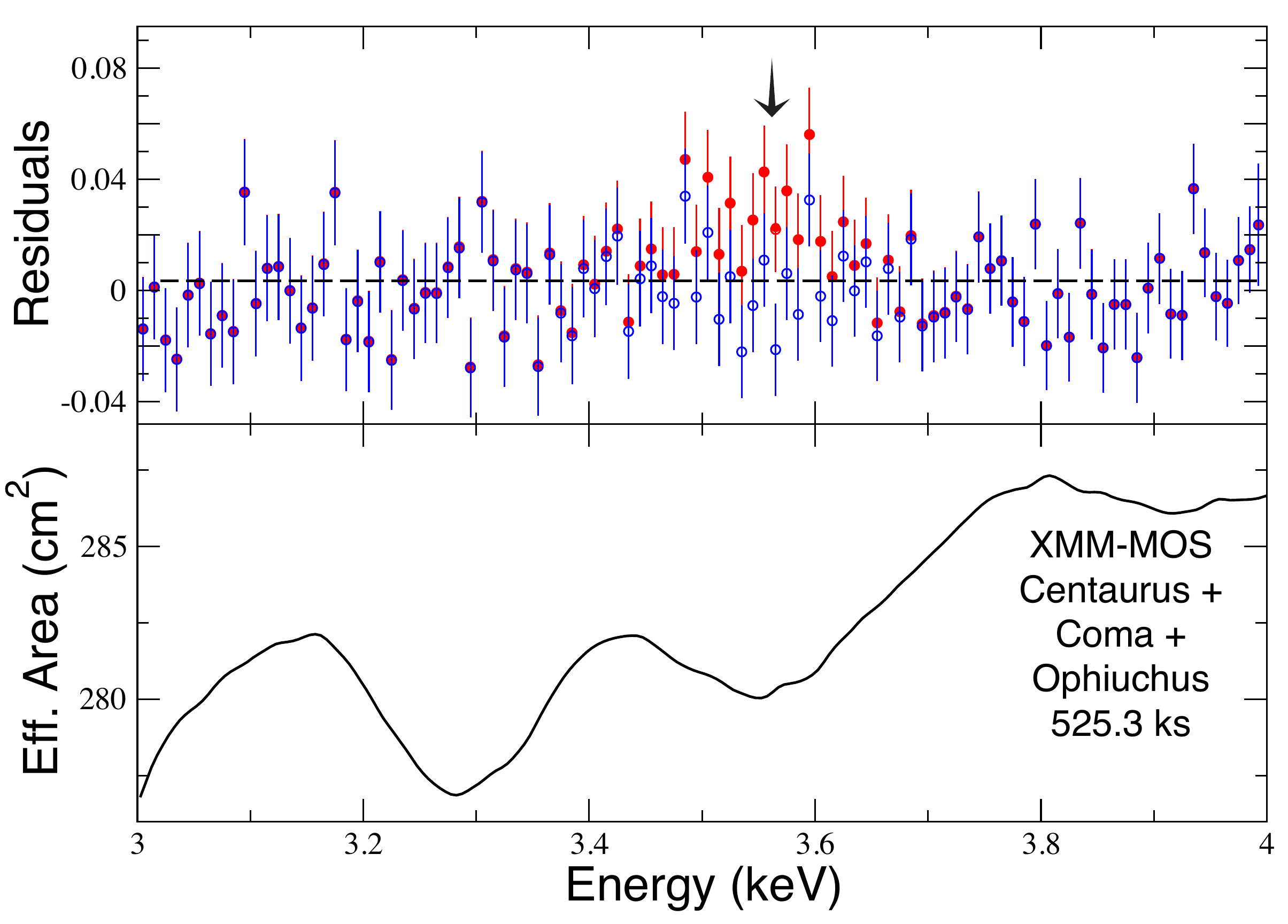}
\includegraphics[width=8.9cm, angle=0]{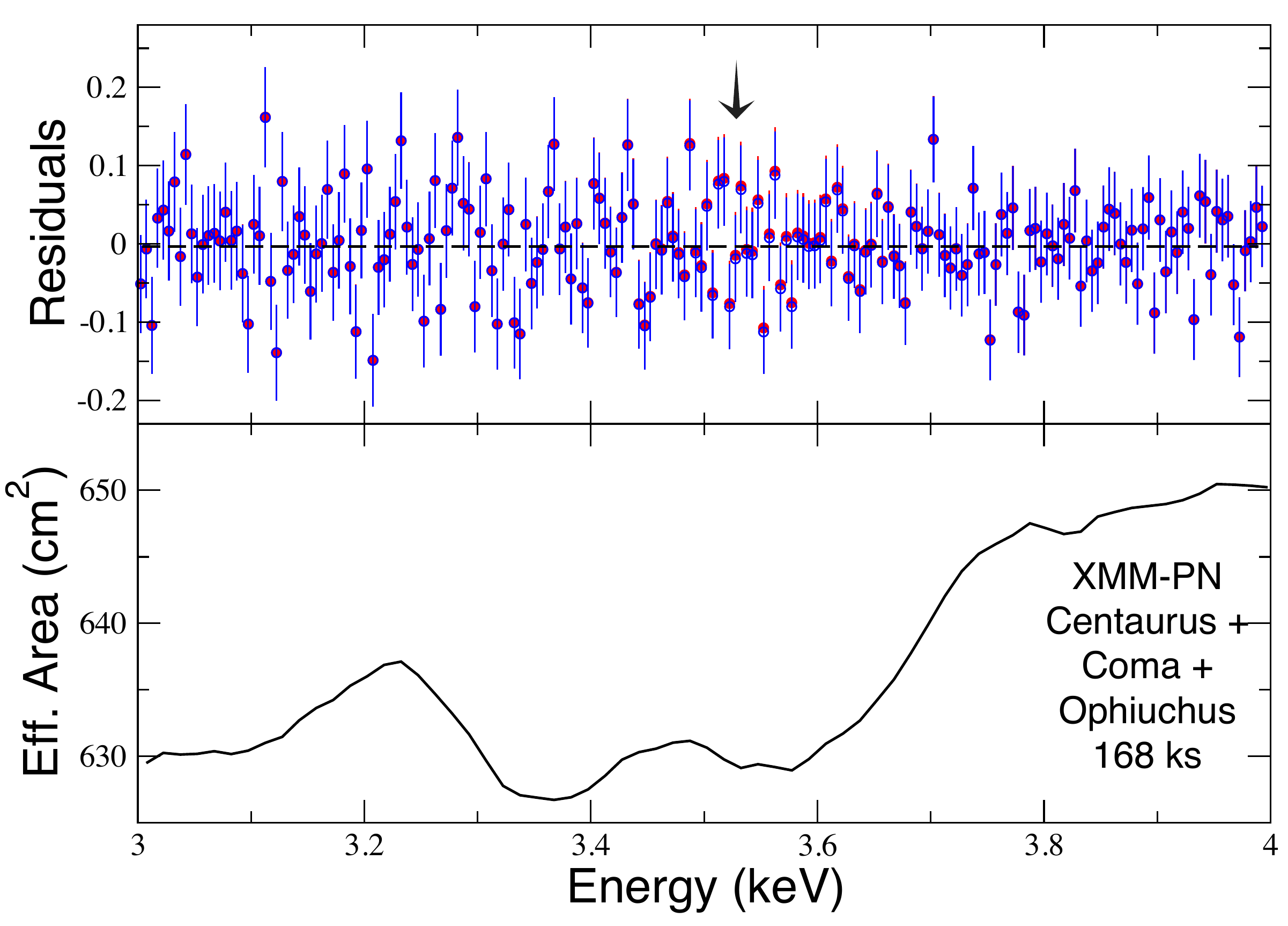}
\hspace{-5mm}\includegraphics[width=8.9cm, angle=0]{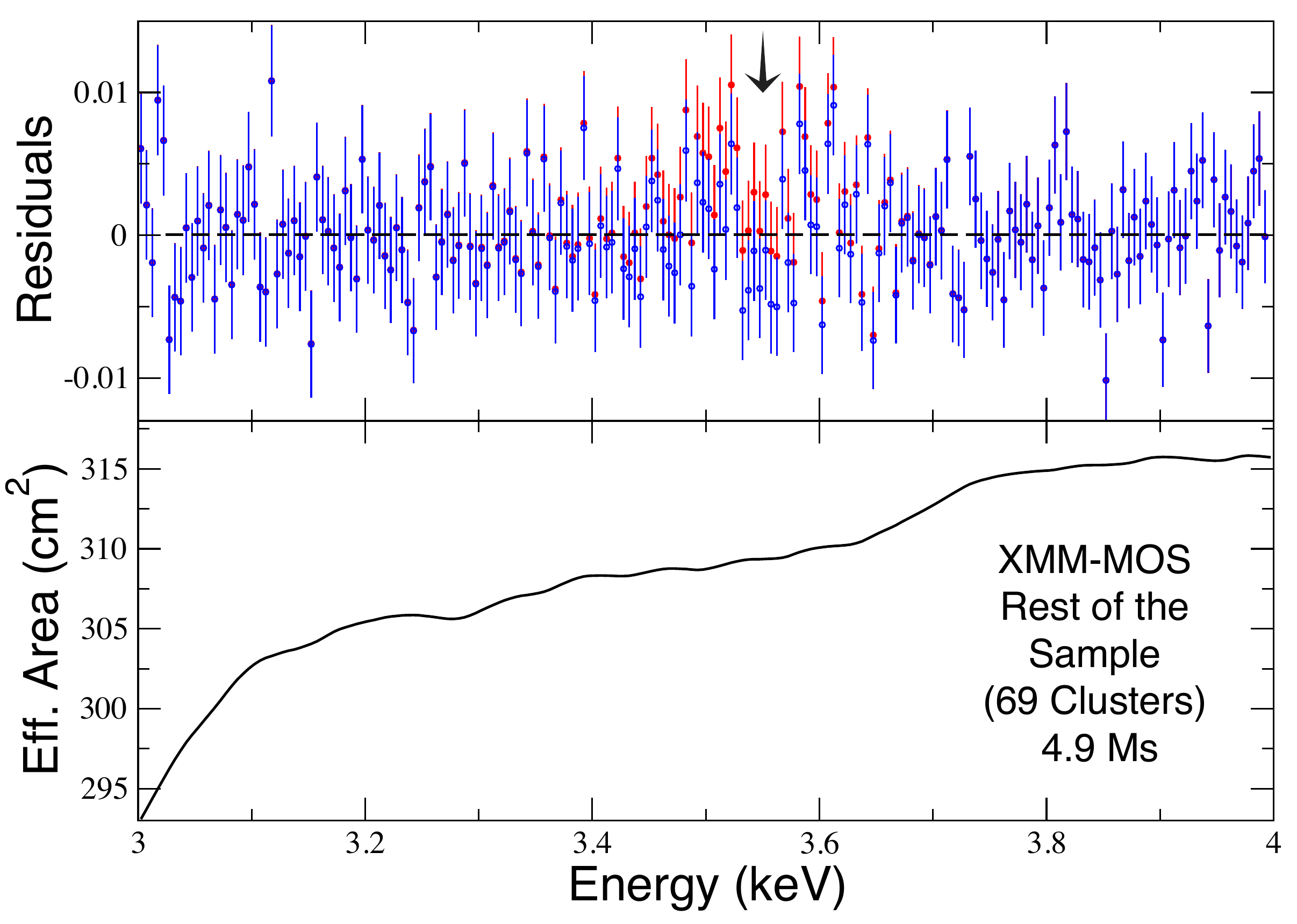}
\includegraphics[width=8.9cm, angle=0]{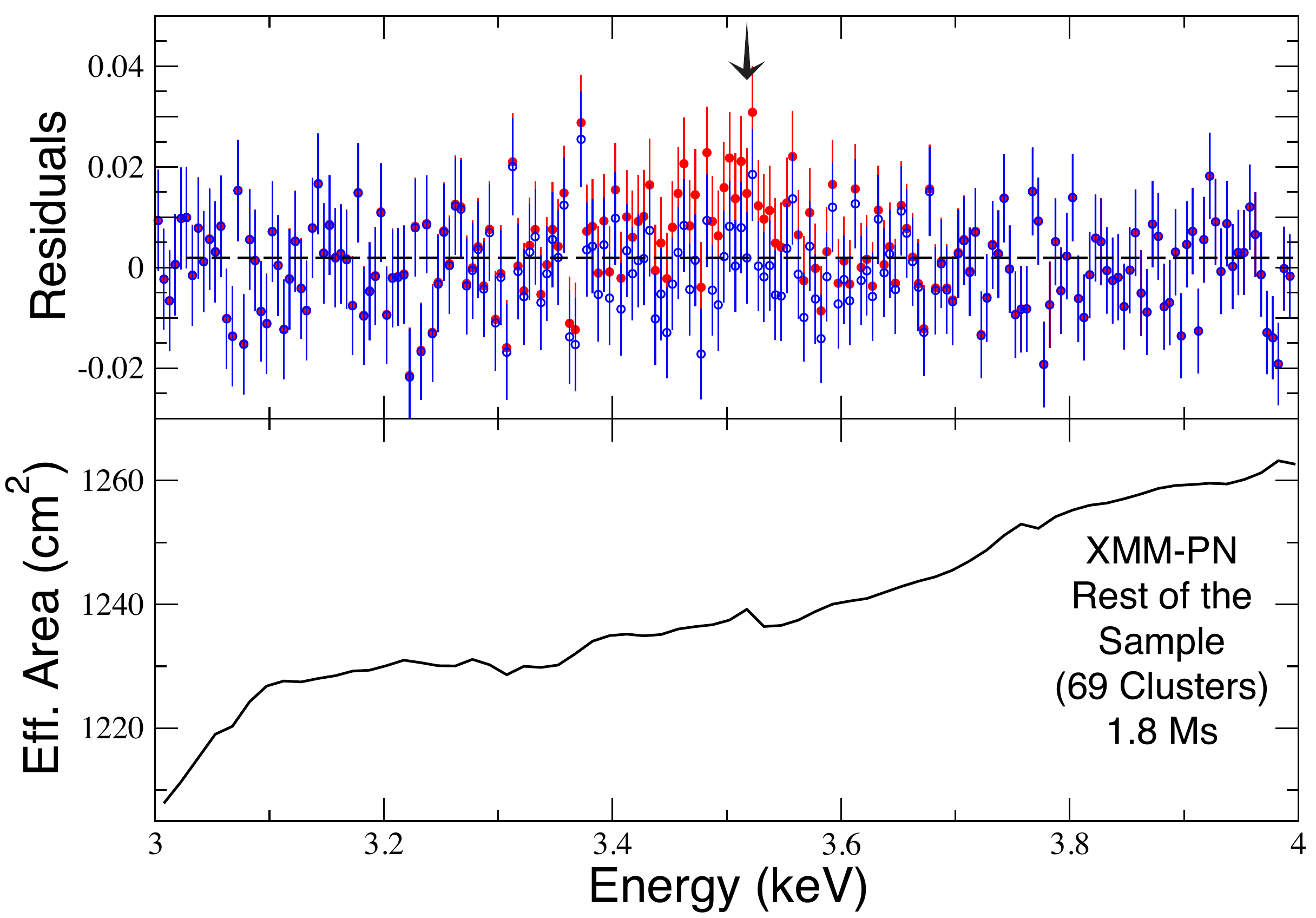}
\caption{Top panels: 3$-$4 keV band of the stacked MOS (left panel) and
  stacked PN (right panel) spectra of the samples. The figures show the
  energy band where the new spectral feature is detected. The
  Gaussian lines with maximum values of the flux normalizations of K
  \textsc{xviii} and Ar \textsc{xvii} estimated using AtomDB were included
  in the models. The red lines in the top panels (shown only for the
  full sample) show the model and the
  excess emission. The blue lines show the total model
  after another Gaussian line is added, representing the new line. Middle panels show the 
  residuals before (red) and after (blue) the Gaussian line is
  added. The bottom panels show the effective area curves (the 
  corresponding ARF). Redshift smearing greatly reduces
  variations of the effective area in  the high-$z$ sample.}
  \vspace{3mm}
\label{fig:epicfits}
\end{figure*}

We also fit the same MOS and PN spectra using the X-ray count-weighted
responses, to check if the detection is dependent on the response
weighting. For MOS, the flux of the detected line was $4.1^{+1.0}_{-0.9}$ ($^{+1.8}_{-1.6}$)
$\times\, 10^{-6} \ \rm{photons}\ \rm{cm}^{-2}\ \rm s^{-1}$; the fit was improved
by $\Delta \chi^{2}$ of 21.8 for 2 dof. For PN, the
line flux was $3.9^{+1.3}_{-1.0}$ ($^{+2.1}_{-2.0})\,\times\, 10^{-6}
\ \rm{photons}\ \rm{cm}^{-2}\ \rm s^{-1}$, while the fit was improved
by $\Delta \chi^{2}$ of 13.8 for 2 dof. This shows that
the detection is robust and the flux is independent of the response scaling.

We will discuss the possible physical interpretations of this emission line
in Section \ref{sec:discussion}. Here we will push forward with one
possible interpretation of the detected line, sterile neutrino
decay, because we need to describe the calculation of 
certain quantities that will be used below for cross-checks and comparison
of the subsamples of our full sample.

For a DM particle decaying radiatively with $E_{\gamma} =m_s/2$,
the detected flux from a clump of matter of a known mass can be converted
into the decay rate.  The
energy of the detected line corresponds to a sterile neutrino particle
mass of $m_s$ = 7.1 $\pm$ 0.07 keV, assuming that the dark matter is solely
composed of sterile neutrinos.  The relation between the flux and mass
implies a mixing angle of
\begin{equation}
\begin{split}
\rm {sin}^{2} (2\theta)= \frac{F_{DM}}{12.76\ \rm{cm^{-2}} \ s^{-1}}
\left( \frac{10^{14}\ M_{\odot}}{M_{DM}^{FOV}}\right) \\
\left( \frac{D_{L}}{100\ \rm{Mpc}}\right)^{2} \left( \frac{1}{1+z}\right)
\left( \frac{1\ \rm{keV}}{m_{s}}\right)^{4}
\end{split}
\end{equation}
\noindent where $F_{DM}$ is the observed DM flux.

\begin{figure*}[ht!]
\centering
\hspace{-5mm}\includegraphics[width=8.9cm, angle=0]{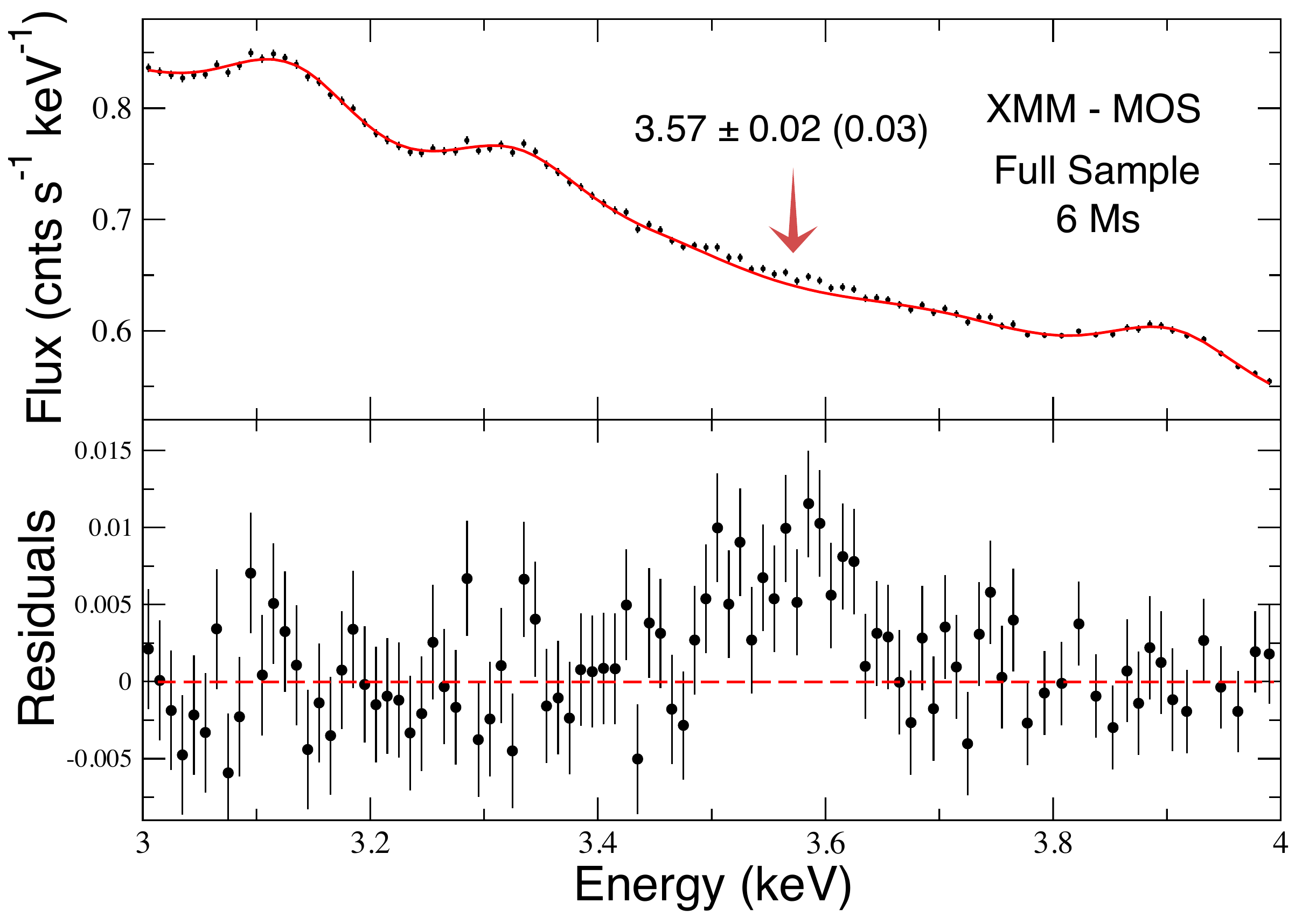}
\includegraphics[width=8.9cm, angle=0]{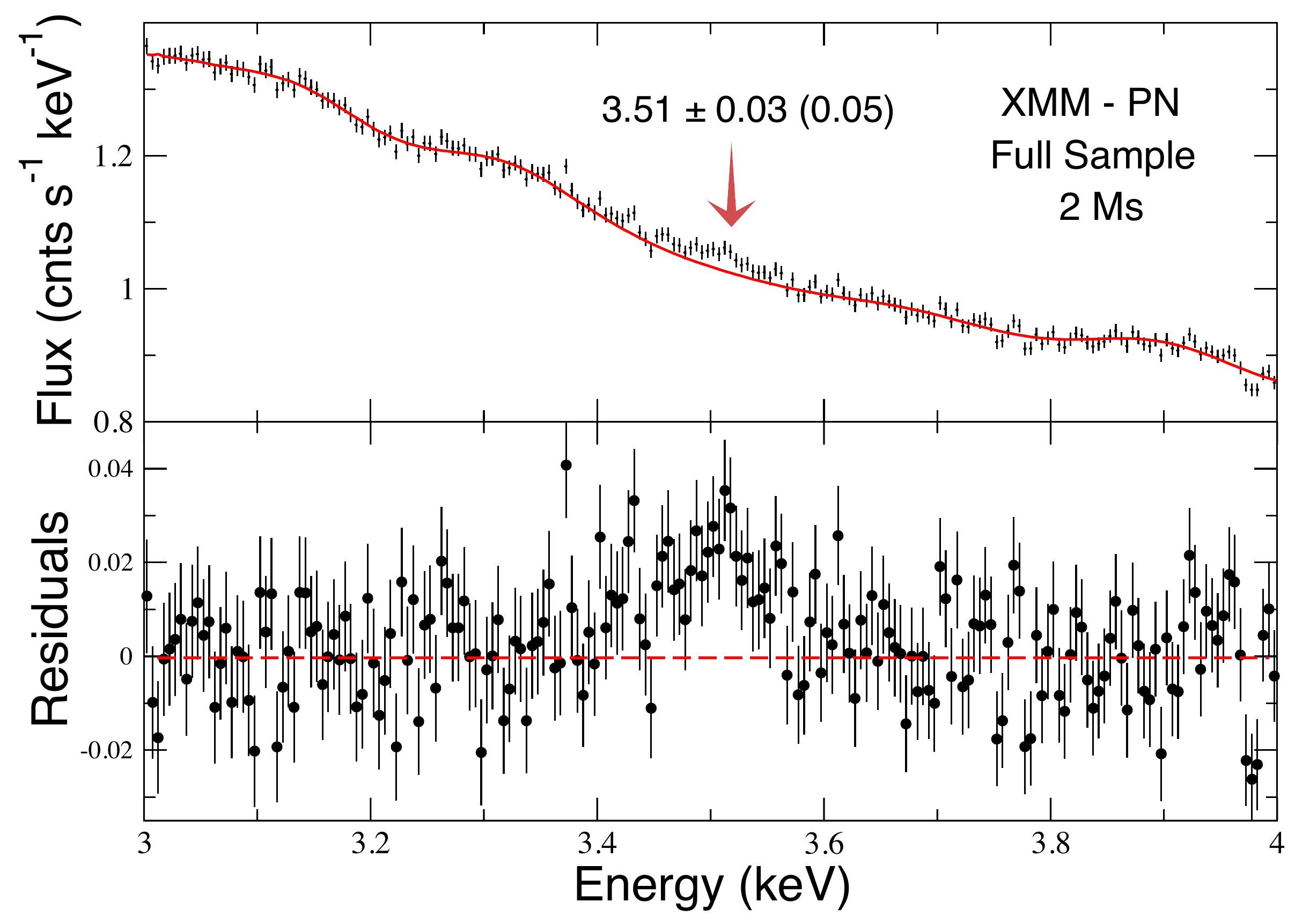}
\hspace{-5mm}\includegraphics[width=8.9cm, angle=0]{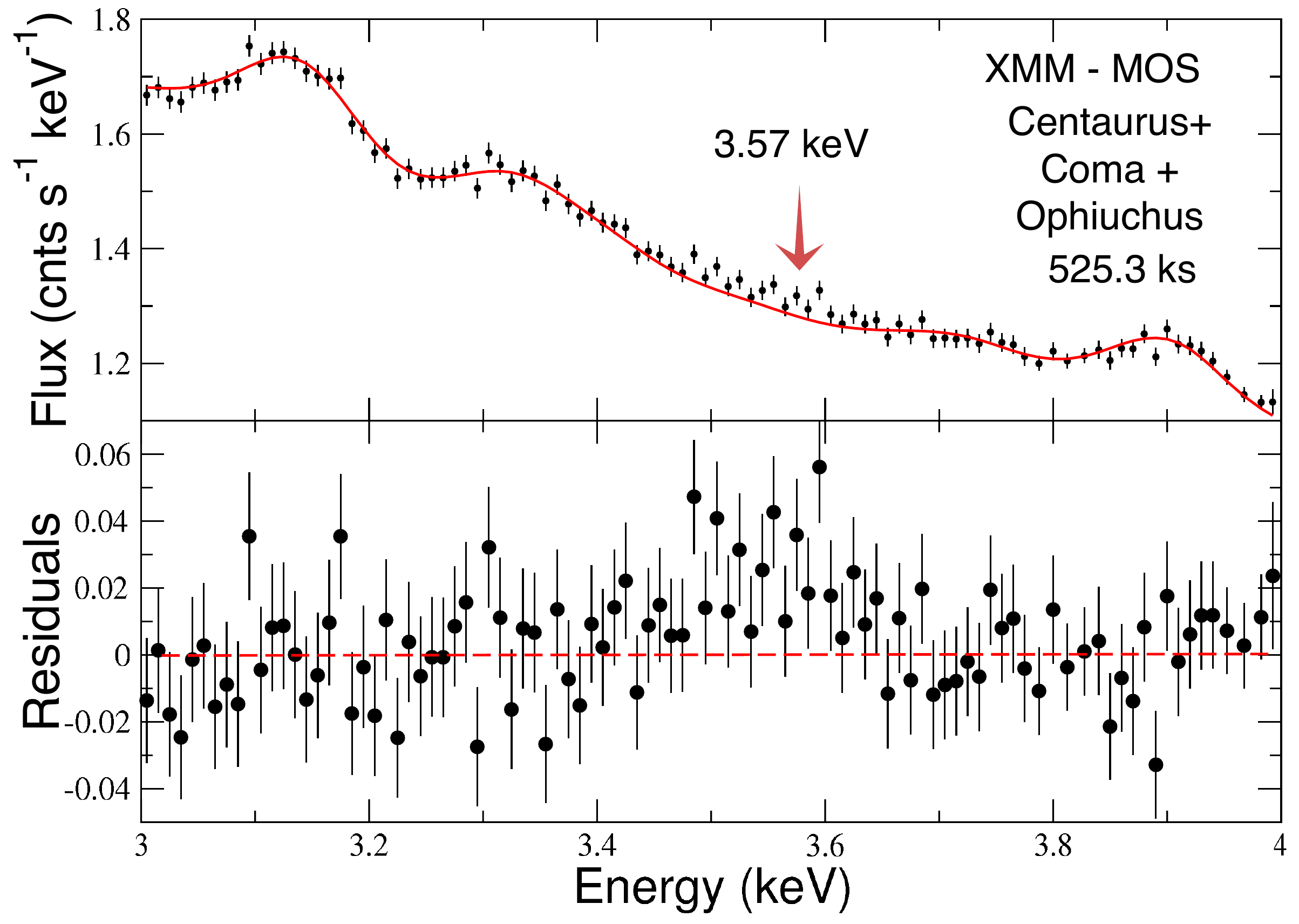}
\includegraphics[width=8.9cm, angle=0]{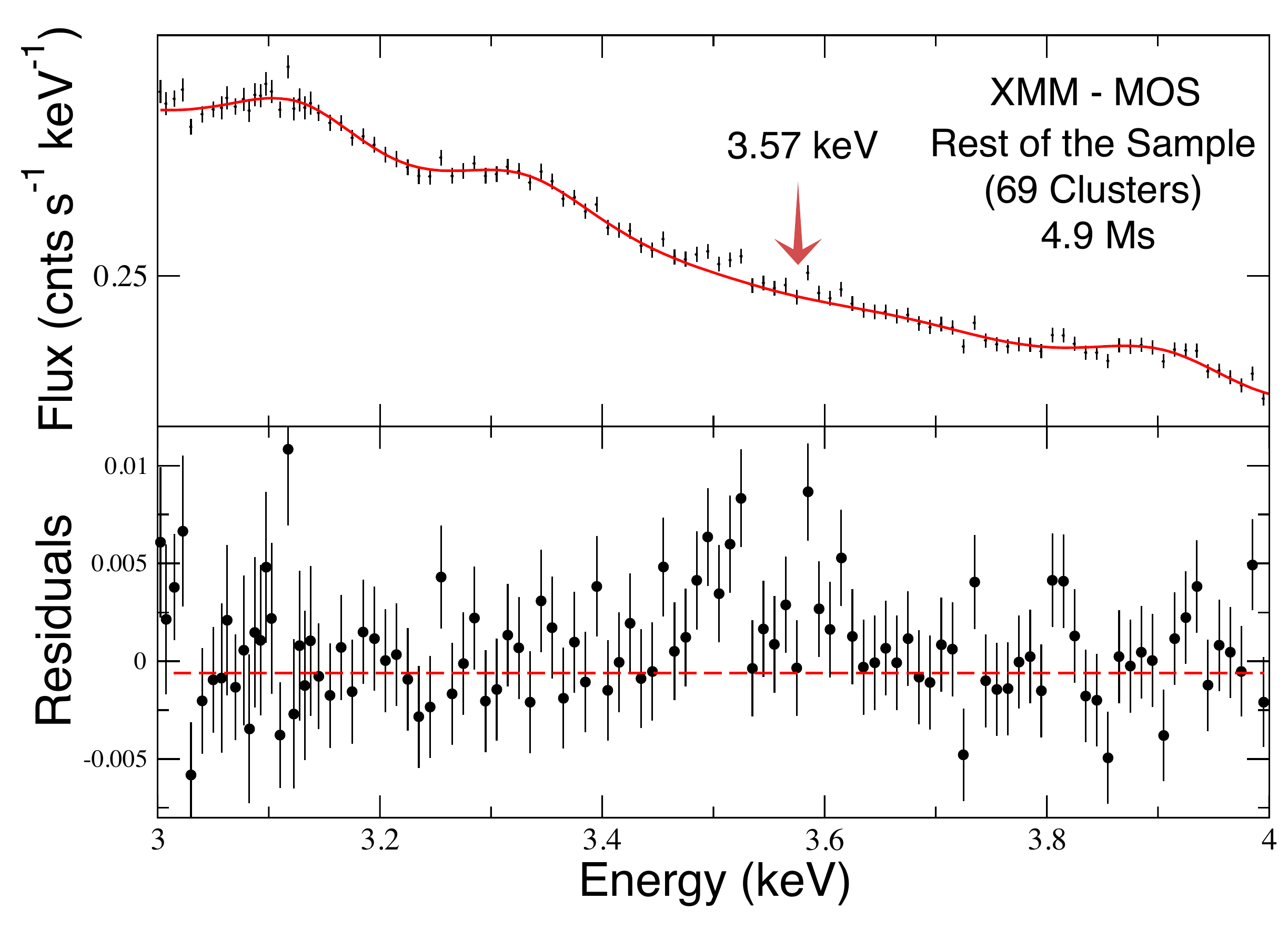}
\includegraphics[width=8.9cm, angle=0]{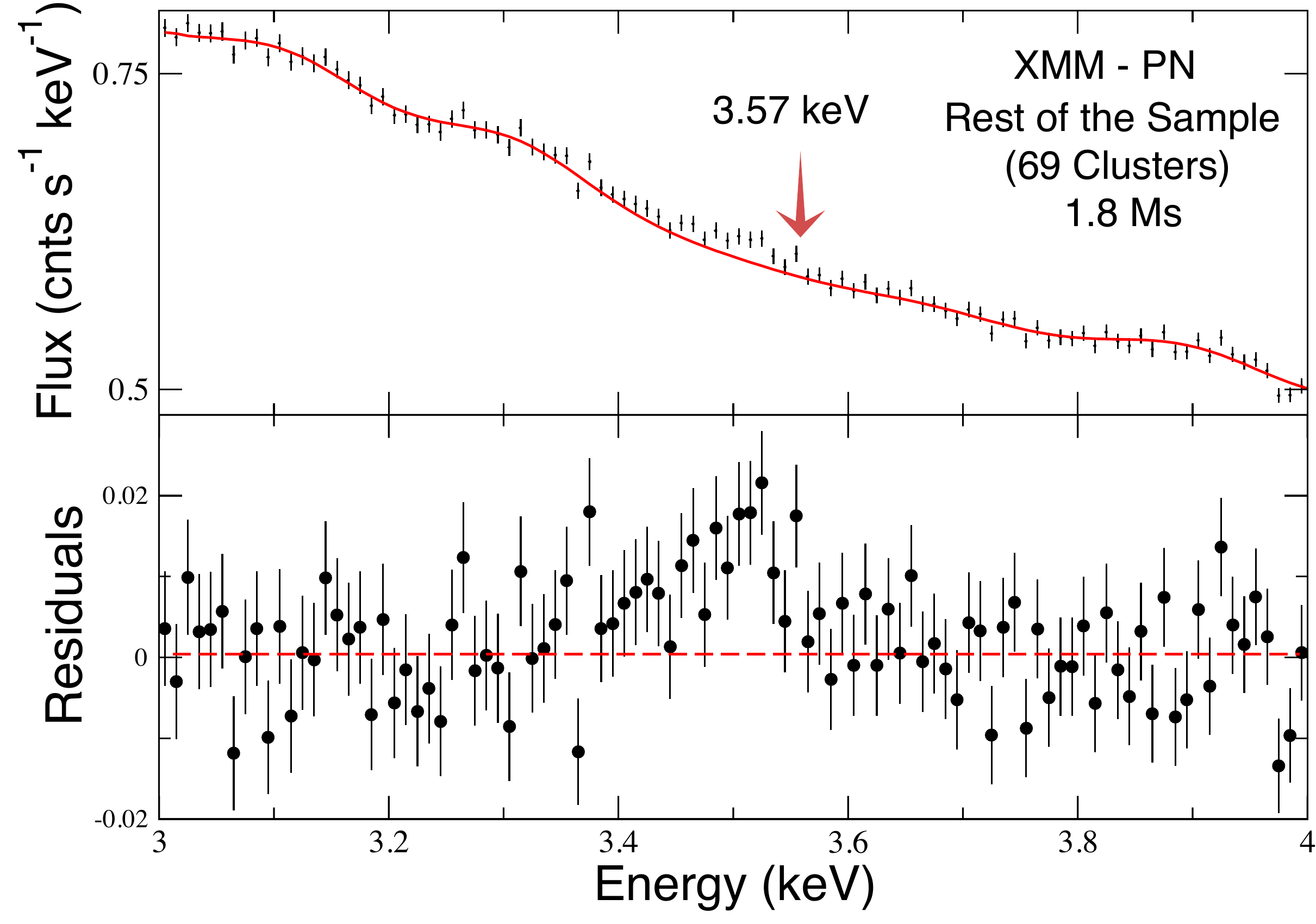}
\caption{3$-$4 keV band of the rebinned \xmm spectra
of the detections.The spectra were rebinned to make the excess at $\sim$3.57 keV
more apparent. \textbf{(APJ VERSION INCLUDES ONLY THE REBINNED MOS SPECTRUM OF THE FULL SAMPLE).}
}
\vspace{3mm}
\label{fig:all-rebin}
\end{figure*}

Using the $\omega_{dm}$ and the projected dark matter masses given in Table
\ref{table:weight}, we find that the weighted projected dark matter mass per
distance squared is $1.82\times\ 10^{10}$ $M_{\odot}/Mpc^{2}$ for the full
sample observed with \xmm MOS. Using Equation (\ref{eqn:dmflux}), one can
calculate the mixing angle for the full MOS cluster sample to be
$\sin^{2}(2\theta)= 6.8^{+1.4}_{-1.4}\ (^{+2.0}_{-3.0})\ \times 10^{-11}$.
The PN observations of the full sample give a mixing angle measurement of
$\sin^{2}(2\theta)= 6.7^{+1.7}_{-1.0}\ (^{+2.7}_{-1.7})\ \times 10^{-11}$
for a weighted mass per distance squared of 1.80 $\times\ 10^{10}$
$M_{\odot}/Mpc^{2}$. These are given in Table \ref{table:flux}.  The PN and
MOS full-sample measurements are consistent with each other and the
constraints placed by previous studies, e.g., the unresolved cosmic X-ray
background (CXB) in the Chandra Deep Fields \citep{abazajian2007} and the
\xmm blank-sky background spectrum \citep{boyarsky2006}, \chandra
observations of the Milky Way \citep{riemer2006}, \chandra observation of
the Bullet Cluster \citep{boyarsky2008}, \chandra observations of the dwarf
galaxy Draco \citep{riemer2009}, and \xmm limits from M31 and Willman 1 and
Fornax dwarf galaxies \citep{boyarsky2010,watson2012}, as showin in Figure\ 
\ref{fig:limitsAbaza}(a). It is in marginal ($\sim$90\% significance)
tension with the most recent \chandra\ limit from M31 \citep{horiuchi2013},
as shown in Figure\ \ref{fig:limitsAbaza}(b).

For the PN flux for the line fixed at the best-fit MOS energy,
the corresponding mixing angle is
$\sin^{2}(2\theta)= 4.3 ^{+1.2}_{-1.0}\ (^{+1.8}_{-1.7})\ \times 10^{-11}$.
This measurement is consistent with that obtained from the
stacked MOS observations at a $1\sigma$ level. Since the
most confident measurements are provided by the highest
signal-to-noise ratio stacked MOS observations of the full sample, we will
use the flux at energy 3.57 keV when comparing the mixing angle
measurements for the sterile neutrino interpretation of this line.

{
{
\begin{table*}[ht!]
\begin{center}
\caption{  \footnotesize Columns (2) and (3) are the measured rest energy and flux of the unidentified line in the units of photons cm$^{-2}$ s$^{-1}$ at the 68\% (90\%) confidence level. The energy's with asterisks are frozen to the indicated values; column (4) and (5) show the $\chi^{2}$ before the line is added to the total model and change in the $\chi^{2}$ when an additional Gaussian component is added to the fit; column (6) is the weighted ratio of mass to distance squared of the samples, and column (7) shows the mixing angle limits measured in each sample. Reported constraining limits are 90\% confidence upper limits. Energies marked with star symbols were held fixed during the model fitting}
\begin{tabular}{ccccccccc}
\hline\hline\\
\scriptsize
&  & (1) & (2) & (3) & (4) & (5) &(6) &(7) \\
Sample && Inst. & Energy & Flux & $\chi ^{2}$ & $\Delta \chi ^{2}$& M$_{DM}^{proj}$/D$^{2}$  & sin$^{2}(2\theta)$\\
&&& (keV)& ($10^{-6}$ phts cm$^{-2}$ s$^{-1}$) & (dof) & ($\Delta$ dof)& ($10^{10}$ M$_\odot$/Mpc$^{2}$)& ( $10^{-11}$ )\\
\\
\hline
\\
&& MOS	 & 3.57 $\pm$ 0.02 (0.03)  &  4.0 $^{+0.8}_{-0.8}$  $(^{+1.8}_{-1.2}$) & 564.8 & 22.8  & 1.82 & 6.8 $^{+1.4}_{-1.4}$  ($^{+2.0}_{-3.0}$)  \\ 
&  &&&& (566)& (2)\\
Full  &\textit{XMM}  \\
 Sample& & PN	 &  3.51 $\pm$ 0.03  ($0.05$)   &  3.9 $^{+0.6}_{-1.0}$ ($^{+1.0}_{-1.6}$)  & 510.5 & 13.9&  1.80 & 6.7 $^{+1.7}_{-1.0}$  ($^{+2.7}_{-1.7}$)\\
& &&&& (564) &  (2) \\
\\
 & & PN	 &  3.57$^{\star}$ &  2.5 $^{+0.6}_{-0.7}$  ($^{+1.0}_{-1.1}$)  & 510.5 & 11.2 &  1.80 & 4.3 $^{+1.2}_{-1.0}$  ($^{+1.8}_{-1.7}$)\\
& &&&& (564)&  (1)\\
\\
\cmidrule{1-9}
\\
& & MOS & 3.57$^{\star}$ & 15.9 $^{+3.4}_{-3.8}$ ($^{+6.7}_{-5.5}$) & 562.3 & 17.1  & 2.68 & 18.2 $^{+4.4}_{-3.9}$  ($^{+12.6}_{-11.5}$)\\
Coma + & &&&& (569) &(1)\\
Centaurus + &\textit{XMM} \\
Ophiuchus&& PN  & 3.57$^{\star}$ &$<$ 9.5  & 377.8  & $-$&$-$ & $<$ 10.9   \\
& &&&& (387)\\
\\
\cmidrule{1-9}
\\

&&MOS & 3.57$^{\star}$ & 21.4 $^{+7.0}_{-6.3}$ ($^{+11.2}_{-10.5}$) & 596.1 & 12.8  & 2.82 & 23.3 $^{+7.6}_{-6.9}$ ($^{+12.2}_{-11.5}$) \\
Perseus & &&&& (574)&(1) \\
(without & \textit{XMM}  \\
the core)&& PN& 3.57$^{\star}$ & $<$ 16.1 &539.1 & $-$&$-$& $<$ 17.6\\
& &&&& (553)\\
\\
\cmidrule{1-9}
\\
 &&MOS & 3.57$^{\star}$ & 52.0 $^{+24.1}_{-15.2}$ ($^{+37.0}_{-21.3}$) & 613.8 & 15.7 & 2.89 & 55.3 $^{+25.5}_{-15.9}$ ($^{+39.3}_{-22.6}$) \\
Perseus& &&&& (574) &(1)\\
 (with& \textit{XMM}  \\
the core) && PN&  3.57$^{\star}$ & $<$ 17.7 & 539.4 & $-$&$-$& $<$ 18.8 \\
 & &&&& (554)\\
\\
\cmidrule{1-9}
\\
 & & MOS	& 3.57$^{\star}$ & 2.1 $^{+0.4}_{-0.5}$  ($^{+0.8}_{-0.8}$)  & 547.2 & 16.5 & 1.08 & 6.0 $^{+1.1}_{-1.4}$  ($^{+2.3}_{-2.3}$)  \\
All& &&&&(573)& (1)\\
Other &\textit{XMM} \\
Clusters && PN	 & 3.57$^{\star}$  & 2.0 $^{+0.3}_{-0.5}$  ($^{+0.5}_{-0.8}$)   & 741.9  & 15.8  & 1.15 &  5.4 $^{+0.8}_{-1.3}$  ($^{+1.3}_{-2.1}$)   \\
& &&&& (751) & (1)\\
\\
\cmidrule{1-9}
\\
& &ACIS-S & 3.56 $\pm$ 0.02 (0.03) &  10.2 $^{+3.7}_{-3.5}$  ($^{+4.8}_{-4.7}$) & 201 & 11.8 & 0.72 &  40.1 $^{+14.5}_{-13.7}$ ($^{+18.9}_{-18.2}$) \\
& &&&& (197) & (2)\\
Perseus& \textit{Chandra} \\
& & ACIS-I & 3.56$^{\star}$   & 18.6 $^{+7.8}_{-8.0}$ ($^{+12.0}_{-16.0}$) & 152.6 & 6.2 & 1.86  & 28.3 $^{+11.8}_{-12.1}$ ($^{+18.2}_{-24.3}$) \\
& &&&& (151) &  (1)\\
\\
\cmidrule{1-9}
\\
Virgo &  \textit{Chandra} &ACIS-I & 3.56$^{\star}$  & $<$ 9.1  & 189.1 & $-$ & 2.41 &  $<$ 10.5 \\
& &&&& (155)\\
\\

\\
\hline\hline
\end{tabular}
\label{table:flux}
\end{center}
\end{table*}
}
%

\subsection{Excluding Bright Nearby Clusters from the Sample}
\label{sec:therest}

We now divide the full cluster sample into three independent
subsamples, in order to check that our line does not originate from any single
object.  The full stacked spectra examined in Section \ref{sec:all} have a
significant contribution of photons from several nearby bright clusters,
e.g., Perseus, Coma, Centaurus, and Ophiuchus.  In order to determine whether
the line detection is dominated by these bright sources, we excluded them
from the sample and stacked the MOS and PN spectra of the remaining 69
fainter galaxy clusters.  We have performed the stacking process following
the same approach described in Section \ref{sec:stacking}. A total of 4.9 Ms of
good stacked MOS and 1.7 Ms good stacked PN exposure were obtained for this
subsample. The weighted mean redshift was 0.06. The stacked MOS and PN
spectra contain 34\% (2.95$\times\,10^{6}$ source counts) and 55\%
(2.79$\times\,10^{6}$ source counts) of the total source counts of the full
cluster sample.

We fit the stacked spectra using the \textit{line-free} \apec model and
additional Gaussian models as described in Section \ref{sec:all} in the 3$-$6 keV
band.  The best-fit temperatures, normalizations of the \textit{line-free}
\apec model, and the fluxes of S \textsc{xvi}, Ca \textsc{xix}, and Ca
\textsc{xx} lines are given in Table \ref{table:temp}.  We then carefully
examined the spectra for any unidentified emission features in the 3.4 $-$
3.7 keV energy interval. Similarly, we determined the maximum fluxes of the
K \textsc{xviii}, K \textsc{xix}, and Ar \textsc{xvii} lines based on the
plasma temperatures and fluxes of hydrogen-like S \textsc{xvi}, helium-like
Ca \textsc{xix}, and hydrogen-like Ca \textsc{xx} lines at 2.63 keV, 3.90
keV, and 4.11 keV, measured from the spectral fits, and AtomDB as described
in Section \ref{sec:all}. As before, the lower and upper limits of the fluxes of K
\textsc{xviii}, K \textsc{xix}, and Ar \textsc{xvii} lines were set to 0.1
to 3 times of the maximum predicted fluxes. The Ar \textsc{xvii} DR line flux at
3.62 keV was allowed to vary between $10^{-3}$ and $10^{-2}$ of the Ar
\textsc{xvii} triplet line at 3.12 keV.

\begin{figure*}[ht!]
\centering
\hspace{-5mm}\includegraphics[width=8.9cm, angle=0]{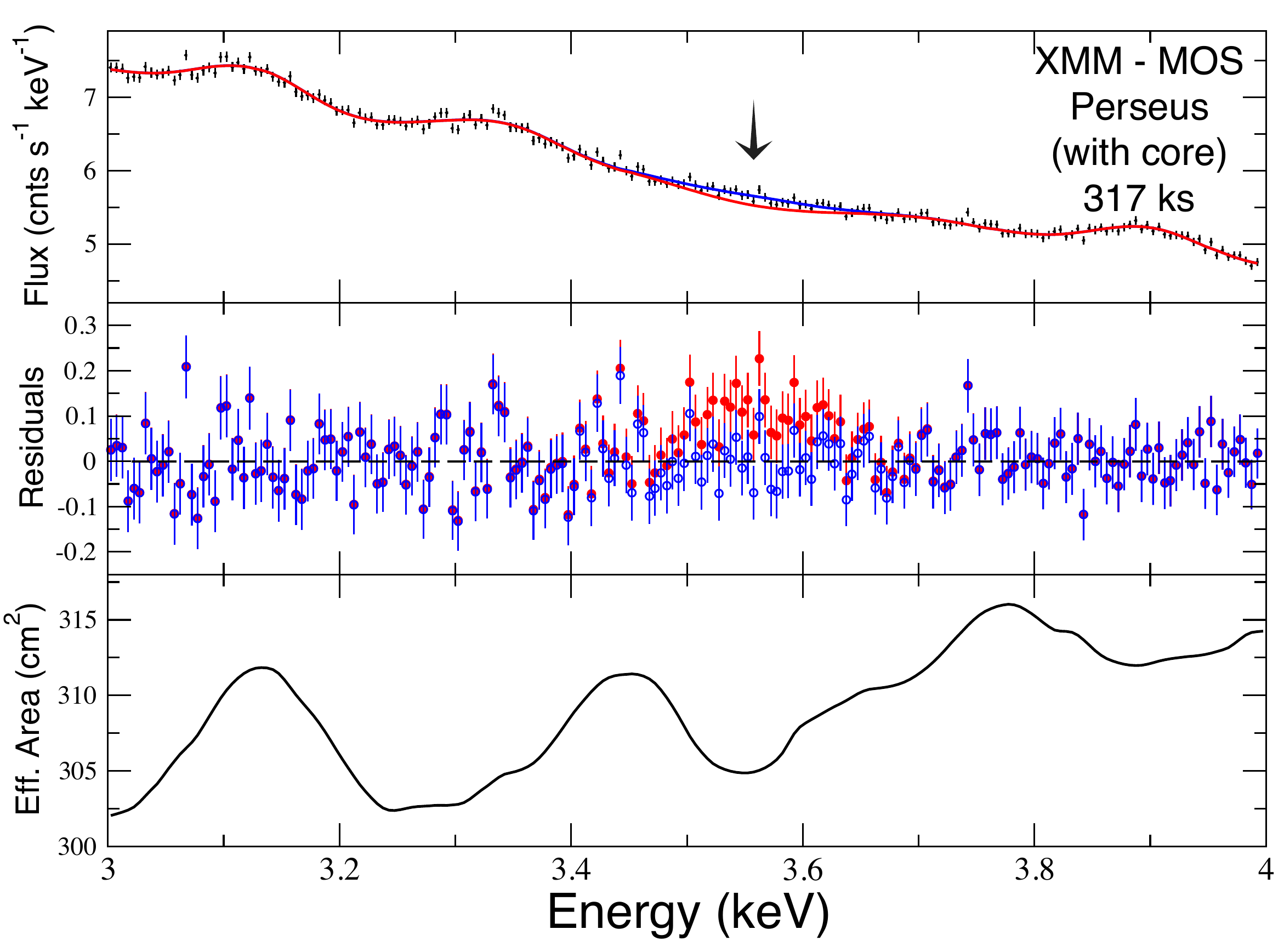}
\includegraphics[width=8.9cm, angle=0]{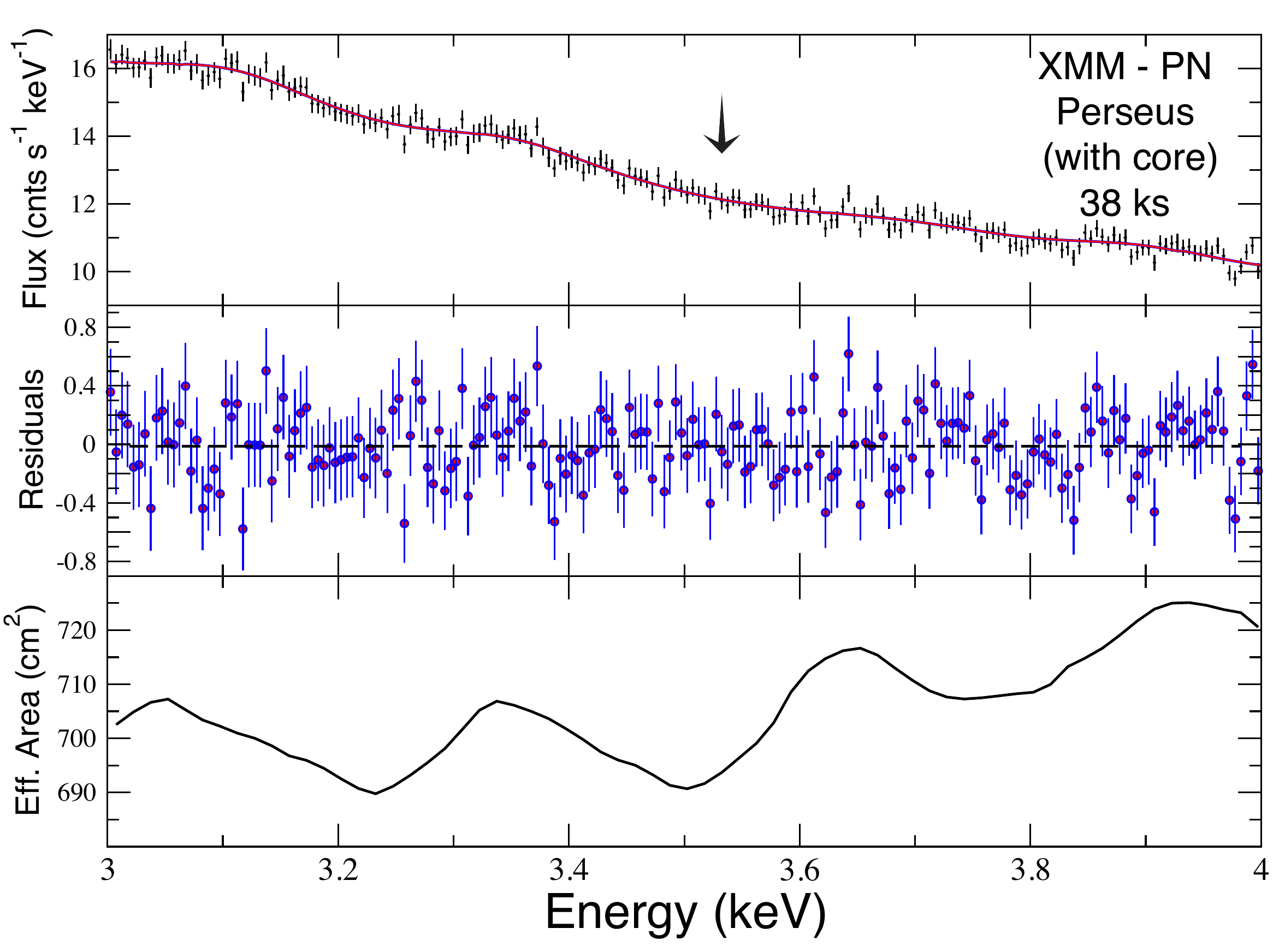}
\caption{3$-$4 keV band of the stacked MOS (left panel) and stacked PN
  (right panel) spectra of the Perseus Cluster.  The figures show the energy
  band, where a new spectral feature at 3.57 keV is detected.  The Gaussian
  lines with peak values of the flux normalizations of K \textsc{xviii} and
  Ar \textsc{xvii} estimated using AtomDB were included in the models. The
  red lines in the top panels show the model and the excess emission in both
  spectra. The blue lines show the total model after a Gaussian line is
  added, indicating that the unidentified spectral line can be modeled with
  a Gaussian. } \vspace{3mm}
\label{fig:perseus-xmm}
\end{figure*}

We obtained an acceptable fit to the stacked MOS spectrum of these 69
clusters. The total $\chi^{2}$ was 557 for 573 degrees of freedom.  Adding
in an extra Gaussian model to the MOS spectrum at 3.57 keV improved the fit
by $\Delta \chi^{2}$ of 16.5 for an additional dof.  We found
that the best-fit flux was 2.1 $^{+0.4}_{-0.5}$ ($^{+0.8}_{-0.8}$) $\times
\ 10^{-6}$ photons cm$^{-2}$ s$^{-1}$. This flux corresponds to a
mixing angle of $\sin^{2}(2\theta)=$ 6.0 $^{+1.1}_{-1.4}$
($^{+2.3}_{-2.3}$) $\times 10^{-11}$, consistent with the mixing angle
estimates obtained from the full sample.

The overall fit to the stacked PN spectrum for these 69 clusters was
acceptable with a total $\chi^{2}$ of 741.9 for 751 dof.
Adding an extra Gaussian line at 3.57 keV improved the fit by $\Delta
\chi^{2}$ of 15.8 for an additional degree of freedom. The PN spectrum
yields the best-fit flux detection of 
2.0 $^{+0.3}_{-0.5}$
($^{+0.5}_{-0.8}$) $ \times 10^{-6}\ \rm{photons}\ \rm{cm}^{-2}\ \rm
s^{-1}$. The mixing angle obtained from the stacked PN observations
$\sin^{2}(2\theta)=$ 5.4 $^{+0.8}_{-1.3}$ ($^{+1.3}_{-2.1}$) $\times\ 
10^{-11}$ is also consistent with the estimates from the full
sample. Bottom panels in Figure \ref{fig:epicfits} show
the residuals before and after a Gaussian line is added at 3.57 keV to MOS and
PN spectral fits.

\subsection{Stacked Spectra of the Nearby Bright Clusters; Centaurus + Coma
  + Ophiuchus} 
\label{sec:brights}

We now check the MOS and PN spectra of the three dominant nearby clusters,
Coma, Ophiuchus, and Centaurus. A total of 525.3 ks of good stacked MOS and
168 ks good stacked PN exposure times were obtained for this subsample. The
total source counts obtained in the MOS and PN spectra were 3.2
$\times\,10^{6}$ and 2.1 $\times\, 10^{6}$, respectively.

We performed the fits as above. The best determinations
for the continuum temperature and normalizations and the fluxes of the
S \textsc{xvi}, Ca \textsc{xix}, and Ca \textsc{xx} are given in Table
\ref{table:temp}. We detected an excess emission feature in the same
band, i.e. 3.4 $-$ 3.7 keV as in the stacked MOS spectra. 
To determine the flux of the emission line at 3.57 keV, we estimated
the maximum fluxes of the K \textsc{xviii}, K \textsc{xix}, and  Ar
\textsc{xvii} lines using the AtomDB and the measured fluxes of S
\textsc{xvi}, Ca \textsc{xix}, and Ca \textsc{xx} as described in Section 
\ref{sec:all}. Using 0.1 and 3 times these fluxes as lower and
upper limits, we found that the unidentified line has a flux of  
1.6$^{+0.3}_{-0.4}$ ($^{+0.7}_{-0.6}$) $\times\, 10^{-5}$  photons
cm$^{-2}$ s$^{-1}$ in the stacked MOS observations. 
Adding this Gaussian to the model improves the fit by
$\Delta \chi^{2}$ of 17.1 for an additional degree of freedom for the
stacked MOS spectrum.  

We then allowed the energy of the additional Gaussian model to vary to test whether the 
energy measured from two different samples is the same. 
The best-fit energy obtained from the stacked MOS observations of
Coma, Centaurus, and Ophiuchus clusters was 3.56 $\pm$ 0.02 (0.03),
with a flux of 1.6$^{+0.52}_{-0.44}$ ($^{+0.81}_{-0.70}$) $\times\,
10^{-5}$  photons cm$^{-2}$ s$^{-1}$. This measurement is consistent
with the energy measured in the MOS observations of the full sample.  The
sterile neutrino mixing angle that corresponds to this flux is
$\sin^{2} (2\theta) =18.2^{+4.4}_{-3.9} \ (^{+12.6}_{-11.5})\ \times
10^{-11} $, consistent at $2\sigma$ with the full-sample value.

The fits to the stacked PN observations did not need an additional
Gaussian line, and resulted in a non-detection. This could be due to the
low count statistics of the stacked PN observations (168 ks clean time). A 90\%
upper limit on the flux of this line at 3.57 keV is 9.5 $\times\,
10^{-6}$  photons cm$^{-2}$ s$^{-1}$ from this spectrum; the upper
limit on the mixing angle from this flux limit is consistent with the
full-sample and MOS detections.

\subsection{Perseus}
\label{sec:perseusxmm}

Initially, we extracted the spectrum of the Perseus Cluster using the entire
MOS and PN FOV.  We have co-added the \xmm MOS and PN observations
of the Perseus Cluster in the cluster's frame.  The total exposure time in
the stacked MOS spectrum was 317 ks with a total of 7$\times 10^{6}$ source
counts in the 2 $-$ 10 keV band and 38 ks total exposure with 2$\times
10^{6}$ source counts in the stacked PN observations.

Following the same approach we used for modeling the full
cluster sample, we first fit the MOS and PN observations with the
\textit{line-free apec} model and additional Gaussian models.
Count-weighted responses were used to fit the plasma emission lines and the
continuum emission. Probing the 3$-$4 keV band the MOS observations revealed
residuals around 3.57 keV, at the same energy band where we detected line
emission in the previous samples. The left panel of Figure
\ref{fig:perseus-xmm} shows the detection in the co-added MOS observations of
the Perseus Cluster. Using the limits on the K and Ar lines (Table
\ref{table:fluxArK}) as above and adding a Gaussian model to the MOS spectrum at the
fixed energy of 3.57 keV improved the fit by
$\Delta\chi^{2}$ of 15.7. The best-fit flux at 3.57 keV was
5.2$^{+2.4}_{-1.5}$ ($^{+3.7}_{-2.1}$) $\times\, 10^{-5}$ photons
cm$^{-2}$ s$^{-1}$. 

This flux corresponds to a mixing angle of $\sin^{2}
(2\theta)=5.5^{+2.6}_{-1.6} \ (^{+3.9}_{-2.3})\ \times 10^{-10}$. This angle
not only is an outlier in our measurements from the other samples, but is
also not consistent with the upper limits on the mixing angle at this
value of $m_s$ from the previous
studies \citep[e.g.,][]{horiuchi2013}. 

We were unable to detect the line in the short (38 ks clean time) PN observation of 
Perseus and placed a 90\% upper limit on the flux of the line of 17.7
photons cm$^{-2}$ s$^{-1}$, which corresponds to an upper limit of 
$\sin^2(2\theta) < 1.9 \times 10^{-10}$, consistent with the MOS
detection. Figure \ref{fig:perseus-xmm} shows both \xmm
Perseus spectra.

\begin{figure}[ht!]
\centering
\hspace{-5mm}\includegraphics[width=8.9cm, angle=0]{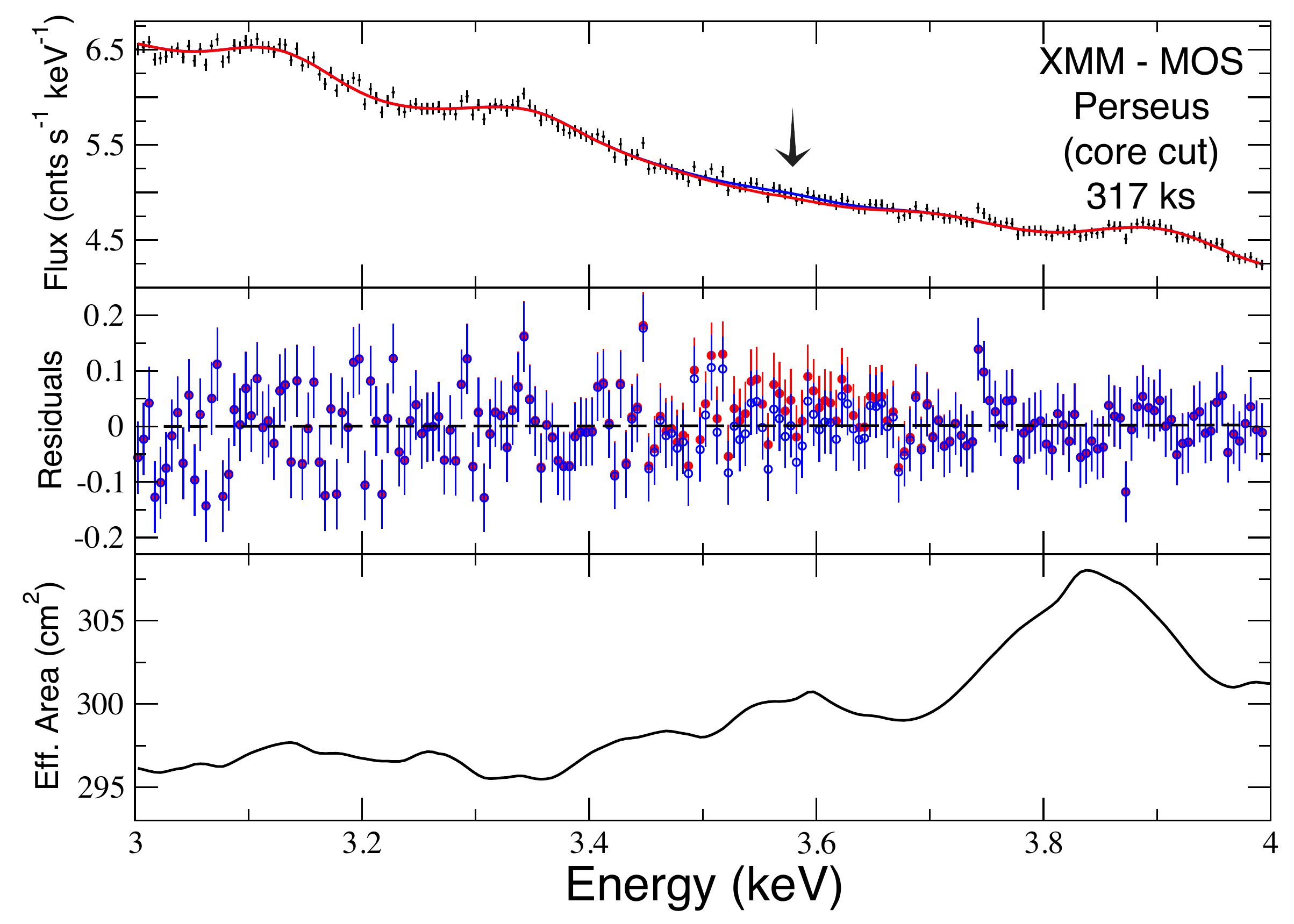}
\caption{3$-$4 keV band of the core-excised stacked MOS spectrum
of the Perseus Cluster. 
The figures show the energy band, where a new spectral feature at 3.57 keV is detected. 
The Gaussian lines with peak values of the flux normalizations 
of K \textsc{xviii} and Ar \textsc{xvii} estimated using AtomDB were included 
in the models. The red lines in the top panels show the model and the excess emission 
in both spectra.
The blue lines show the total model after a Gaussian line is added, indicating 
that the unidentified spectral line can be modeled with a Gaussian. 
}
\vspace{3mm}
\label{fig:perseus-mos-corecut}
\end{figure}

Since this is a single-cluster spectrum, we first check whether the Perseus
signal is not an artifact of our blueshifting procedure. For this we fit
the original, redshifted MOS spectrum with a \textit{line-free apec} model.
We obtained a best-fit $\chi^{2}$ of 463 for 385 dof. Adding a Gaussian line at
3.57 keV (rest energy) improved the fit by $\Delta \chi^{2}$ of 16 for an
additional dof. The best-fit flux was 5.3 $\pm$ 1.2 (2.0)
$\times\ 10^{-5}$ photons cm$^{-2}$ s$^{-1}$, which is in agreement
with the flux obtained from the blue-shifted spectrum. We conclude that our
detection is independent of shifting the spectrum.

\begin{figure*}[ht!]
\centering
\hspace{-5mm}\includegraphics[width=8.9cm, angle=0]{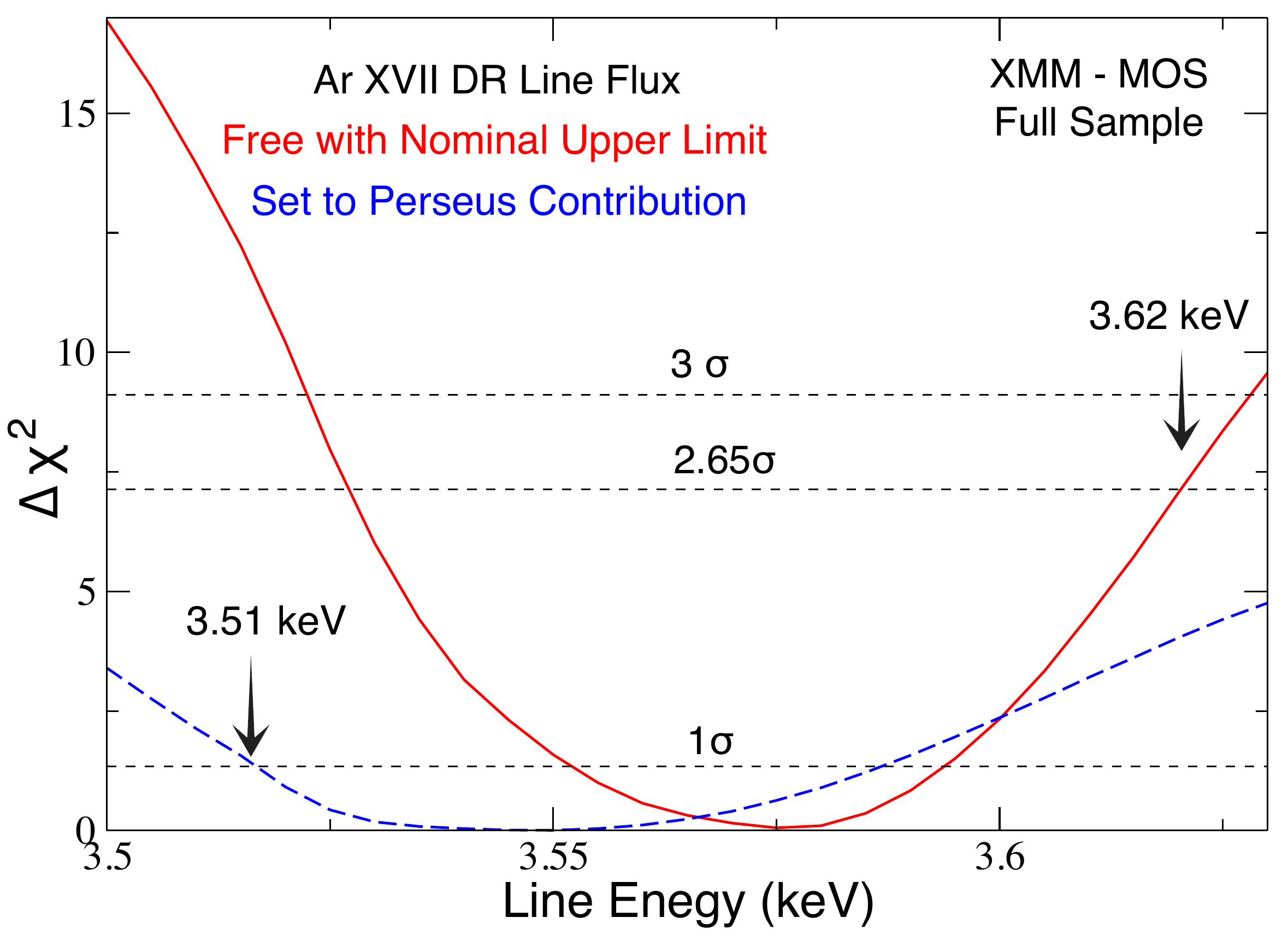}
\includegraphics[width=9.2cm, angle=0]{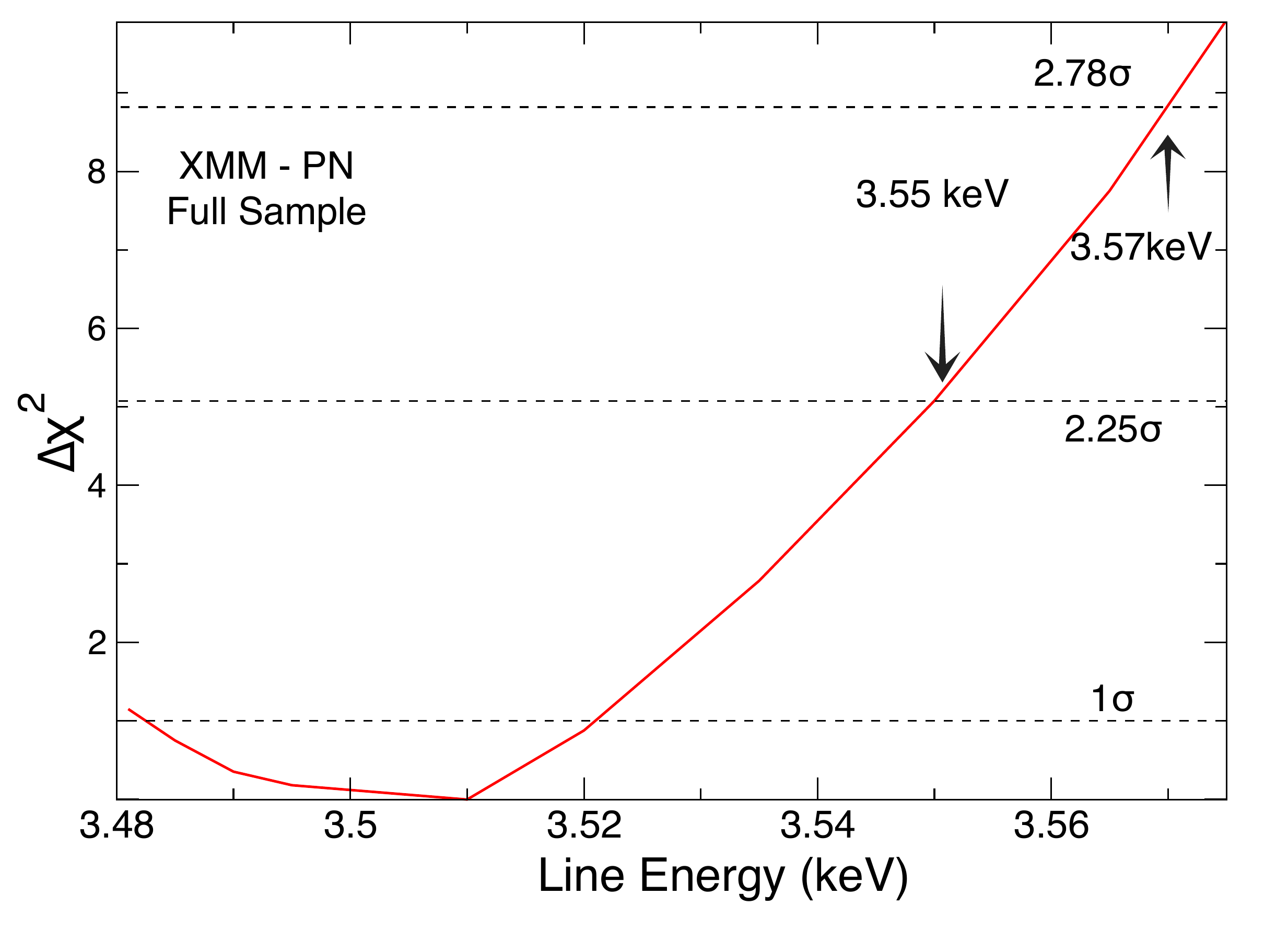}
\caption{Left Panel: change in the goodness-of-fit statistics as 
a function of the detected line energy at 3.55--3.57 keV obtained from the stacked MOS observations of
the full sample. The red solid line shows the confidence of the line energy
when the flux of the Ar \textsc{xvii} DR line at 3.62 keV was left free to vary within the 
AtomDB predicted boundaries. In this case, the detected line is $2.65 \sigma$ from the Ar \textsc{xvii} DR
line at 3.62 keV line. The blue dashed line shows the confidence curve of the line energy
when the flux of the Ar \textsc{xvii} DR line at 3.62 keV was fixed at the 
maximum DR contribution from the Perseus Cluster. In this case the line energy is consistent with the PN detection. Right Panel: change in the goodness-of-fit statistics as 
a function of the line energy obtained from the stacked PN observations of
the full sample. The line energy is $2.2\sigma$ and $2.7\sigma$ away from the MOS detections.}
\vspace{3mm}
\label{fig:steppar}
\end{figure*}

Not ready to abandon the sterile neutrino explanation based on the line flux
incorrectly scaling with cluster mass that we see for Perseus, we tried to
investigate possible astrophysical reasons behind the excess of the line
flux in Perseus. First, we investigated the dependence of the energy and
flux of this unidentified line on the AtomDB predicted fluxes of nearby
lines, i.e., the K \textsc{xviii} line at 3.51 keV and the Ar \textsc{xvii}
DR line at 3.62 keV.  Allowing the energy of the Gaussian component to vary
produced a best fit for an energy of 3.56 $^{+0.01}_{-0.02}$
($^{+0.02}_{-0.03}$) keV, with a flux of 6.0$^{+1.8}_{-1.4}$
($^{+2.4}_{-1.7}$) $\times\, 10^{-5}$ photons cm$^{-2}$ s$^{-1}$ ($\chi^{2}$
of 598.1 for 572 dof). The best-fit energy is consistent with the energy
measured from the MOS observations of the full sample.  However, the fluxes
of the nearby K \textsc{xviii} line at 3.51 keV and the Ar \textsc{xvii} DR line
at 3.62 keV  were at their allowed upper limits predicted from
AtomDB. Relaxing the upper limits has shifted the line energy higher, to 
3.59 $^{+0.01}_{-0.03}$ ($^{+0.02}_{-0.04}$) keV with a flux
of 5.5$^{+1.7}_{-0.8}$ ($^{+3.7}_{-1.5}$) $\times\, 10^{-5}$ photons
cm$^{-2}$ s$^{-1}$ giving  a slightly better fit ($\chi^{2}$ of 594.5 for 572
dof). We note that the line energy of this extra line gets close to the Ar
\textsc{xvii} DR line at 3.62 keV. So we removed the extra Gaussian line
and re-fit the Perseus spectrum removing the upper limits on the Ar
\textsc{xvii} DR line. We obtained only a slightly worse fit than the previous
case, with a $\chi^{2}$ of 598.8 (574 dof). The measured flux of the Ar
\textsc{xvii} DR line at 3.62 keV in this case was 4.8$^{+0.7}_{-0.8}$
($^{+1.3}_{-1.4}$) $\times \, 10^{-5}$ photons cm$^{-2}$ s$^{-1}$,
which is a 
factor of 30 above the predicted maximum flux of the Ar \textsc{xvii} DR
line based on the measured flux of the Ar \textsc{xvii} line at $\sim$3.12
keV and AtomDB line rates. The predicted maximum flux of the Ar \textsc{xvii} DR line
for the Perseus spectrum was 1.6 $\times \, 10^{-6}$ photons cm$^{-2}$
s$^{-1}$ ($<0.01$ times the flux of the Ar \textsc{xvii} triplet at
$\sim$3.12 keV).

This test showed that the line detected in the
Perseus Cluster could also be interpreted as an abnormally bright Ar
\textsc{xvii} DR line. We note, however, that obtaining such a
bright DR line relative to the He-like triplet at 3.12 keV is problematic.
The emissivity of the satellite line peaks at kT=1.8 keV, and declines
sharply at lower temperatures, in addition to the change in the ionization
balance which reduces the Ar$^{+17}$ content of the plasma. The emissivity ratio
for the DR/3.12 keV has its maximum value of 0.04 at kT=0.7 keV, but the
emissivity of both lines is weak here, so any hotter component will dominate
and lead to a lower ratio being observed.

To avoid cool gas in the Perseus core contaminating the flux of the nearby
Ar and K lines, we also tried excising the central region with 1$^\prime$ radius of the
cluster and performed the fit on the core-excised co-added MOS spectrum. We
found that adding an extra Gaussian line at 3.57 keV has improved the fit by
$\Delta \chi^{2}$ of 12.8 for an additional degree of freedom with a
best-fit flux of 2.1 $^{+0.7}_{-0.6}$ ($^{+1.2}_{-1.1}$) $\times \, 10^{-5}$
photons cm$^{-2}$ s$^{-1}$ (see Figure \ref{fig:perseus-mos-corecut}). 
Excising the innermost $1^{\prime}$ reduced the
flux of the detected line by a factor of two, indicating that the most of the
flux of this emission originates from the cool core. The mixing angle
that corresponds to the line flux from the core-excised Perseus
spectrum is consistent within
$1\sigma-2\sigma$ with those for the bright clusters
(Centaurus+Coma+Ophiuchus) and the full sample, respectively (Table
\ref{table:flux}).

We also note that some scatter of the dark matter decay
signal between individual clusters is naturally expected. For example,
one can imagine a filament of dark 
matter along the line of sight in the direction of Perseus, which may boost the flux 
of the detected line and cause tension between the Perseus Cluster and the 
full sample. However, such a filament would have to be rather extreme.

{
{\renewcommand{\arraystretch}{1.2}
\begin{table*}[ht!]
\begin{center}
\caption{  \footnotesize Summary of \chandra Observations of 
the Perseus Cluster and Virgo Cluster used in This Work. The columns list
(1) primary detector array used; (2) 
observation number (3) and (4) right ascension and 
declination of the pointing (J2000); (5) good exposure time 
in ks after filtering.}
\begin{tabular}{ccccccc}
\hline\hline\\
	(1) & (2) & (3) & (4) & (5) \\
Cluster & Detector &  Obs ID & RA & DEC &  Exposure (ks)& Redshift \\
	&	&	& &			  \\ \hline\\
Perseus &ACIS-I & 11713 & 03 19 31.8 & +41 37 49.0 & 113.0 & 0.017 \\ 
&ACIS-I & 11714 & 03 19 42.6 & +41 34 07.0 & 92.3 & 0.017 \\
&ACIS-I & 11715  & 03 19 44.2 & +41 25 18.0 & 73.6 & 0.019  \\
&ACIS-I & 11716 & 03 19 44.2 & +41 25 18.0 & 39.4 & 0.017\\
&ACIS-I & 12025 & 03 19 31.8 & +41 37 49.0 & 17.6 & 0.017 \\
&ACIS-I & 12033 & 03 19 31.8 & +41 37 49.0 & 18.6 & 0.018\\
&ACIS-I & 12036 & 03 19 31.8 & +41 37 49.0 & 47.7& 0.018\\
&ACIS-I & 12037 & 03 19 44.2 & +41 25 18.0 & 85.0&  0.018\\
\\
Perseus &ACIS-S & 4289 & 03 19 47.6 & +41 30 37.0 & 95.4 & 0.018\\
&ACIS-S & 3209 & 03 19 47.6 & +41 30 37.0 & 95.7 & 0.018\\
&ACIS-S & 4946 & 03 19 48.2 & +41 30 42.2 & 23.6&0.018\\	
&ACIS-S & 6139 & 03 19 48.2 &+41 30 42.2 & 56.4&0.018\\
&ACIS-S & 4947 & 03 19 48.2 & +41 30 42.2 & 29.7&0.018\\	
&ACIS-S & 6145 & 03 19 48.2 & +41 30 42.2 & 85.0&0.018\\
&ACIS-S & 4948 & 03 19 48.2 & +41 30 42.2 & 118.6 &0.018\\
&ACIS-S & 4949 & 03 19 48.2 & +41 30 42.2 & 29.4&0.018\\	
&ACIS-S & 6146 & 03 19 48.2 & +41 30 42.2 & 47.1&0.018\\
&ACIS-S & 4951 & 03 19 48.2 & +41 30 42.2 & 96.1 &0.018\\
&ACIS-S & 4952 & 03 19 48.2 & +41 30 42.2 & 164.2 &0.018\\
&ACIS-S & 4953 & 03 19 48.2  & +41 30 42.2 &30.1&0.018\\
\\
Virgo & ACIS-I & 5826 & 12 30 49.5 & +12 23 28.0 & 127.5 & 0.0040 \\
& ACIS-I  & 5827 & 12 30 49.5 & +12 23 28.0 & 157.6 & 0.0038 \\
& ACIS-I & 5828 & 12 30 49.5 & +12 23 28.0 & 33.2 & 0.0036\\
& ACIS-I & 6186 & 12 30 49.5 & +12 23 28.0 & 50.8 & 0.0040 \\
& ACIS-I & 7210 & 12 30 49.5 & +12 23 28.0 & 31.1& 0.0033 \\
& ACIS-I & 7211 & 12 30 49.5 & +12 23 28.0 & 15.5& 0.0038\\
& ACIS-I & 7212 & 12 30 49.5 & +12 23 28.0 & 	65.3& 0.0036\\

\\
\hline\hline
\end{tabular}

\label{table:sample2}
\end{center}
\end{table*}
}
%

\subsection{Refitting Full Sample with Anomalous 3.62 keV Line}
\label{sec:perseusafter}

With the knowledge that the 3.62 keV line can be anomalously high
(at least in Perseus),
we should now try to re-fit the stacked MOS spectrum of the full 
sample to see if the line in the full sample 
is affected by the 3.62 keV excess from Perseus, which is
part of the full sample. We set the 
flux of the 3.62 keV line to the Perseus contribution of the Ar
\textsc{xvii} DR line to the full-sample spectrum ($2.3\, \times\, 
10^{-6}$ photons cm$^{-2}$ s$^{-1}$), assuming that all the new line flux
in Perseus originates from the abnormally bright
DR line. We note that this flux was already a factor of 30 above the
predicted upper limits by AtomDB. Adding an extra Gaussian component, representing the new line,
to a model with the anomalous 3.62 keV line
still improves the fit by $\Delta \chi^{2}$ of 6.52 for 2 dof. The best-fit energy and  
flux were 3.55 $\pm$ 0.03 (0.05) and 2.2$^{+1.6}_{-0.9}$
($^{+2.2}_{-1.5}$) $\times \, 10^{-5}$ photons cm$^{-2}$ s$^{-1}$,
respectively. The new line is still required with 2.5$\sigma$ in the
full sample; however, the
energy of this line gets lower and its confidence interval wider. The
line energy comes into agreement with the energy detected in
the PN full sample (see Figure \ref{fig:steppar}, left panel). If we completely
free the normalization of the 3.62 keV line in the full-sample MOS
spectrum, it becomes lower than the Perseus contribution that we
considered above.

\section{Chandra Observations of Perseus and Virgo} 
 
Due to the potential significance of the discovery of an emission line
due to the decay of sterile neutrinos in clusters, 
it is necessary to confirm it with another instrument.
Pending a full stacking analysis of the \chandra and \suzaku cluster archives (which is a current
work in progress), we analyze two \chandra observations of the Perseus Cluster
and Virgo Cluster, which 
have over 1 Ms and 500 ks of total \chandra exposure, respectively. A
summary of the \chandra observations  
used in this work to confirm the detection is given in Table \ref{table:sample2}.

The \chandra ACIS-I and ACIS-S data were processed following \citet{vikhlinin2005},
 using CIAO 4.5 and CALDB 4.5.7. Each event list was filtered for high background 
periods. After this filtering the total good times were 487 ks 
and 883 ks for the ACIS-I and ACIS-S observations of the Perseus Cluster core, respectively.
We have extracted the ACIS-S spectra from the full S3 chip excluding the 1$^\prime$ region surrounding the cluster centroid
and one of the observations (ID: 4950) with a background flare. The ACIS-I spectrum was extracted
using a circular region covering the full ACIS-I FOV.
The filtered ACIS-I good time for the Virgo Cluster core was 481 ks. The radius of
500$^{\prime\prime}$ was used for extraction of ACIS-I spectra of the Virgo Cluster.
Analysis steps include image creation, point-source detection with
\textit{wavdetect}, and their removal.

Background corrections were made using the blank-sky background fields, 
including the ``period-E" background files. 
For each target event file, a corresponding background event file 
was generated and normalized by the ratio of counts in the 9.0$-$12.0 keV energy range
\citep{hickox2006}. Because we are interested in the high-energy part of the spectrum, 
modeling of the soft sky CXB is not relevant.

\subsection{Chandra Data Analysis}
\label{sec:chandraAnalysis}

 \begin{figure*}
 \centering
\hspace{-4mm}\includegraphics[width=9.cm, angle=0]{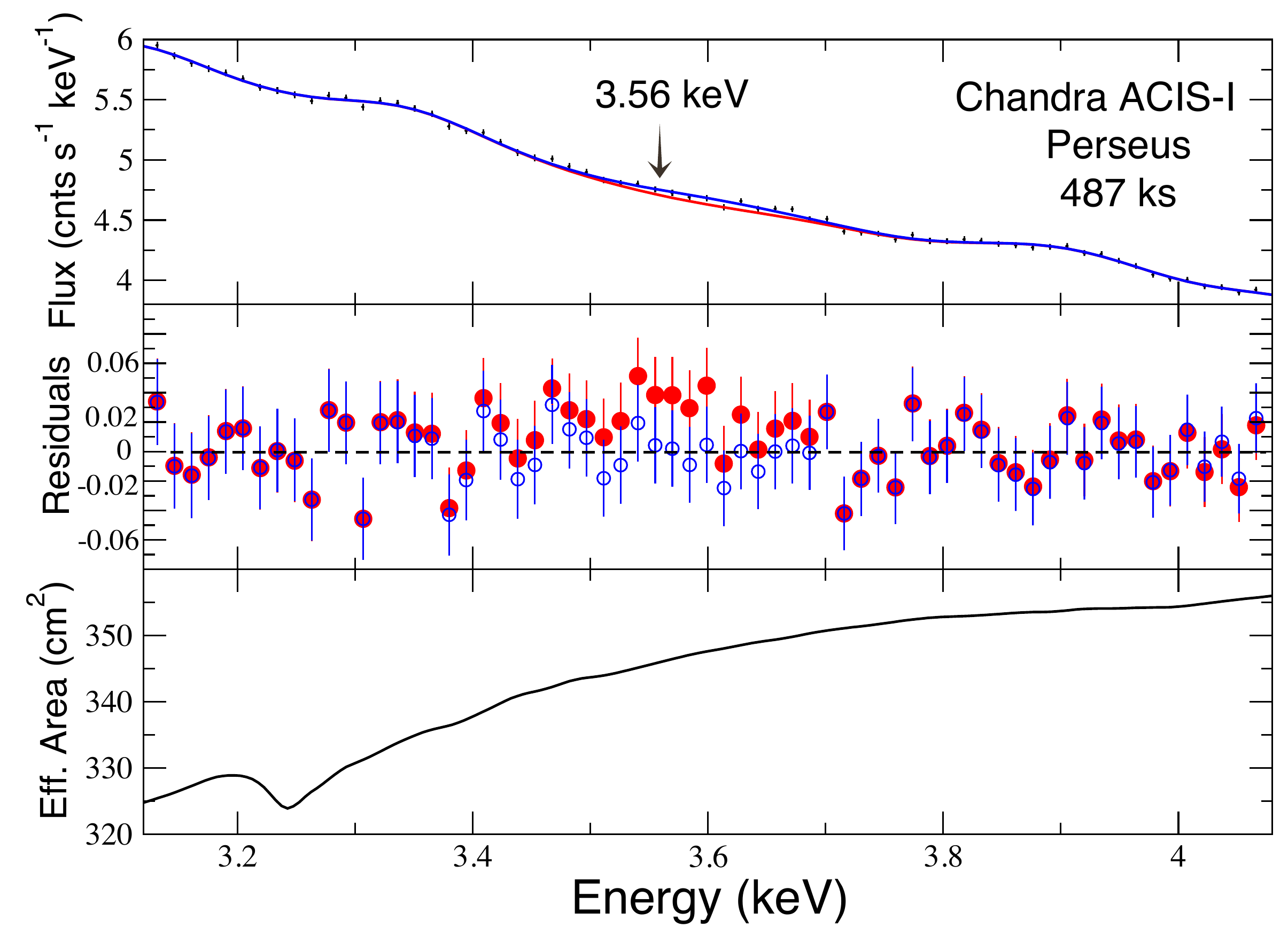}
\includegraphics[width=9.cm, angle=0]{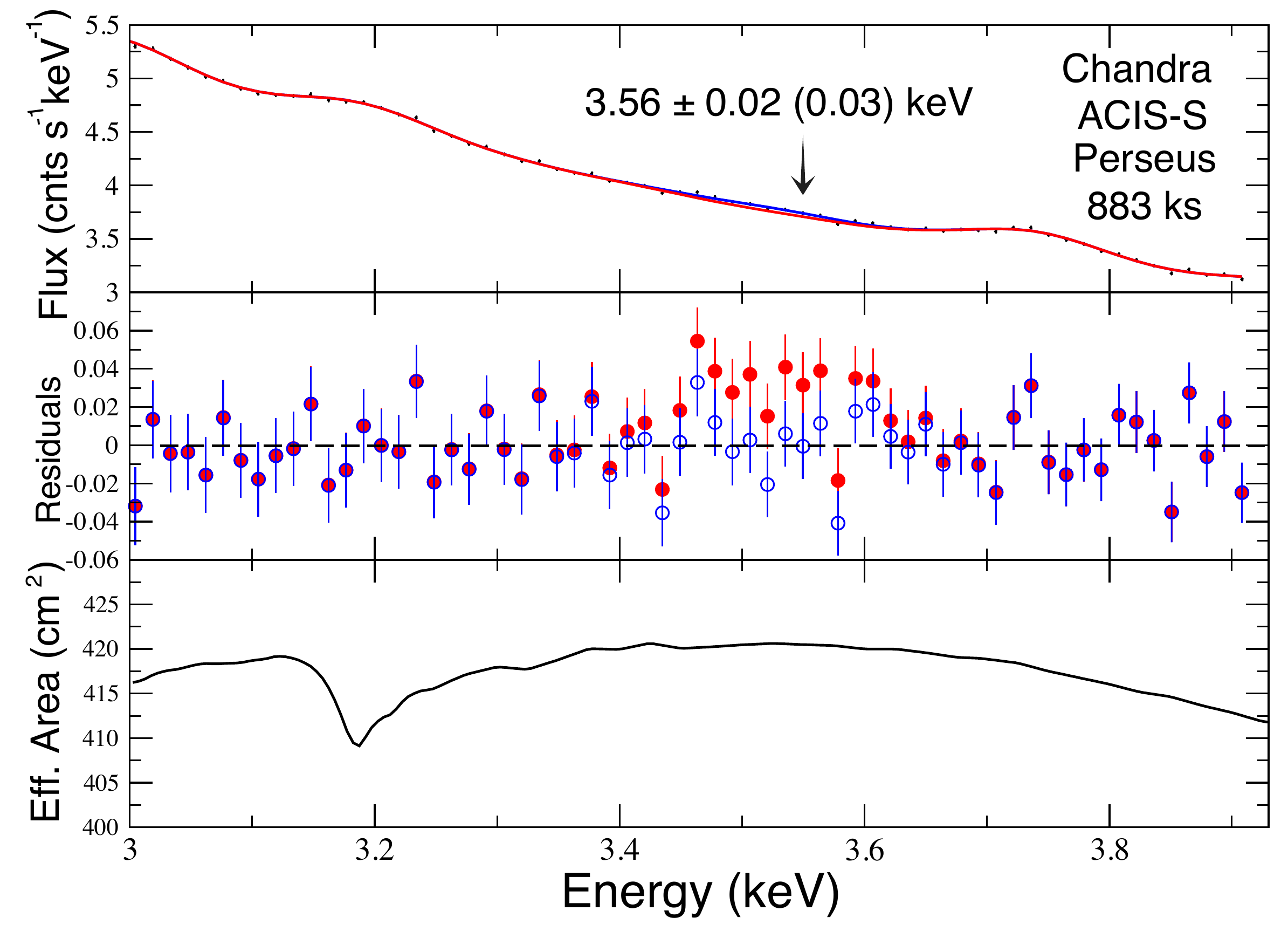}
\caption{Close-up 3.1 $-$ 4.1 keV energy interval of the co-added spectrum obtained 
from \chandra ACIS-I and ACIS-S observations of the Perseus Cluster. The continuum emission 
was fit with the 2T \textit{line-free} \apec model, while emission lines were 
modeled with additional Gaussian components. The
K \textsc{xviii} (3.51 keV), Ar \textsc{xvii} (3.62 keV), and  Ar \textsc{xvii} (3.68 keV) lines are also 
included in the total model shown as a red line on the top panel. The blue line shows the total
model after a Gaussian component is added to the total model, indicating that the weak residual 
can be modeled with a Gaussian. The bottom panels show the weak residual 
before and after the Gaussian is added to the total model.
}
\label{fig:chandra}
\end{figure*}

Each spectrum was fitted using a standard multi-temperature 
\textit{apec} model as described in Section \ref{sec:stacking} to determine 
the best-fit X-ray redshift of each observation, shown in Table \ref{table:sample2}. Each event file was then
blue-shifted to the cluster's source frame using these best-fit redshifts. 
The source and background spectra in 
the source's frame were obtained by
generating spectra using the scaled event energy values in the event files.
The ARFs and RMFs were remapped based on the estimates of the best-fit redshifts. 
The RMFs and ARFs were weighted by 
only the exposure time of each observation. 
The scaled source and background spectra were co-added using 
the FTOOL \textit{mathpha}, 
whereas ARFs and RMFs were merged using the FTOOLS \textit{addarf} and \textit{addrmf} 
tools, respectively. 

 \begin{figure}
 \centering
\hspace{-4mm}\includegraphics[width=9.cm, angle=0]{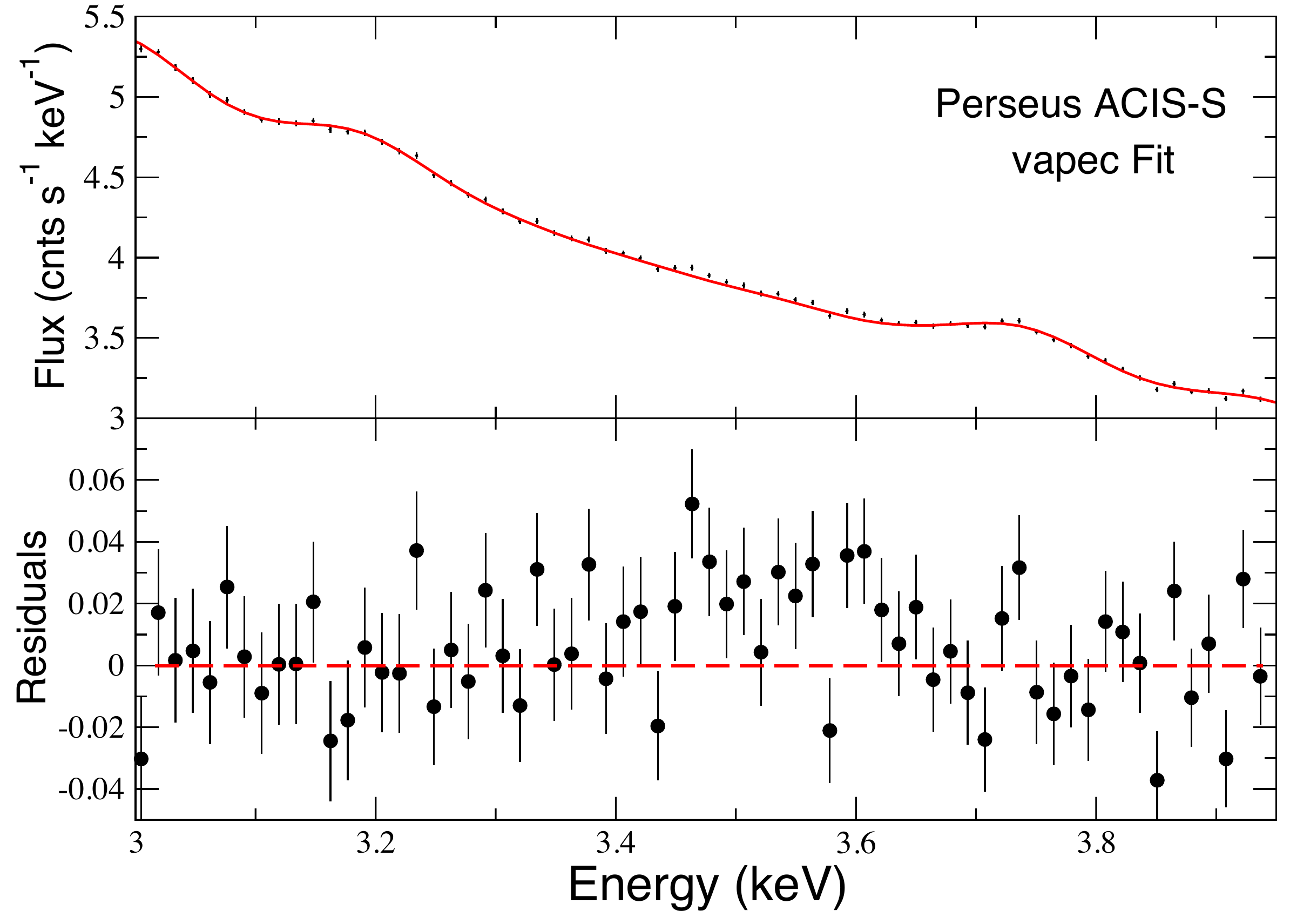}
\caption{The 3 $-$ 4 keV energy interval for the co-added spectrum obtained 
from \chandra ACIS-S observations of the Perseus Cluster. The continuum emission  
was fit with two \textit{vapec} models. This detection demonstrates that the detected line is not a 
fitting artifact.
}
\label{fig:perseusVapec}
\end{figure}
{
{
\begin{table*}[ht!]
\begin{center}
\caption{Best-fit Temperature and Normalizations of \textit{line-free apec} Model Fit to the Co-added Chandra Observations of the Perseus and Virgo Clusters. Fluxes of the S \textsc{xvi}, Ca \textsc{xix},  Ca \textsc{xx} at the rest energies 2.63 keV, 3.90 keV, 4.11 keV are given. }
\renewcommand{\arraystretch}{1.5}
\begin{tabular}{lcc|cc}
\hline\hline\\
& Perseus &&Virgo &\\
Model Paramaters         &ACIS-I & ACIS-S  & ACIS-I\\
\\\hline
\\
kT$_{1}$ (keV) & 4.58 $\pm$ 0.07  & 2.77 $\pm$ 0.18 & 1.18  $\pm$ 0.07 \\
$N_{1}$  ($10^{-1}$ cm$^{-5}$)  & 1.20 $\pm$ 0.01 & 3.21 $\pm$ 0.35 & 2.22 $\pm$ 0.10\\
kT$_{2}$  (keV) & 5.34 $\pm$ 0.02  & 4.79 $\pm$ 0.18 & 5.08 $\pm$ 0.37 \\
$N_{2}$ ($10^{-1}$ cm$^{-5}$) & 3.83 $\pm$ 0.13 & 2.31 $\pm$ 0.24 & 1.02 $\pm$ 0.23\\
Flux of S \textsc{xvi} ($10^{-4}$ pht cm$^{-2}$ s$^{-1}$) & 3.94 $\pm$ 0.15  &  3.72 $\pm$ 0.85 & 6.18 $\pm$ 0.13  \\
Flux of Ca \textsc{xix} ($10^{-4}$ pht cm$^{-2}$ s$^{-1}$) &  1.29 $\pm$ 0.08   & 1.05 $\pm$ 0.08 & 1.22 $\pm$ 0.88\\
Flux of Ca \textsc{xx} ($10^{-4}$ pht cm$^{-2}$ s$^{-1}$) & 1.10 $\pm$ 0.05 &  1.07 $\pm$ 0.05 & 0.35 $\pm$ 0.05\\
\\
\hline\hline
\end{tabular}
\label{table:chandraTemp}
\end{center}
\end{table*}
}

\subsection{Chandra Detection of the Emission Line in Perseus}

Following the same method as described in Section \ref{sec:all}, the continuum
emission was fit using the \textit{line-free} \apec model with additional 
Gaussian models to represent the strong emission lines. 
The best-fit temperature from the 2.0 $-$ 6.0 keV band and 
normalizations of the \textit{line-free} \apec model, 
fluxes, and equivalent widths of S \textsc{xvi}, Ca \textsc{xix}, 
and Ca \textsc{xx} lines at 2.63 keV, 
3.90 keV, and 4.11 keV are given in Table \ref{table:chandraTemp}. 
We have searched especially the 3.0 $-$ 4.0 keV interval where the 3.57 keV line emission
was detected in the stacked \xmm observations.
The measured fluxes of S \textsc{xvi}, Ca \textsc{xix}, 
and Ca \textsc{xx} lines from the ACIS-I and ACIS-S spectra with the 
AtomDB fluxes yielded the maximum
predicted fluxes of K \textsc{xviii} lines at 3.47 keV and 3.51 keV, Ar \textsc{xvii} 
line at 3.68 keV, and K \textsc{xviii} line at 3.71 keV as described in detail in Section \ref{sec:all}. The triplet emission line at Ar \textsc{xvii} 3.12 keV was used to determine the maximum allowed flux of the Ar \textsc{xvii} DR line at 3.62 keV at any plasma temperature as described above.
The predicted fluxes of these lines are given in Table
\ref{table:acisFlux}. Using 0.1 and 3 times of the upper bound of these estimates as lower and upper limits for K \textsc{xviii} andAr \textsc{xvii}, and $10^{-3}$ -- $10^{-2}$ times the flux of the Ar \textsc{xvii} triplet for the lower and upper bounds for the Ar \textsc{xvii} DR line, we determined the best-fit flux of the weak residual
around 3.57 keV. 

{
{
\begin{table*}[ht!]
\begin{center}
\caption{Best-fit Temperature and Normalizations of \textit{line-free apec} Model Fit to the Co-added Chandra Spectra of Perseus and Virgo Clusters. (1) and (2) are the estimated fluxes of K \textsc{xviii} at the rest energy 3.51 keV and  Ar \textsc{xvii} at the rest energy 3.68 keV lines obtained from AtomDB in the units of photons cm$^{-2}$ s$^{-1}$; (3) and (4) are the estimated energy in keV and flux of the unidentified emission line in the units of photons cm$^{-2}$ s$^{-1}$; (5) is the measured equivalent width of the spectral feature, indicates the ratio of line flux to continuum flux in the units of keV.}
\renewcommand{\arraystretch}{1.5}
\begin{tabular}{ccccccc}
\hline\hline
Cluster& Inst. &  Flux & Flux& Flux &Flux & Flux \\
&& K \textsc{xviii}  & K \textsc{xviii} & Ar \textsc{xvii}  &Ar \textsc{xvii}&K  \textsc{xix} \\
&& (3.47 keV)   & (3.51 keV) & (3.62 keV) & (3.68 keV) & (3.71 keV)\\
&& ($10^{-6}$)  &($10^{-6}$) & ($10^{-7}$)  &($10^{-5}$) &($10^{-6}$)     \\
\hline
\\

& ACIS-I & 3.4 $\pm$ 2.7 & 3.1 $\pm$ 2.4 & 4.3 $\pm$ 3.5 & 0.8 $\pm$ 0.4 &2.6 $\pm$ 2.1  \\
Perseus\\	
& ACIS-S &4.5 $\pm$ 2.3 &4.15 $\pm$ 2.2   & 5.8 $\pm$ 2.8 & 1.3 $\pm$ 1.0 & 3.4 $\pm$ 2.7 \\
\\
\cmidrule{1-7}
\\
Virgo & ACIS-I& 2.0 $\pm$ 0.5 & 3.6 $\pm$ 1.0   & 38.2 $\pm$ 10.6  & 1.7 $\pm$ 0.5  & 1.8 $\pm$ 0.6\\

\\
\hline\hline
\end{tabular}
\label{table:acisFlux}
\end{center}
\end{table*}
}

An additional Gaussian model improves the fit by $\Delta \chi^{2}$ of 11.8 
for an additional 2 dof. The line was unresolved and consistent
with broadening by the instrument response in the Perseus Cluster spectra. 
The Perseus ACIS-S spectra yield a best-fit energy of 3.56 $\pm$ 0.02 (0.03) keV for an 
additional Gaussian model, 
given in Table \ref{table:flux}. The flux of the detected signal is
1.02 $^{+0.4}_{-0.4}$ ($^{+0.5}_{-0.5}$) $\times\,10^{-5}$ $\rm{photons}\ \rm{cm}^{-2}\ \rm{s}^{-1}$.
This detection corresponds to a false detection probability of 0.5\% in the co-added ACIS-S spectrum.
The right panel of the Figure \ref{fig:chandra} shows the signal in the \chandra ACIS-S observations of the Perseus Cluster
before and after the Gaussian model 
is added to the fit.  

To further demonstrate that the detected flux is independent of the spectral modeling, we fit the ACIS-S spectrum of the Perseus Cluster with a two-temperature \textit{vapec} model with abundances of trace elements set to that of Fe. We obtained an acceptable fit in the 3 $-$ 6 keV energy band with $\chi^{2}$ of 182.1 for 147 dof. An additional Gaussian model at 3.56 keV (rest energy) improved the fit by $\Delta \chi^{2}$ of 16 for an extra degree of freedom. The best-fit flux of the line is 1.09 $\pm$ 0.26 (0.42) $\times\,10^{-5}$ photons cm$^{2}$ s$^{-1}$, which is consistent with the flux measured in the \textit{line-free apec} model fit with additional Gaussian models. This test shows that the detection is robust and independent of the method used in the spectrum modeling. The Perseus co-added spectrum 
fit with a two-temperature \textit{vapec} model is shown in Figure \ref{fig:perseusVapec}.

We then performed the same search in the co-added ACIS-I spectrum of the Perseus Cluster.
Fitting the 2.5--6 keV band of the ACIS-I spectrum with a  \textit{line-free apec} model with additional Gaussian
lines as described above produced a good fit overall with a total $\chi^{2}$ of 158.7 for 152 dof. Adding a Gaussian line
at 3.56 keV, the energy where the line was detected in the co-added ACIS-S spectrum of the Perseus Cluster,
improved the fit by $\Delta \chi^{2}$ of 6.2 for an additional degree of freedom. The flux of the detected signal was 1.9 $^{+0.8}_{-0.8}$ ($^{+1.2}_{-1.6}$) $\times\,10^{-5}$ $\rm{photons}\ \rm{cm}^{-2}\ \rm{s}^{-1}$ in the co-added ACIS-I spectrum. The left panel of the Figure \ref{fig:chandra} shows the ACIS-I spectrum of the Perseus Cluster before and after an additional Gaussian model is added to the total model, to demonstrate the detection of the line.

The mixing angle $\sin^{2}(2\theta)$ estimate from the co-added \chandra ACIS-S observations of the Perseus Cluster is 4.0 $^{+1.5}_{-1.4}$ ($^{+1.8}_{-1.8}$) $\times \ 10^{-10}$, which is consistent with the angle obtained from the co-added ACIS-I and \xmm MOS observations of the Perseus Cluster at the 1$\sigma$ level. Since the ACIS-S chip covers the central 4$^{\prime}$region of the Perseus core, higher flux measured from ACIS-S observations also indicates that this emission is concentrated in the core, confirming the results from the \xmm observations of the Perseus core.

 \begin{figure}
 \centering
\hspace{-6mm}\includegraphics[width=8.7cm, angle=0]{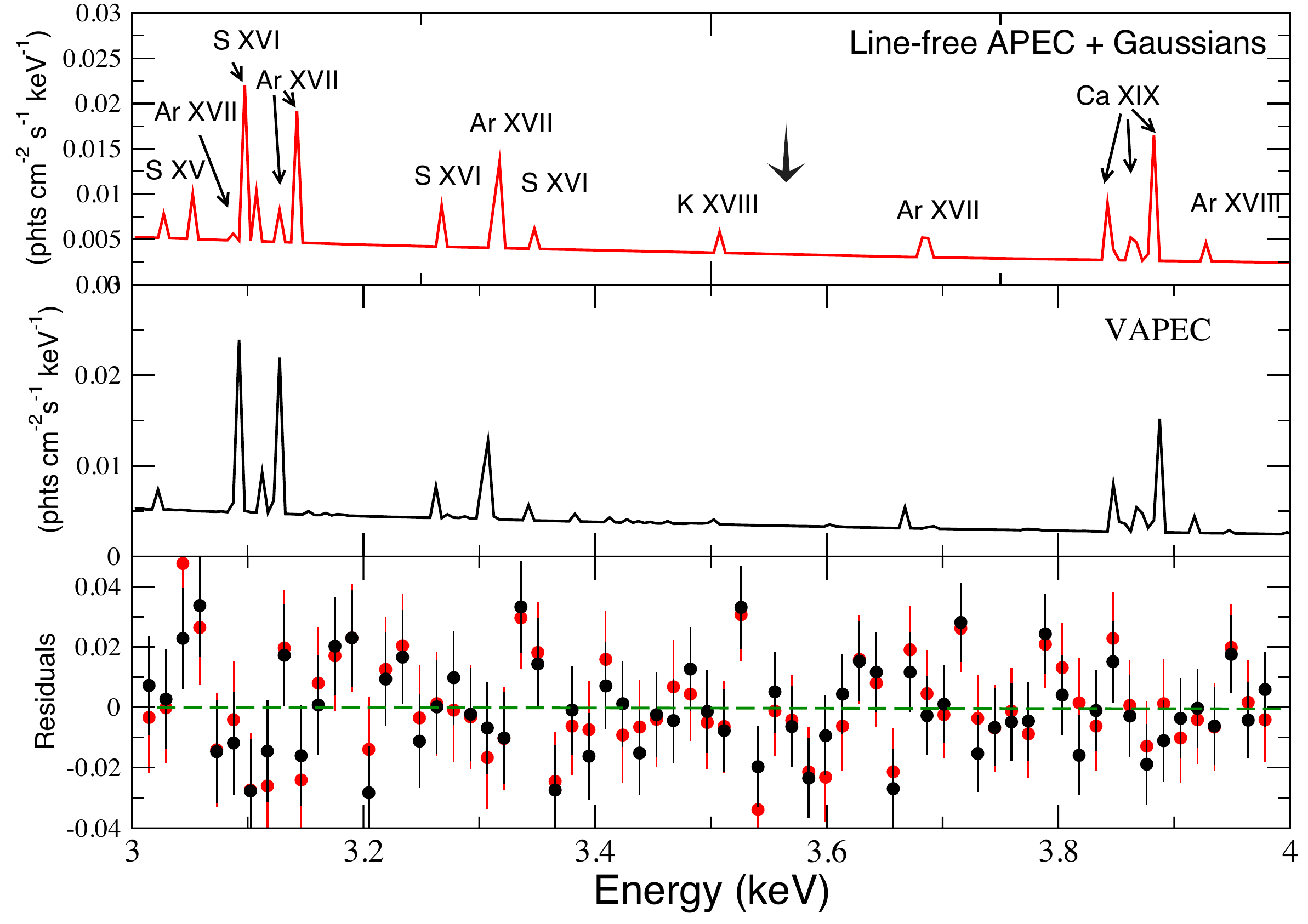}
\caption{3 $-$ 4 keV energy interval for the co-added spectrum obtained 
from \chandra ACIS-I observations of the Virgo Cluster at the redshifted frame. The continuum emission  
was fit with the \textit{line-free apec} model with Gaussians components (upper panel) and two-temperature \textit{vapec} models (middle panel). The lower panel shows the differences in the residuals for the two models. Red data points show the residuals of the \textit{line-free apec} model with Gaussians components and black data points show the residuals of the \textit{vapec} model. The energy where the line is detected in the \chandra ACIS observations of the Perseus Cluster is indicated with an arrow. }
\label{fig:virgoVapec}
\end{figure}

%
\subsection{Chandra Upper Limit on the Line from Virgo}

We have performed the same fitting strategy described above to the
co-added spectra of the Virgo Cluster, e.g., \textit{line-free apec}
model with additional Gaussian lines. We used the lower and upper
limits to the K and Ar line in the 3.4 $-$ 3.7 keV band based on the
upper limits estimated from AtomDB (given in Table
\ref{table:acisFlux}). The overall fit was acceptable with a total
$\chi^{2}$ of 82.5 for 62 dof. 
Unlike the Perseus Cluster, the co-added Virgo Cluster did not show
any residuals around 3.57 keV 
in the fit with the \textit{line-free apec} model. Adding a Gaussian
line did not significantly improve the fit.  
We were able to place an upper limit of 9.1$\times\, 10^{-6}$
$\rm{photons}\ \rm{cm}^{-2}\ \rm{s}^{-1}$ 
 at the 90\% confidence level. This limit corresponds to an upper
 limit on the mixing angle of $\sin^{2}(2\theta)\ <$ 1.1 $\times$
 10$^{-10}$.

We also fit the 2.5$-$4.0 keV band of the Virgo spectrum using a two-temperature standard \textit{vapec} model. The fit has a total $\chi^{2}$
obtained from the \textit{vapec} model of 91.7 for 82 dof. We overall obtained a better fit with the standard \textit{vapec} model than the fit with the \textit{line-free apec} model. The best-fit model also did not require the addition of a line at 3.56 keV. The 90\% upper limit to the flux of this line is  $<$ 6.2 $\times\ 10^{-6}$ photons cm$^{-2}$ s$^{-1}$. The differences in the modeling approaches used in the ACIS-I spectrum fits of the Virgo Cluster (\textit{line-free apec} with Gaussians and \textit{vapec}) are demonstrated in Figure \ref{fig:virgoVapec}. The factor of two difference in the upper limits on the flux measurements indicates that the systematical uncertainties in the flux measurements can be as large as a factor of two depending on the modeling method used in this analysis.

\section{Discussion}
\label{sec:discussion}

Stacking X-ray spectra of galaxy clusters from different redshifts can make
it possible to detect weak emission features. This method was successfully
used before; for example, \citet{SandersFabian11} the stacked \xmm Reflection Grating Spectrometer (RGS)
observations of 62 clusters to find evidence of O{\sc vii}\ in cluster
cores. (The RGS energy coverage is limited to $E<2$ keV.) In this work, we
stacked the \xmm MOS (6 Ms) and PN (2 Ms) spectra of 73 nearby ($z<0.35$)
well-exposed galaxy clusters to look for any unknown emission lines at $E>2$
keV. The redshift difference in our full cluster sample corresponds to an
energy difference of up to 1.2 keV at $E\sim 3.5$ keV and still more at
higher energies. Thus, when the cluster spectra are stacked in the source
reference frame, any uncalibrated detector response features or instrumental
and background lines are smeared by this difference in energy, minimizing
their effects and making it possible to detect the lines in the source
spectrum that for any single object would be below the level of detector
artifacts.

We have detected a weak line at the rest energy of $3.57 \pm 0.02$ keV (68\%
statistical confidence interval) in \xmm MOS observations. A similar
emission line has been found independently in the stacked PN spectrum of the
full sample, although its best-fit energy was lower, $3.51\pm 0.03$ keV.
There is tension between these energies at a $2.8\sigma$ level, including
only statistical errors; they become consistent once we introduce another
degree of freedom in the model.

The best-fit fluxes of 4.0$^{+0.8}_{-0.8}$ $\times\, 10^{-6}$ $\rm{photons}\ 
\rm{cm}^{-2}\ \rm{s}^{-1}$ and
3.9$^{+0.6}_{-1.0}$ $\times\, 10^{-6}$ $\rm{photons}\ \rm{cm}^{-2}\ 
\rm{s}^{-1}$ obtained from the stacked MOS and PN
observations of the full sample are consistent with each other. This
emission line has been found in a blind search in $\sim70$ independent
energy bins. Even taking into account the ``look-elsewhere'' effect (the
probability of finding a line in {\em any}\/ of those bins), the statistical
probability of a false detection of such a $4-5\sigma$ line at the same
energy in two statistically independent data sets is negligibly small.

To see if the signal originates from one of the dominant nearby clusters in
the sample, we divided the full sample
into three subsamples --- Perseus, Centaurus + Coma + Ophiuchus, and all
others --- and detected the line in all three. In the Centaurus + Coma
+ Ophiuchus MOS spectrum, the 
line was found with a flux of 1.6$^{+0.3}_{-0.4}$ $\times\, 10^{-5}$ photons
cm$^{-2}$ s$^{-1}$ at 3.57 keV. The lower signal-to-noise ratio (128 ks total) PN
spectrum yielded a non-detection, with a 90\% upper limit of 9.5 $\times\,
10^{-6}$ photons cm$^{-2}$ s$^{-1}$.

The stacked MOS and PN observations in the rest frame of the fainter
69 ``all other'' clusters exhibited the emission line at 3.57 keV line with the best-fit flux of
2.1$^{+0.4}_{-0.5}$ $\times\, 10^{-6}$ and 2.0$^{+0.3}_{-0.5}$ $\times\,
10^{-6}$ photons cm$^{-2}$ s$^{-1}$ at the energy 3.57 keV. Thus, the
significant detection of the line does not originate from one or a few
dominant clusters, but is produced by all the clusters in subsamples.

We investigated the spectrum of the Perseus Cluster in greatest detail,
because it contributes a large fraction of the total number of
photons, and because the line in Perseus significantly deviates from
scaling by cluster mass (implied by one of our explanations for this
line). The full-FOV  
MOS spectrum of Perseus shows a line with the best-fit energy of 3.57 keV
line and a flux of 5.2$^{+2.4}_{-1.5}$ $\times\, 10^{-5}$ photons cm$^{-2}$
s$^{-1}$; the addition of a line improves the fit by $\Delta\chi^{2}=15.7$
for an additional degree of freedom. We note that the flux of the detected
line is dependent on the predicted fluxes of the nearby K \textsc{xviii} triplet
at 3.51 keV and an Ar \textsc{xvii} DR line at 3.62 keV.
In the spectral fits of the Perseus Cluster, the fluxes of these nearby
lines were at their allowed upper limits. Relaxing these upper limits shifts
the best fit line energy to 3.59 keV, suggesting that the detected line
could be the Ar \textsc{xvii} DR line at 3.62 keV. To test this, we removed
the line model component at 3.57 keV, instead letting the 3.62 keV line have arbitrary
normalization, and found that the Perseus spectra could be represented
without an additional line.  However, in this case, the implied flux of the
Ar \textsc{xvii} DR line had to be significantly increased from the AtomDB
estimate, which is 1\% of the strength of the Ar \textsc{xvii} triplet,
to 30\%. Physically, it is difficult to create such a bright Ar \textsc{xvii} DR line
relative to the Ar \textsc{xvii} He-like triplet at 3.12 keV. The emissivity
ratio for the Ar \textsc{xvii} DR line to the Ar \textsc{xvii} triplet at
3.12 keV has its maximum value of 4\% at kT=0.7 keV. Since the emissivity of
both lines is weak at this temperature, any hotter temperature
components (which produce the bulk of the cluster emission)
will dominate the spectra, leading to an even lower observed normalization
ratio. To explain our line, the required flux of the Ar \textsc{xvii} DR
line has to be increased from the AtomDB
estimate by an implausible factor $\sim 30$. This possibility is
further discussed in Section \ref{sec:unknownline}.  

To further investigate the origin of this excess emission, we excluded
the central bright and complex region with 1$^\prime$ radius of the Perseus core. The best-fit flux
of  $2.1^{+0.7}_{-0.6}$ $\times\, 10^{-5}$ photons cm$^{-2}$ s$^{-1}$
at 3.57 keV decreased to half of the full cluster flux. This decrease
indicates that the emission is concentrated in the immediate cool core
of the Perseus Cluster.
 
In addition, we investigated the \chandra ACIS-S and ACIS-I spectra of
Perseus to confirm that the detected signal is not an \xmm detector feature. 
An independent search of the 3.0 $-$ 4.0 keV interval of the ACIS spectra
revealed a positive detection of the feature at a significance of
$3.4\sigma$ and 2.5$\sigma$ for
the line energy given by \xmm fits. 
The measured best-fit energy of 3.56 $\pm$ 0.02 keV in the ACIS-S
spectrum (with the 1$^{\prime}$ central region of the core excised) is
consistent with the best-fit energy obtained from the stacked \xmm  
observations of the full sample. The observed  
flux of the detected feature is $1.0^{+0.4}_{-0.4}$ $\times\, 10^{-5}$ 
photons cm$^{-2}$ s$^{-1}$. The same feature was also observed in the
co-added ACIS-I spectrum of the Perseus Cluster with the best-fit flux of  $1.8^{+0.8}_{-0.8}$ $\times\, 10^{-5}$ 
photons cm$^{-2}$ s$^{-1}$, with a lower significance ($\Delta
\chi^{2}$= 6.2 for 1 dof). However, the feature was not detected in
the \chandra ACIS-I observations of the Virgo Cluster, for which we
placed a 90\% upper limit of  $9.1\times\,
10^{-6}$ photons cm$^{-2}$ s$^{-1}$.

\begin{figure*}
\centering \hspace{-5mm}\includegraphics[width=18cm, angle=0]{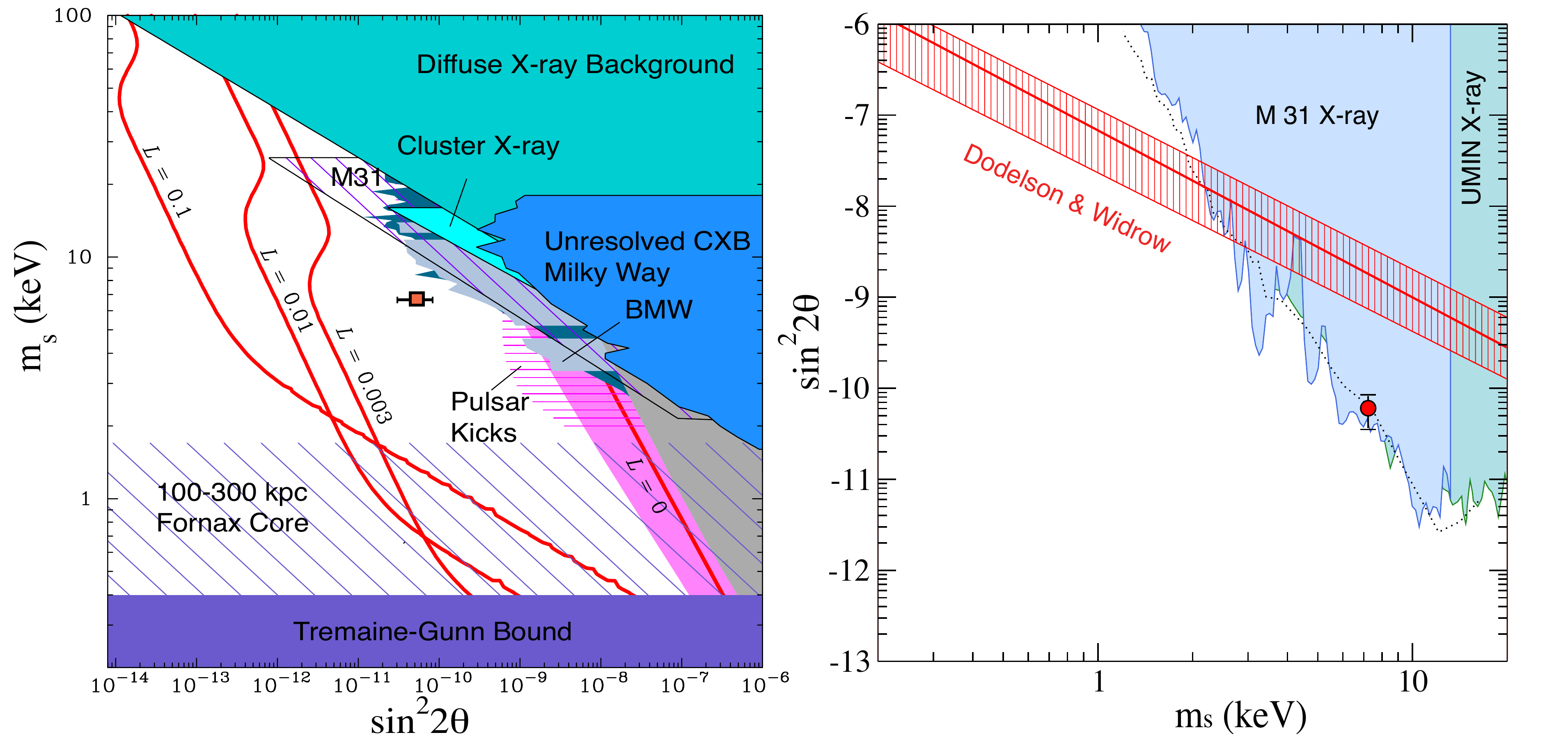}
\caption{Constraints on the sterile neutrino model from the literature. Our
  full-sample MOS line detection (assuming that the line is from sterile
  neutrino and that all dark matter is in sterile neutrino) is shown by red
  symbols in both panels; error bar is statistical 90\%.  {\em Left:}\/
  historic constraints from \cite{abazajian2009}. Red curves show
  theoretical predictions for the Dodelson-Widrow mechanism assuming sterile
  neutrinos constitute the dark matter with lepton numbers L=0, 0.003, 0.01,
  0.1. See \cite{abazajian2009} for explanation of the various observational
  constraints that come from \cite{tremaine1979, bode2001, boyarsky2006,
    strigari2006, abazajian2007}.  {\em Right:}\/ most recent X-ray
  constraints (reproduced from \cite{horiuchi2013}), based on deep \chandra\ 
  \citep{horiuchi2013} and \xmm\ \citep{watson2012} observations of M31 and
  \suzaku\ observations of Ursa Minor \citep{loewenstein2009}. The red band
  marked ``Dodenson \& Widrow'' is same as the $L=0$ curve in left panel.
  Our measurement lays at the boundary of the constraints from M31.}
\label{fig:limitsAbaza}
\end{figure*}
%

\subsection{Unknown Plasma Emission Line?}
\label{sec:unknownline}

One possible interpretation is that the detected line is an
unknown plasma emission line. The flux of the line corresponds to a
maximum emissivity of 3.3 $\times\, 10^{-18}$ photons cm$^3$ s$^{-1}$,
derived using the emission measure appropriate for the lowest
temperature (4.36 keV) component as described in Section \ref{sec:all}. For
comparison, this is similar to the maximum emissivity of the Ca
\textsc{xx} Ly$\alpha$\ line at 4.1 keV.  Given that the Ca
\textsc{xx} line was previously observed in individual galaxy cluster
spectra, including the Perseus Cluster \citep[e.g.][]{tamura2009},  a
line as strong at $\sim$ 3.57 keV 
would have been observed had it been expected.
However, there is no likely candidate for an atomic transition near 3.57
keV.  The emission lines of strong hydrogen- and helium-like ions are
well known, and none fall in this band.  
The only candidate emission line from such an ion would be the
 He-like Cl {\sc xvi} $n=5\rightarrow1$\ transition at 3.52 keV, but if 
 this is the cause it would imply the presence of even stronger lines from 
 the $n=3\rightarrow1$\ and  $n=4\rightarrow1$\ transitions at 3.27 and 3.44 keV, 
 respectively, but these are not seen. Emission lines from \textit{L}-shell ions form a
far more complex pattern.  However, the binding energy of Li-like Zn
($Z=30$) is only 2.782 keV, so the transition lines of all lighter
elements or less ionized species must be at lower energies than this. 
If this line is a K-shell fluorescence transition, it must be from an 
element whose neutral and Li-like K-shell fluorescent line energies
bound 3.57 keV.  The only such atoms are argon and potassium, but in
this case the relevant Ar K-shell fluorescence transition is simply
another name for the Ar {\sc xvii}\ DR line discussed in detail above.
The neutral potassium K$\alpha$\ fluorescence line is at 3.313 keV,
while neutral K$\beta$ is at 3.59 keV, so there must be transitions at
the relevant energy.  In this case, the best matches are the K$\alpha$
transitions of K  {\sc XVI} through K  {\sc XIV} ions, which occur at $\sim
3.57$\,keV \citep{palmeri2012}. However, since at any temperature
above 1 keV potassium will have at most 2 bound electrons, any such
line would have to be originating from an unknown source of
photoionized potassium in clusters. Thus this scenario is very
unlikely, since the compact sources (e.g., active galactic nuclei) are not strong enough
to photoionize the low-density ICM.

Although a complete analysis was not shown, adding an Ar  {\sc XVII} DR line at 3.62 keV with unconstrained flux into all of our spectra would significantly impact both the fit results and detection level of a line at 3.57 keV.  We have constrained this line to be at most 1\% of the strength of the unresolved Ar {\sc XVII} triplet at 3.12 keV, but must consider the physical situation required to maximize the 3.62 keV DR line. In thermal equilibrium, the maximum strength of this line is 4\% of the Ar {\sc XVII} triplet, albeit at a temperature where the expected emission is negligible.  One might also consider an extreme non-equilibrium situation with cold electrons that are unable to collisionally excite any Ar  {\sc XVII} lines, but dielectronic recombination is still possible.  Examining the satellite line data in AtomDB, taken from \citet{VS80}, shows that even in this case the maximum ratio is only 7\%, as there are DR satellite lines at the energies of the Ar  {\sc XVII} triplet as well and these lines would also be excited in such a case.  While not physically impossible if there was a significant and unexpected error in the atomic physics calculations, we have no reason to believe this has occurred.

One other possibility is a radiative recombination continuum (RRC) edge feature. The S  {\sc XVI} recombination edge lies at 3.494 keV, and if it was bright enough, it might fill in some of the flux in this region. However, we note that the use of the no-line model has already included the RRC feature under equilibrium conditions. Producing a stronger RRC would require a sharp drop in the electron temperature, while retaining a large, fully stripped S$^{16+}$ population. Also, if the temperature drops below $\approx 0.1$keV, the RRC feature becomes very narrow and will be an order of magnitude less powerful at 3.57 keV compared with right at the edge: this shape is not consistent with our observations. Similarly, at hotter temperatures, the RRC becomes almost constant with energy once above the edge. If the RRC was really there, we would expect to see a residual at about 3.6 keV, which we do not. Finally, we note that the edge is 50-80eV from our proposed line, which makes an unlikely source of the line.

We also note that our assumptions regarding relative line strengths have assumed that the ICM is in thermal equilibrium or close to it. Charge exchange (CX) between highly ionized ions and neutral hydrogen or helium could also create X-ray emission lines with different ratios \citep{Smith12}. This could affect our assumption of equilibrium line ratios, although we have included a substantial range around the equilibrium values.  It is important to note that these CX lines are not new, but rather the same lines occurring in different ratios.  Due to its large cross section relative to electron excitation rates, astrophysical CX can occur only in a thin sheet where ions and neutrals interact directly, limiting its total emission relative to the large ICM volume.  In certain cases, such as the core of the Perseus Cluster where many neutral filaments are known, it is possible that CX could be large enough to create a small fraction of the total X-ray emission, although it would not create or enhance a line at 3.57 keV or the DR line at 3.62 keV.  CX could not dominate the overall emission, however, as it would also create Fe {\sc XVII} and other lines that are not detected.

\subsection{Sterile neutrino decay line?}

An interesting interpretation of the line is the decay signature of the
sterile neutrino, a long-sought dark matter particle candidate
(\citet[e.g.,][]{boyarsky2009}, see our Section 1).
The mass of the sterile neutrino would be double the decay photon energy,
$m_s=$7.1 keV. The line flux detected in our full sample corresponds
to a mixing angle for the decay $\sin^{2}(2\theta)\sim 7 \times10^{-11}$. This
value is below the upper limits
placed by the previous searches, shown in Figure \ \ref{fig:limitsAbaza}. Our
detections from the stacked \xmm MOS observations of galaxy clusters are
shown with a star in red in that figure. Figure \ref{fig:limits} shows
the detections and upper limits
we obtained from our various subsamples we used in this work (based on
the included cluster masses and distances), as well as
a comparison with the previous upper limit placed using the Bullet cluster
by \citet{boyarsky2008} at 3.57 keV, which is the most relevant
earlier constraint for us. Since the mixing angle is a universal
quantity, all the subsample measurements must agree.

The line in the subsample of  69 fainter clusters (full sample sans Perseus,
Coma, Ophiuchus and Centaurus) corresponds to a mixing angle that is
consistent with the full sample; the same is seen (though with a mild
$1.5\sigma$ tension) for the subsample of bright nearby clusters
Coma+Centaurus+Ophiuchus. However, the brightness of the new line in the
\xmm spectrum of Perseus corresponds to a significantly higher mixing angle
than that for the full sample (by factor of eight in
terms of the line flux for the MOS spectrum), which
poses a problem in need of further investigation.  In
  principle, an
  enhanced flux of the detected line in the Perseus 
  spectra may be due to a dark matter filament along the
  line of sight, though it would have to be rather extreme, so it is
  wise to look for more plausible explanations.

We tried to excise the central 1$^{\prime}$ region of the Perseus
cluster, to see if the flux originates in the cool core of the
cluster. Indeed, this decreased the flux in the line in half and removed
most of the tension with the other measurements. However, this suggests that
either some of the line flux is astrophysical in origin (at least in
Perseus), or the
cool gas in the core of the cluster affects our
ability to measure the continuum and the fluxes of the nearby K
\textsc{xviii} and Ar \textsc{xvii} lines, in the end resulting in an
overestimate of the flux of our detected line. It appears that in
Perseus, there is an anomalously strong line at the position of the Ar
\textsc{xvii}
DR line at 3.62 keV.

With this knowledge, we have tried to add this anomalous 3.62 keV line
in the model for the full sample,
where we have the most statistically significant line detection. The
additional line is still required, albeit at a lower significance and
a slightly lower energy of $3.55\pm0.03$ keV. Note that the
sample of bright clusters is dominated by the emission from the cool
cores of the Ophiuchus and Centaurus Clusters, if this Ar 3.62 keV line
anomaly is typical of cool cores, they may also be affected. However,
freeing the flux of the 3.62 keV line in the MOS full-sample fit did
not require additional contribution from clusters other than Perseus,
though the constraints are obviously weak.

The radial distribution of the flux of this line should be investigated
further in the nearby bright clusters, including those with and
without cool cores.
 \begin{figure}
 \centering
\hspace{-4mm}\includegraphics[width=9.cm, angle=0]{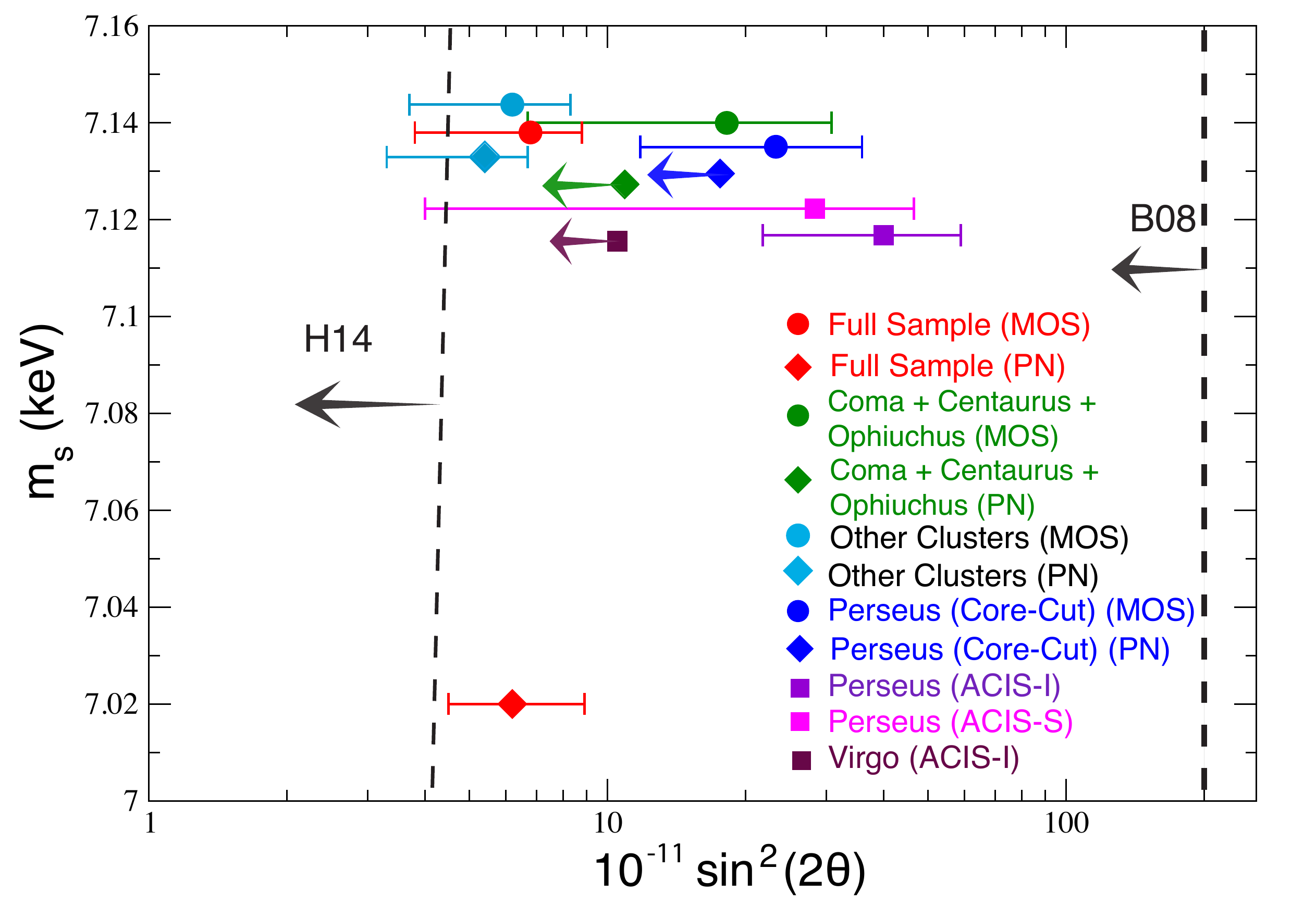}
\caption{Sterile neutrino mass and mixing angle measurements and 
upper limits obtained from the different samples used in this study. 
The comparison of our stacking method with the limits placed by the 
single well-exposed Bullet Cluster at 3.57 keV \cite{boyarsky2008}
and \cite{horiuchi2013}
are also shown and marked with ``B08'' and ``H14" in the figure, respectively. 
The error bars and upper limits are in the 90\% confidence level.}
\label{fig:limits}
\end{figure}

We note that even if the sterile neutrino interpretation of the
emission line is correct, this detection would not necessarily imply
that all dark matter is composed of these particles. Assuming a
standard cosmological history below a temperature of a few hundred
MeV, sterile neutrinos would be produced by oscillations with active
neutrinos at an abundance determined by the mass and mixing angle
\citep[e.g.][]{dodelson1994, kusenko2009}. Accounting for the
increase in mixing angle that would be inferred for a dark matter
fraction in sterile neutrinos less than unity, we find that this
fraction is $\sim$13\%-19\% based on the methods in \citet{abazajian2006} and \citet{asaka2007}
-- and cannot exceed 26\% based on the absolute lower
bound distorted wave production estimate in  \citet{asaka2007}.

This implies that either (1) sterile neutrinos are a subdominant
component of dark matter, (2) sterile neutrinos are predominantly
produced by some other mechanism, or (3) the emission line originates
from some other radiatively decaying light dark matter candidate such
as moduli dark matter \citep{kusenko2013}. The Shi-Fuller mechanism
is one of the possible production mechanisms for the sterile neutrino
dark matter interpretation of this detection. The implications of the
detection for structure formation in cosmological small scales are
discussed in detail in \citep{abazajian2014}.

They may also be produced by means that do not involve oscillations,
such as inflaton or Higgs decay \citep{kus06,st06,pk08,kusenko2009}, although there may still
be sufficient mixing to provide an observable radiative decay
signal. This detection is consistent with 100\% of dark matter composed
of sterile neutrinos produced by these mechanisms, as well as by the
split seesaw mechanism \citep{kty10}. Even in
this case, some sterile neutrinos would be produced by non-resonant
oscillations. However, based again on the calculations in \citet{abazajian2006} and \citet{asaka2007}, only 
$\sim$1\% -3\% of the sterile neutrino
abundance (with an upper limit of 7\%) would be accounted for in this
way for a sterile neutrino with mass of 7.1 keV and a mixing angle
corresponding to $\sin^{2}(2\theta)\ \sim 7\, \times\, 10^{-11}$.

 \begin{figure}
 \centering
\hspace{-4mm}\includegraphics[width=9.cm, angle=0]{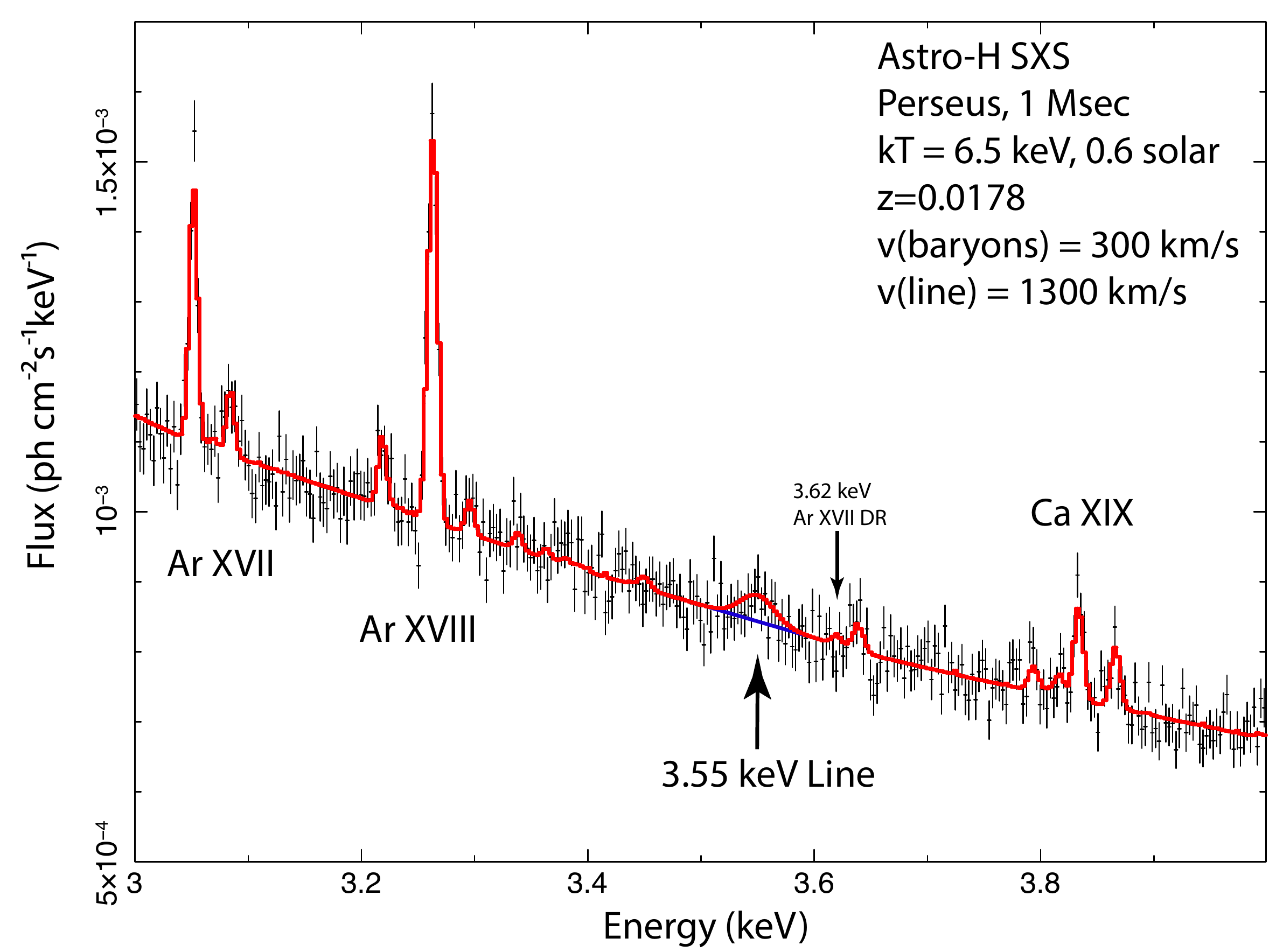}
\caption{1 Ms Astro-H Soft X-ray Spectrometer (SXS) simulations of the Perseus Cluster. The line width corresponds to line of sight velocity dispersion of 1300 km s$^{-1}$. The figure shows that the decaying dark matter line broadened by the virial velocities of dark matter particles will easily be distinguished from the plasma emission lines which are broadened by turbulence in sufficiently
  deep observations of the Perseus Cluster.}
\label{fig:sims}
\end{figure}

Our result must be verified using a variety of X-ray instruments,
X-ray emitting dark matter dominated objects, methods of data
reduction, background subtraction, and statistical techniques to
investigate the interpretation of this line. The future
high-resolution Astro-H observations will be able to measure the
broadening of the line, which will allow us to measure its velocity
dispersion. To detect a dark matter decay line, which is much weaker
than the plasma lines will require a significantly long exposure. We
performed 1 Ms Astro-H SXS simulations of the Perseus Cluster assuming
that the width (15 eV) of the dark matter decay line is determined by
the virial velocities of dark matter particles of 1300 km
s$^{-1}$. Figure \ref{fig:sims} shows that the broader dark matter line
will be easily distinguished from the plasma emission lines, which
are only broadened by the turbulence in the X-ray emitting gas.

\section{Caveats}
\label{sec:caveats}

As intriguing as the dark matter interpretation of our new line is, we
should emphasize the significant systematic uncertainties affecting the line
energy and flux in addition to the quoted statistical errors. The line is
very weak, with an equivalent width in the full-sample spectra of only $\sim
1$ eV. Given the CCD energy resolution of $\sim 100$ eV, this means that our
line is a $\sim 1$\% bump above the continuum.  This is why an accurate
continuum model in the immediate vicinity of the line is extremely
important; we could not leave even moderately significant residuals
unmodeled. To achieve this, we could not rely on any standard plasma
emission models and instead had to let all the tabulated lines free
(including their fluxes, energies and widths, within reasonable bounds), as
described in Section 3.

This approach results in a very large number of parameters to fit
simultaneously, among which are the line energies and widths that
notoriously cause problems for the statistic minimization algorithms. It was
difficult to make \xspec find absolute minima; the convergence of all of the
reported fits had to be verified by manually varying key parameters and
refitting using different minimization algorithms. Nevertheless, it is not
inconceivable that some of our fits did not find an absolute $\chi^2$
minimum.

Another consequence of the very low line flux is illustrated by the lower
panels in Figure\ \ref{fig:epicfits}, showing the effective area curves. There
are variations in these curves at a $\sim 1$\% level, some of which may be
at the position of our line. If these features are not accurately
calibrated, they may cause artifacts that would be falsely detected as
spectral features. This is where our stacking approach makes a big
difference --- those effective area variations are almost completely smeared
out in the ``all other'' sample, for which the instrument response was
averaged over a large range of cluster redshifts (lowest panels), and the
line is still present in those spectra. Nevertheless, for the nearby
subsamples, these variations represent a systematic uncertainty.

Finally, there is a discrepancy, albeit at a marginal statistical
significance, between the line energy obtained for the full sample by MOS
and PN (see Table\ \ref{table:flux}).
The energies come into agreement if we allow the flux of the Ar 3.62 keV
line in the MOS full-sample spectrum to increase many times above its
theoretical upper limit, to a value corresponding to that observed in
Perseus (which is part of the full sample); see Figure\
\ref{fig:steppar}. (The Perseus
contribution to the PN full sample is smaller, and the 3.62 keV line would
not affect the fit.) The new line in MOS becomes slightly fainter and less
statistically significant, with a lower energy. However, the energy of both
MOS and PN lines then becomes consistent with that of the weak K line at
3.515 keV. The expected flux of the K line (Table \ref{table:flux}) is
10--20 times lower than the observed line fluxes for both detectors; if
something is wrong by that much in the thermal plasma model, this would
provide an alternative explanation for our line. The MOS full-spectrum line
is also consistent with just the 3.62 keV line at the anomalously high flux
that accounts for Perseus, if the new line is at the higher end of its error
interval.  However, this would not explain the PN line.

Disentangling these possibilities is impossible at the present energy
resolution and has to wait until the launch of \astroh.  The other
systematic uncertainties mentioned above also have the low-energy
resolution as their root cause.

In the meantime, we are planning to analyze other independent
stacked samples from \chandra\ and \suzaku\ and their various
subsamples.  In particular, it would be very interesting to pinpoint the origin, if not the
nature, of the anomalous Ar 3.62 keV line.  We will also divide the
clusters spatially into annuli, to study the radial distribution of
the emission and determine if the flux is proportional to the mass
density squared, in order to distinguish the thermal plasma emission
from a dark matter decay line.

Soon after this manuscript has been submitted and posted on the
  preprint archive, a preprint by \cite{boyarsky2014} was posted, which
  reported an independent detection of a line at the same energy in the \xmm
  spectra of M31 and the outskirts of the Perseus Cluster. They excluded the
  central Perseus pointing that we used in our \xmm analysis; thus, their
  dataset has no overlap with ours. Their X-ray line flux, when converted to
  the sterile neutrino decay rate, is in agreement with ours, providing an
  independent confirmation of the line existence (and perhaps even of its
  dark matter origin).

\section*{Acknowledgements}

The authors thank Alexey Vikhlinin for extensive discussions, cross-checking
the results and spotting several errors; Douglas Finkbeiner and Serkan Cabi for useful
discussions; Christine Jones, Signe Riemer-Sorensen, Alexander Kusenko, and
the anonymous referee for useful comments on the draft; and Kevork
Abazajian, and Shunsaku Horiuchi for providing the limits from their M31
analysis. E.B. also thanks Adrian Batu Gerard for patiently waiting for the submission 
to be born. 
EB was supported in part by Chandra grant AR0-11020X and
GO1-12104X. A.F. and R.S. were supported in part by NASA ADAP grant NNX12AF44G.
S.W.R. was supported by the Chandra X-ray Center through NASA contract
NAS8-03060, and the Smithsonian Institution.


\end{document}